\documentclass[preprint,10pt,5p,twocolumn]{elsarticle} 

\usepackage{amssymb}
\usepackage{amsmath}
\usepackage{url}
\usepackage{rotating}
\usepackage{mathtools, cuted}

\usepackage{textcomp}

\journal{Computer Physics Communications}

\newcommand{\vb}[1]{\mathbf{#1}}
\newcommand{\vortex}[1]{\textsc{vortex}}
\newcommand{\vortexp}[1]{\textsc{vortex-p}}
\newcommand{\dd}[0]{\mathrm{d}}
\usepackage{esint}

\begin{document}

\begin{frontmatter}

\title{\vortexp{}: a Helmholtz-Hodge and Reynolds decomposition algorithm for particle-based simulations}

\author[inst1]{David Vall\'{e}s-P\'{e}rez} \corref{cor1} \ead{david.valles-perez@uv.es} 
\author[inst1,inst2]{Susana Planelles}
\author[inst1,inst2]{Vicent Quilis}
\author[inst3]{Frederick Groth}
\author[inst3]{Tirso Marin-Gilabert}
\author[inst3,inst4]{Klaus Dolag}
\cortext[cor1]{Corresponding author}

\address[inst1]{Departament d'Astronomia i Astrof\'{\i}sica, Universitat de
  Val\`encia, E-46100 Burjassot (Val\`encia), Spain}
\address[inst2]{Observatori Astron\`omic, Universitat de Val\`encia, E-46980
  Paterna (Val\`encia), Spain}
\address[inst3]{Universitäts-Sternwarte, Fakultät für Physik, Ludwig-Maximilians-Universität München, Scheinerstr.1, 81679 München, Germany} 
\address[inst4]{Max-Planck-Institut für Astrophysik, Karl-Schwarzschild-Straße 1, 85741 Garching, Germany}

\begin{abstract}
Astrophysical turbulent flows display an intrinsically multi-scale nature, making their numerical simulation and the subsequent analyses of simulated data a complex problem. In particular, two fundamental steps in the study of turbulent velocity fields are the Helmholtz-Hodge decomposition (compressive+solenoidal; HHD) and the Reynolds decomposition (bulk+turbulent; RD). These problems are relatively simple to perform numerically for uniformly-sampled data, such as the one emerging from Eulerian, fix-grid simulations; but their computation is remarkably more complex in the case of non-uniformly sampled data, such as the one stemming from particle-based or meshless simulations. In this paper, we describe, implement and test \vortexp{}, a publicly available tool evolved from the \vortex{} code, to perform both these decompositions upon the velocity fields of particle-based simulations, either from smoothed particle hydrodynamics (SPH), moving-mesh or meshless codes. The algorithm relies on the creation of an ad-hoc adaptive mesh refinement (AMR) set of grids, on which the input velocity field is represented. HHD is then addressed by means of elliptic solvers, while for the RD we adapt an iterative, multi-scale filter. We perform a series of idealised tests to assess the accuracy, convergence and scaling of the code. Finally, we present some applications of the code to various SPH and meshless finite-mass (MFM) simulations of galaxy clusters performed with \textsc{OpenGadget3}, with different resolutions and physics, to showcase the capabilities of the code.
\end{abstract}


\end{frontmatter}

\section{Introduction}
\label{s:intro}

Astrophysical flows on vastly different scales, ranging from the interior of stars to galaxy clusters, are turbulent in nature, in the sense of them being highly irregular both in space and time (see, e.g., \citep{Brandenburg_2011} for a review on astrophysical turbulence). Turbulent motions in highly-compressible media manifest in the emergence of compressive and solenoidal velocity modes spanning several orders of magnitude in scales.

As a particular example, turbulent motions arise during the assembly of cosmic structures and, in particular, throughout the formation and evolution of galaxies and galaxy clusters, emerging as a central phenomenon for the understanding of the physics of the baryonic component in such structures \citep{Dolag_2005, Subramanian_2006, Miniati_2014, Iapichino_2017}. In galaxy clusters, the emergence of turbulence in the intracluster medium (ICM) is tightly linked to the presence of mergers and smooth accretion \citep{Vazza_2017, Valles-Perez_2021_MNRAS}, but can also be triggered by the sloshing of a cool core \citep{Markevitch_2001}, by galaxy motions \citep{Faltenbacher_2005, Ruszkowski_2011}, or by AGN outflows \citep{Quilis_2001, Gaspari_2015}, amongst other mechanisms. In turn, once present in the ICM, turbulent motions are responsible for providing a significant contribution to the non-thermal pressure resisting gravitational collapse \citep{Dolag_2005, Subramanian_2006, Vazza_2018, Angelinelli_2020} and may hence bias mass measurement relying on the hydrostatic equilibrium assumption \citep{Biffi_2016, Bennett_2022}. Additionally, small-scale plasma motions are responsible for the amplification of magnetic fields \citep{Porter_2015, Beresnyak_2016, Vazza_2018_magnetic}, the acceleration of cosmic rays \citep{Fujita_2003, Cassano_2005}, and may impact processes at even smaller scales, such as star formation \citep{Federrath_2013_sfr, Federrath_2016}.

Due to their intrinsic multi-scale character, properly resolving turbulent flows in astrophysical computational fluid dynamics (CFD) applications demands high numerical resolutions. While the most consistent approach for accurately resolving the turbulent cascade involves the usage of uniform, fix grids \citep[e.g.,][]{Kritsuk_2007, Kritsuk_2011}, other more computationally feasible strategies are usually followed, either by means of Eulerian Adaptive-Mesh Refinement (AMR) codes \citep{Iapichino_2008a, Iapichino_2008b, Vazza_2010, Vazza_2011}, where the grid resolution can be increased locally based on different criteria, or with particle-based methods, whose Lagrangian nature makes them particularly attractive for structure formation simulations. While traditional \citep{Gingold_1977, Lucy_1977} and improved versions \citep{Beck_2016} of the Smoothed Particle Hydrodynamics (SPH) method are customarily used to study the properties of turbulent flows \citep{Dolag_2005, Sijacki_2006, Valdarnini_2011, Valdarnini_2019}, in the recent years, there has been a growing interest in meshless finite volume (MFV) and finite mass (MFM) methods \citep{Hopkins_2016, Groth_2023}, which combine some of the advantages of AMR and SPH codes.

Despite the fact that these alternative approaches alleviate to a considerable extent the computational burden associated to having to sample the whole domain at constant resolution, they also pose serious complications for posterior analyses and post-processing of the simulation data. One of the main computationally-expensive analyses corresponds to the process of splitting the (turbulent) velocity field in its compressive and solenoidal components, formally known as the Helmholtz-Hodge decomposition (HHD). In the context of galaxy cluster physics, this is exceedingly interesting since these two components differ significantly in their spectra \citep{Federrath_2013, Miniati_2014, Miniati_2015}, their spatial distribution \citep{Vazza_2017, Federrath_2009, Iapichino_2011}, and their role. Regarding the latter, the small-scale solenoidal velocity is responsible, e.g., for part of the amplification of primordial magnetic fields through the action of dynamo mechanisms \citep{Porter_2015, Vazza_2018_magnetic, Dominguez-Fernandez_2020}; while compressive supersonic motions play a central role in shaping the thermodynamic properties of the ICM \citep{Miniati_2000, Planelles_2013, Porter_2015}. Both, in grid-based and in particle-based simulations, this decomposition is customarily carried out in Fourier space \citep[e.g.,][]{Vazza_2017, Valdarnini_2019}, where it reduces to performing a linear projection and the overall cost scales as $\mathcal{O}(N \log N)$ by employing the Fast Fourier Transform (FFT) algorithm. Nevertheless, this comes at the cost of assigning the velocity field onto a fix grid, potentially losing signal in high-density regions.

Another fundamental step in the analysis of a simulation of turbulent flows may be the Reynolds decomposition (RD), i.e., the process of extracting the turbulent part of the velocity field. While earlier works in the literature used a fix-length low-pass filter to define a bulk velocity field \citep{Dolag_2005, Vazza_2009}, from whose residual the turbulent velocity was defined, it soon became obvious that the complexity of the ICM (hosting substructures, internal shock waves, etc.) makes the problem more complicated. With this motivation, using fix-grid simulations, \citet{Vazza_2012, Vazza_2017} proposed an iterative algorithm to locally constrain the outer scale of turbulence, which was subsequently extended to SPH \citep{Valdarnini_2019} or AMR \citep{Valles-Perez_2021_MNRAS}.

In \citet{Valles-Perez_2021_CPC} we introduced, tested and publicly released \vortex{}, a post-processing code for patch-based AMR simulations for performing the HHD of an input vector field. Subsequently, in \citet{Valles-Perez_2021_MNRAS}, we added a module for optionally performing a RD following the algorithm of \citet{Vazza_2012}. 

In this paper, we introduce \vortexp{}, a public tool for performing the HHD and the RD of vector fields of particle-based simulations, which is applicable not only to SPH simulations, but also to MFM, moving-mesh or any kind of meshless or non-uniform mesh simulation data, by means of the creation of an ad-hoc hierarchy of AMR grids covering the domain of interest.

The rest of the manuscript is organised as follows. In Sec. \ref{s:method}, we describe the algorithmic details of the method (especially focusing on what is new with respect to the original version of \vortex{}, as well as a description of the input data, the output formats and the free parameters). Our implementation of \vortexp{} is thoroughly tested in Sec. \ref{s:tests}, while different applications of the code to various simulations of galaxy clusters are presented in Sec. \ref{s:application}. Finally, we further summarise and discuss our conclusions in Sec. \ref{s:conclusion}.

\section{Description of the method}
\label{s:method}

The basic idea behind the algorithm presented here relies on the interpolation of the smoothed velocity field described by the particle distribution onto an ad-hoc AMR mesh hierarchy, from which the decomposition can be obtained as in the \vortex{} code (summarised in Sec. \ref{s:method.reference}). Subsequent sections cover the required input data for the code (Sec. \ref{s:method.input}), the mesh creation process (Sec. \ref{s:method.mesh}), the velocity interpolation procedure (Sec. \ref{s:method.interpolation}), the shock flagging scheme necessary for the RD (Sec. \ref{s:method.shock}) and the results of the process (Sec. \ref{s:method.output}). An OpenMP-parallelised version of the code is publicly available and documented.\footnote{The code can be found in \url{https://github.com/dvallesp/vortex-p}. The documentation can be accessed in \url{https://vortex-particles.github.io}.} Table \ref{tab:parameters} contains a summary of the free parameters of the code.

\begin{table*}[h]
    \small 
    \centering
    \caption{Summary of the main parameters that can be tuned to run \vortexp{}. The different blocks in the table contain the parameters that determine the creation of the mesh, the assignment of velocities from particles onto the grid, the solution of the elliptic equations yielding the Helmholtz-Hodge decomposition, and the Reynolds decomposition.}
    \begin{tabular}{lc|l}
        \hline
         Parameter & (Symbol) & Description and remarks  \\ 
         \hline
         \multicolumn{3}{c}{Mesh creation} \\ \hline
         Region of interest & & Specified as a orthohedron by its lower and upper limits along \\
         & & $\quad$ the three spatial dimensions.  \\
         Base grid size & $N_x$ & Typically set to $N_x \sim \sqrt[3]{N_\mathrm{part}}$. \\
         Number of refinement levels & $n_\ell$ & Peak resolution will be $L/(N_x\times 2^{n_\ell})$, typically set to match \\
         & & $\quad$ the best resolution of the simulation. \\
         Refinement threshold & $n_\mathrm{part}^\mathrm{refine}$ & Number of particles to flag a cell as `refinable'.\\
         Minimum size of the patch & $N_\mathrm{min}^\mathrm{patch}$ & Only patches with minimum dimension above this are kept.\\
        
         \hline
         \multicolumn{3}{c}{Velocity interpolation} \\ 
         \hline
         Number of neighbours & $N_\mathrm{ngh}$ & Minimum number of particles within the kernel around a cell. \\ 
         Kernel & & Either Monaghan's M$_4$, or \citeauthor{Wendland_1995}'s C$^4$ or C$^6$. \\
         
         \hline 
         \multicolumn{3}{c}{Poisson solver} \\ 
         \hline
         SOR precision parameter & $\epsilon_\mathrm{SOR}$ & SOR is considered to converge when the maximum relative \\
         & & $\quad$variation of $\phi$ falls below this parameter. \\
         Max. num. of iterations & $N_\mathrm{SOR} $& Preventive stopping condition in case SOR does not converge. \\ 
         Border for AMR patches & & Patches are extended with this number of ghost cells to avoid \\
         & & $\quad$boundary effects. (Default: 2)\\

         \hline
         \multicolumn{3}{c}{Multi-scale filter} \\ 
         \hline
         Apply the multi-scale filter & & Boolean (yes/no).\\ 
         Filter tolerance and growing step & $\Delta_\mathrm{tol}, \, \chi$ & See \citet{Valles-Perez_2021_MNRAS}, \S2.2. \\
         Strong shock threshold & $\mathcal{M}^\mathrm{thr}$ & A shocked cell with $\mathcal{M} > \mathcal{M}^\mathrm{thr}$ will stop the iterations to \\
         & & $\quad$determine $\mathcal{L}(\vb{x})$. If specified, $\alpha^\mathrm{thr}$ and $(\nabla \cdot \vb{v})^\mathrm{thr}$ are ignored. \\
         Threshold on artificial viscosity & $\alpha^\mathrm{thr}$ & If $\mathcal{M}$ is not supplied, strong shocks are located by imposing \\ 
         Threshold on velocity divergence & $(\nabla \cdot \vb{v})^\mathrm{thr}$ & $\quad$ $\alpha>\alpha^\mathrm{thr}$ and $\nabla \cdot \vb{v} < (\nabla \cdot \vb{v})^\mathrm{thr}$.
 \\ 
         \hline
    \end{tabular}
    \label{tab:parameters}
\end{table*}

\subsection{The reference version of \vortex{}}
\label{s:method.reference}

In \citet{Valles-Perez_2021_CPC}, we introduced and publicly released a new algorithm for performing the HHD of a vector field defined on a patch-based AMR grid hierarchy.\footnote{\url{https://github.com/dvallesp/vortex}.} The algorithm solves for a scalar potential, $\phi$, and a vector potential, $\vb{A}$, from which the compressive and solenidal components are derived, respectively, by taking the gradient and the curl,

\begin{equation}
    \vb{v_\mathrm{comp}} = -\nabla \phi, \qquad \vb{v_\mathrm{sol}} = \nabla \times \vb{A}.
    \label{eq:definition_potentials}
\end{equation}

The scalar potential, as well as each Cartesian component of $\vb{A}$ (under the Coulomb gauge, $\nabla \cdot \vb{A} = 0$), are found as the solutions to elliptic partial derivative equations, whose sources are the negative divergence and curl of the input velocity field,

\begin{equation}
    \nabla^2 \phi = - \nabla \cdot \vb{v}, \qquad 
    \nabla^2 \vb{A} = - \nabla \times \vb{v}.
\end{equation}

\vortex{} uses a combination of FFT methods for the base grid, and iterative solvers (in particular, the successive overrelaxation method, SOR) for the refinement patches, where the boundary conditions are set by the coarser levels.

Subsequently, in \citet{Valles-Perez_2021_MNRAS}, we extended \vortex{} to include the possibility to perform a Reynolds decomposition (RD), i.e., extracting the turbulent part of the velocity field, prior to the HHD. This was achieved by extending to non-uniform resolutions the multi-scale turbulent filter initially introduced by \citet{Vazza_2012, Vazza_2017}, which determines a local filtering length by iteratively constraining the outer scale of turbulence and by the distance to the nearest influential shock.

\subsection{Input data}
\label{s:method.input}

In order to perform the HHD of the velocity field, \vortexp{}  needs to be fed with a set of $N_\mathrm{part}$ particles (for SPH or MFM simulations) or mesh-generating points (for moving-mesh codes outputs), with given positions $\{\vb{x_i}\}_{i=1}^{N_\mathrm{part}}$, velocities $\{\vb{v_i}\}_{i=1}^{N_\mathrm{part}}$, and smoothing lengths $\{h_i\}_{i=1}^{N_\mathrm{part}}$. Except where stated otherwise, all the processes regarding the HHD and the RD do not depend on the input units. 

Additionally, if the particle mass is not uniform and the velocity field is to be mass-weighted (volume-weighted), particle masses (masses and densities) also need to be supplied. Finally, for performing the RD, some additional variable to flag strong shocks is needed, whose possibilities are discussed below in Sec. \ref{s:method.shock}.

\vortexp{} allows to select a particular domain for performing the computations, in such a way that only particles in this domain are considered. This can be used, for instance, to restrict the decomposition to a specific object within a larger simulation domain. However, since the HHD appears as the solution of elliptic equations, care must be taken with the boundary conditions, which are always assumed to be periodic due to the usage of FFTs for the base grid. In practical terms, this only biases the decomposed velocities in a few cells around the boundaries, implying that it is generally advisable to specify a domain reasonably larger than the object of interest (see \ref{s:appendix.periodicity}).

\subsection{Mesh creation}
\label{s:method.mesh}

\vortexp{} employs a custom patch-based AMR implementation, following the general description of \citet{Berger_1989} and inherited from that included in the hydrodynamical code \textsc{MASCLET} \citep{Quilis_2004}. The domain of interest, with longest dimension $L$, is mapped with a uniform (base) grid of $N_x^3$ cells with resolution $\Delta x = L / N_x$. Regions of high particle number density are recursively refined in rectangular patches, up to a maximum number of refinement levels $n_\ell$, each time halving the cell sidelength. Although they are free parameters, a common choice is to set $N_x$ to the closest power of 2 so that $N_x = \sqrt[3]{N_\mathrm{part}}$, in such a way that the base grid matches the mean interparticle separation; while $n_\ell$ can be set so that the peak AMR resolution matches the smallest smoothing length (or moving-mesh cell size).

The process of generation of the AMR mesh hierarchy for a given particle distribution is generally similar to that implemented in the halo finder \textsc{ASOHF} \citep{Valles-Perez_2022}, and it is based on a single threshold on the particle count, $n_\mathrm{part}^\mathrm{refine}$. Cells with higher particle count are flagged as refinable, and the mesh-creation routine covers these regions as efficiently as possible with a set of possibly-overlapping\footnote{We note for the reader that, unlike octree-based AMR implementations, patches at a given refinement level may overlap. This serves a double purpose in \vortexp{}. On the one hand, this refinement strategy is well suited for structure formation simulation data (which were one of the original application focuses of the code): using a patch-based approach, refinement domains can be centred on the structures of interest. On the other hand, despite implying a less trivial data structure, a patch-based approach reduces considerably the emergence of errors due to the interpolation of boundary conditions in coarse-fine interfaces (which are geometrically more complex in an octree description), since large, contiguous refinement domains reduce the surface-to-volume ratio.} rectangular cuboids, that will become the refinement patches at level $\ell=1$ with resolution $\Delta x_\ell = \Delta x / 2^\ell$. For reasons of memory layout and efficiency, not all these regions are accepted, but only the ones whose number of child (refined) cells along the smallest of their dimensions exceeds a lower threshold $N_\mathrm{min}^\mathrm{patch}$. Each of these patches constitutes an independent domain, subject to the boundary conditions imposed by the immediately coarser grids. The process is subsequently iterated up to the specified number of refinement levels, keeping an approximately constant particle count per (non-refined) cell. For a more thorough description of the mesh-creation algorithm given a set of refinable cells, we refer the reader to the description of the equivalent procedure in \textsc{MASCLET} \citep[][their \S 3.1]{Quilis_2004}.

\subsection{Velocity assignment onto the grid}
\label{s:method.interpolation}

On top of the original \vortex{} implementation, and besides the mesh generation, which sets the level of detail with which the computations on the velocity field can be ultimately performed, the core of \vortexp{} relies on the method used to assign the velocity field defined by the particle velocities $\{\vb{v_i}\}_{i=1}^{N_\mathrm{part}}$ at locations $\{\vb{x_i}\}_{i=1}^{N_\mathrm{part}}$ to the AMR grid. The desired decomposition must comply with the following precepts:

\begin{itemize}
    \item It must represent the values of the underlying velocity field at cell centres. This has to be the case because the HHD algorithm solves for the velocity components using finite differences.
    \item It must be smooth, since Helmholtz decomposition theorem requires the vector field to be of class $\mathcal{C}^1$. While actual velocity fields developed in simulations might develop discontinuities (e.g., in shocks), this does not pose a serious problem since there is always a continuous representation of the discrete velocity field. However, when assigning the velocity field to the grid, it is important to avoid velocity jumps and fluctuations due to sampling noise, i.e., due to the particle distribution. This is especially important since, by using a patch-based AMR implementation, it might happen that a refinement patch ends up containing a region of very low particle density.
    \item At the same time, it needs to preserve as much detail as possible from the original velocity field.
\end{itemize}

The solution adopted in \vortexp{}, fulfilling these requirements, is to construct the velocity field, at the location of a cell centre $\vb{x}$ in a level $\ell$ grid, as an average of the particle individual velocities, $\vb{v}_i$, over a length $h(\vb{x}) = \max (l_{N_\mathrm{ngh}}, \Delta x_\ell)$, where $l_{N_\mathrm{ngh}}$ is the distance to the $N_\mathrm{ngh}$-th nearest neighbour of the cell centre $\vb{x}$ ($N_\mathrm{ngh}$ being a free parameter), while $\Delta x_\ell$ is the cell size at the AMR level considered:

\begin{equation}
    \vb{v}(\vb{x}) = \frac{\sum_{i \in \mathrm{ngh}} \vb{v}_i \, W(|\vb{x} - \vb{x}_i|, h(\vb{x}))}{\sum_{i \in \mathrm{ngh}} W(|\vb{x} - \vb{x}_i|, h(\vb{x}))}.
\end{equation}

Here, $W(\cdot, \cdot)$ is a smoothing kernel. Currently implemented in \vortexp{} there are the cubic spline (M$_4$) kernel \cite{Monaghan_1985},

\begin{equation}
    W_{M_4}(q) = \left\{ \begin{aligned}
    &1 - \frac{3q^2}{2} + \frac{3q^3}{4} {\equiv \frac{1}{4}(2-q)^3 - (1-q)^3}, &0 \leq q \leq 1 \\
    &\frac{1}{4}(2-q)^3,  & 1 \leq q \leq 2
    \end{aligned}
    \right.,
    \label{eq:kernel_M4}
\end{equation}

\noindent and \citeauthor{Wendland_1995}'s \citep{Wendland_1995} C$^4$, 

\begin{equation}
    W_{C^4}(q) = 
    \left(1 - \frac{q}{2} \right)^6 \left(\frac{35q^2}{12} + 3q + 1 \right)
    ,
    \label{eq:kernel_C4}
\end{equation}

\noindent and C$^6$ kernels,

\begin{equation}
    W_{C^6}(q) = 
    \left(1 - \frac{q}{2} \right)^8 \left(4q^3 + \frac{25q^2}{4} + 4q + 1 \right)
    ,
    \label{eq:kernel_C6}
\end{equation}

\noindent where $q \equiv \frac{|\vb{x}-\vb{x_i}|}{h(\vb{x})/2}$ and $W(q\geq 2) =0$ in all three previous cases. Normalisations are unimportant here, and are deliberately omitted. {Additionally, also higher order kernels from the B-spline family, namely the M$_5$ quartic kernel,

\begin{strip}
\begin{equation}
    W_{M_5}(q) = \left\{ \begin{aligned}
    & \left(\frac{5}{2} - q \right)^4 - 5\left(\frac{3}{2} - q \right)^4 + 10 \left(\frac{1}{2} - q \right)^4, &0 \leq q \leq \frac{1}{2} \\
    & \left(\frac{5}{2} - q \right)^4 - 5\left(\frac{3}{2} - q \right)^4,                                      &\frac{1}{2} \leq q \leq \frac{3}{2} \\
    & \left(\frac{5}{2} - q \right)^4,                                                                         &\frac{3}{2} \leq q \leq \frac{5}{2}
    \end{aligned}
    \right.,
    \label{eq:kernel_M5}
\end{equation}
\end{strip}

\noindent with $q \equiv \frac{|\vb{x}-\vb{x_i}|}{2 h(\vb{x})/5}$, and the M$_6$ quintic,

\begin{equation}
    W_{M_6}(q) = \left\{ \begin{aligned}
    &(3-q)^5 - 6 (2-q)^5 + 15 (1-q)^5, &0 \leq q \leq 1 \\
    &(3-q)^5 - 6 (2-q)^5, &1 \leq q \leq 2 \\
    &(3-q)^5, &2 \leq q \leq 3 \\
    \end{aligned}
    \right.,
    \label{eq:kernel_M6}
\end{equation}

\noindent with $q \equiv \frac{|\vb{x}-\vb{x_i}|}{h(\vb{x})/3}$, are included, although we will concentrate the tests on the first three ones.
}

We note for the reader that, by construction, this procedure for interpolation from particles to the grid is not conservative, in the sense that the sum of a given quantity over the particles does not necessarily equate to the sum over the grid. The violation of the conservativeness is in the order of a few percents, as discussed in more detail in \ref{s:appendix.conservativeness}. However, since the quantity being interpolated is precisely the velocity field, which is not a conserved quantity (nor in the evolution sense, neither when varying resolution), this does not pose any actual conceptual nor practical problem for the algorithm. By this procedure, \vortexp{} gets a velocity assignment onto the AMR grid that ensures smoothness because of the overlapping of the kernels associated to contiguous cells, even in regions of low particle density. A more thorough discussion on the choice of different kernels or neighbour numbers is given through the Tests in Sec. \ref{s:tests}.

Given a distribution of $N_\mathrm{part}$ particles, the refinement criterion based on particle counts will generate\footnote{{This $\mathcal{O}(N_\mathrm{part})$ is to be seen only as a typical, order of magnitude scaling relation. In particular, the multiplicative constant will depend on several free parameters of the mesh creation, such as $n_\mathrm{part}^\mathrm{refine}$ and $N_\mathrm{min}^\mathrm{patch}$, and on the specific problem (which conditions the geometry of the refinable regions).}} $\mathcal{O}(N_\mathrm{part})$ cells, implying that a naive implementation of density assignment would prohibitively scale as $\mathcal{O}(N_\mathrm{part}^2)$. To bring this down to practical computing costs, the  current version of \vortexp{} makes use of a space-partitioning kd-tree implementation from the \textsc{Coretran} library\footnote{\url{https://github.com/leonfoks/coretran}.}. Building the tree implies an initial $\mathcal{O}(N_\mathrm{part} \log N_\mathrm{part})$ overhead, but the density assignment cost then drops to $\mathcal{O}(N_\mathrm{part} N_\mathrm{ngh} \log N_\mathrm{part})$, which is much cheaper than the naive implementation in virtually all practical situations.

\subsection{Multi-scale filter and strong shock identification}
\label{s:method.shock}

The multi-scale filtering algorithm to perform the RD, as introduced in Sec. \ref{s:method.reference} and thoroughly described in \cite{Valles-Perez_2021_MNRAS}, determines the local filtering scale, $\mathcal{L}(\vb{x})$, seeking convergence on the turbulent velocity field,

\begin{equation}
    \delta \vb{v}(\vb{x}) = \vb{v}(\vb{x}) - \langle \vb{v}\rangle_{\mathcal{L}(\vb{x})}(\vb{x})
    \label{eq:turbulent_velocity}
\end{equation}

\noindent when iteratively increasing the filtering scale, where $\langle \vb{v}\rangle_{\mathcal{L}(\vb{x})}(\vb{x})$ is the bulk velocity field at position $\vb{x}$,

\begin{equation}
    \langle\vb{v}\rangle_{\mathcal{L}(\vb{x})}(\vb{x}) = \frac{\iiint_{|\vb{x'}-\vb{x}|<\mathcal{L}(\vb{x})} w(\vb{x'},|\vb{x}-\vb{x'}|) \vb{v}(\vb{x'}) \dd^3 \vb{x'}}{\iiint_{|\vb{x'}-\vb{x}|<\mathcal{L}(\vb{x})} w(\vb{x'},|\vb{x}-\vb{x'}|) \dd^3 \vb{x'}}
    \label{eq:bulk_velocity}
\end{equation}

\noindent and $w(\cdot, \cdot)$ is an optional weighting function that could depend on the distance $|\vb{x}-\vb{x'}|$ and other properties (e.g., on the density at $\vb{x'}$, if the average is mass-weighted). While it is easy to implement any weighting scheme, currently available in \vortexp{} we give the option to use a mass-weighted bulk velocity computation, $w(\vb{x'}) = \rho(\vb{x'})$, or to set $w=1$, so that the bulk velocity is computed by locally applying a spherical top-hat smoothing,

\begin{equation}
    \langle\vb{v}\rangle_{\mathcal{L}(\vb{x})}(\vb{x}) = \frac{\iiint_{|\vb{x'}|<\mathcal{L}(\vb{x})} \vb{v}(\vb{x}+\vb{x'}) \dd^3 \vb{x'}}{\iiint_{|\vb{x'}|<\mathcal{L}(\vb{x})}\dd^3 \vb{x'}}.
\end{equation}

\noindent We refer the reader to \ref{app:filter.weight} for a comparison of both weighting schemes.

However, in the presence of abrupt gradients in the velocity field such as the ones generated by strong shock waves, this average may not converge when increasing $\mathcal{L}(\vb{x})$, because the shock surface separates two dynamically unconnected regions (in the sense that no information can propagate upstream of the shock). This is why \citet{Vazza_2012, Vazza_2017} suggested to limit the iterative procedure when a shocked cell, with Mach number above a certain threshold $\mathcal{M} > \mathcal{M}_\mathrm{thr} \sim 2$, enters the integration domain. This requires the code to be fed with information from a shock finder, or either to use some alternative method for flagging strong shocks.

In the original, grid-based version of \vortex{}, the former was the only available option. In \vortexp{}, while we recommend the option of feeding the code with Mach number data (if available, for instance because the simulation has been run together with an on-the-fly shock-finding scheme, e.g. \cite{Beck_2016_shock}, or because shocks have been identified by a dedicated tool in post-processing), we also offer the option to use a simplified, ad-hoc scheme to flag strong shocks if the former methods are not available. In the latter case, strong shocks are detected in a computationally-cheap way (only in SPH simulations) by establishing a lower threshold on artificial viscosity, $\alpha^\mathrm{thr}$, and an upper threshold on velocity divergence, $(\nabla \cdot \vb{v})^\mathrm{thr}$. While this procedure is not as accurate and introduces the only two dimensional free parameters, it overcomes the necessity of performing a more expensive shock-finding process within the code. A test on the validity of this simplified scheme can be found in \ref{app:filter.flag}.

\subsection{Output possibilities}
\label{s:method.output}

In general, there are two possibilities to save and use \vortexp{} results, that can be chosen via runtime flags.

\begin{itemize}
    \item Use the gridded data: the results of the decomposition ($\vb{v_\mathrm{comp}}$ and $\vb{v_\mathrm{sol}}$, or their small-scale filtered versions) on the AMR grid hierarchy can be written, together with additional variables describing the AMR mesh, the overlaps between grids, etc. To handle these data, the \vortexp{} package includes a \textsc{Python} module suited with functions to transparently read and handle these outputs, together with some example \textsc{Jupyter} notebooks.
    \item Reinterpolate the decomposed velocities back to particles: in this case, it must be borne in mind that the interpolated values correspond to the smoothed velocities at the location of each particle, rather than the original particle velocities. Additionally, while the compressive (solenoidal) velocity field defined on the grid is explicitly curl-free (divergence-free), once reinterpolated back to particles, the divergence- or curl-free condition is not exactly held in the SPH sense, but one must expect errors in the same order as the ones involved in the grid assignment (cf. Tests 1 and 2 in Sec. \ref{s:tests}).
\end{itemize}

\subsection{Code dependencies}
\label{s:method.dependences}

As stated previously, \vortexp{} requires the \textsc{Coretran} library to be installed in the system and properly linked in order to use its kd-tree implementation.

Additionally and in an optional manner, the FFTW library \citep{Frigo_2005} can be used to perform the Fourier transforms for the base grid in an efficient, parallel way. This can be omitted at compilation time, in which case \vortexp{} resorts to a less-efficient, serial implementation of the FFT algorithm, which still works in reasonable times (a few seconds) for base grid sizes up to $256^3$.

\section{Tests}
\label{s:tests}

The algorithm proposed in Sec. \ref{s:method} for the HHD and its implementation in \vortexp{} have been validated with a series of tests aimed to assess the robustness and convergence of the results of our method. The four tests presented in the following sections (Secs. \ref{s:tests.test1}-\ref{s:tests.test4}) are similar to the ones applied to the original code \cite{Valles-Perez_2021_CPC} and are aimed to analyse different aspects of the procedure, but the set-up needs to be substantially changed (Sec. \ref{s:tests.setup}) with respect to the original ones. Furthermore, the convergence of the method and its computational scalability are discussed in Secs. \ref{s:tests.convergence}
and \ref{s:tests.scalability}, respectively. We do not present here tests around the RD, since this decomposition is not unique and there is not a ground solution to be compared to. Instead, for the RD, we refer the reader to Sec. \ref{s:application} and \ref{app:filter}, where its results are shown in several applications.

\subsection{Test set-up}
\label{s:tests.setup}

The input vector fields for the test in the following sections need to be seeded on a particle distribution on a cubic domain of arbitrary side length $L=1$. In order to provide a situation that triggers a reasonably high level of refinement by the mesh-creation strategy (Sec. \ref{s:method.mesh}), we consider a set of $N_\mathrm{part}$ particles, divided in two groups:

\begin{itemize}
    \item A quarter of the particles constitute a homogeneous background, generated by drawing $\lfloor N/4 \rfloor$ random numbers from a uniform distribution in the interval $(0,L)$ for each spatial dimension.
    \item The remaining $\lceil 3N/4 \rceil$ particles are distributed in $n_\mathrm{blobs}=100$ blobs, whose centres are randomly placed through the central region of the domain ($x, \, y, \, z \in [L/4, 3L/4]$). Each blob is then realised by sampling the coordinates relative to the centre from a three-dimensional Gaussian distribution with standard deviation $\sigma = 0.02$ (which determines an effective blob radius). 
\end{itemize}

\begin{figure}
    \centering 
    \includegraphics[width=0.5\textwidth]{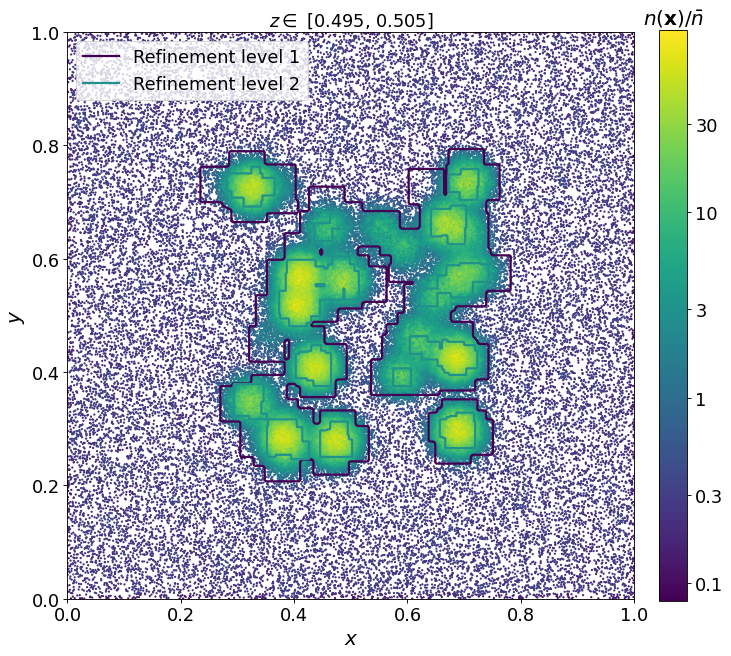}
    \caption{Thin slice showing the distribution of particles generated for the tests, corresponding to the realisation with $N_\mathrm{part}=16 \times 10^6$ particles. Dots represent the location of individual particles, colour-coded by their local density $n(\vb{x})$ in units of the mean particle density $\bar n$ according to the colour-scale on the right. Purple and turquoise contours indicate the regions refined up to level $\ell=1$ and $\ell=2$, respectively.}
    \label{fig:test_setup}
\end{figure}

These values, although arbitrary, are chosen so that the average number density within the $1\sigma$ radius of the blobs is $\approx 178$ times the background number density, similar to the overdensity of a virialised dark matter halo. This situation is enough to trigger the creation of at least two refinement levels covering the high-density regions, as exemplified in Fig. \ref{fig:test_setup}, where a thin slice through the realisation with $N = 16 \times 10^6$ particles is shown, together with an indication of the refined regions crossing the slice where resolution is recursively increased when running \vortexp{}. The fact that blobs are only placed in the central octant allows us to use simple test cases which may not respect the periodic boundary conditions (as it is the case of Tests 1 and 2), since the boundaries are far away enough from the region of interest. To assess the results of each test, we obtain a distribution of particle-wise errors by comparing the input fields with known decomposition and the results decomposed by \vortexp{}. To this aim, and to prevent the effects of non-periodic boundaries, we do not consider the particles closer than $d = 0.05 = L/20$ to any of the domain faces.

Through all tests below, the input velocity fields are assigned to particles by evaluating their analytic expressions stated in Eqs. \ref{eq:test1_input}, \ref{eq:test2_input}, \ref{eq:test3_input_comp}, \ref{eq:test3_input_sol}, \ref{eq:test4_compressive} and \ref{eq:test4_solenoidal} on the particle positions. That is to say, if $\vb{v}(\vb{x})$ is the analytic expression of the velocity field, the velocity of the $i$-th particle, located at $\vb{x_i}$, is $\vb{v_i} = \vb{v}(\vb{x_i})$.

\subsection{Test 1: constant divergence field}
\label{s:tests.test1}

\begin{figure*}
    \centering 
    \includegraphics[width=0.5\textwidth]{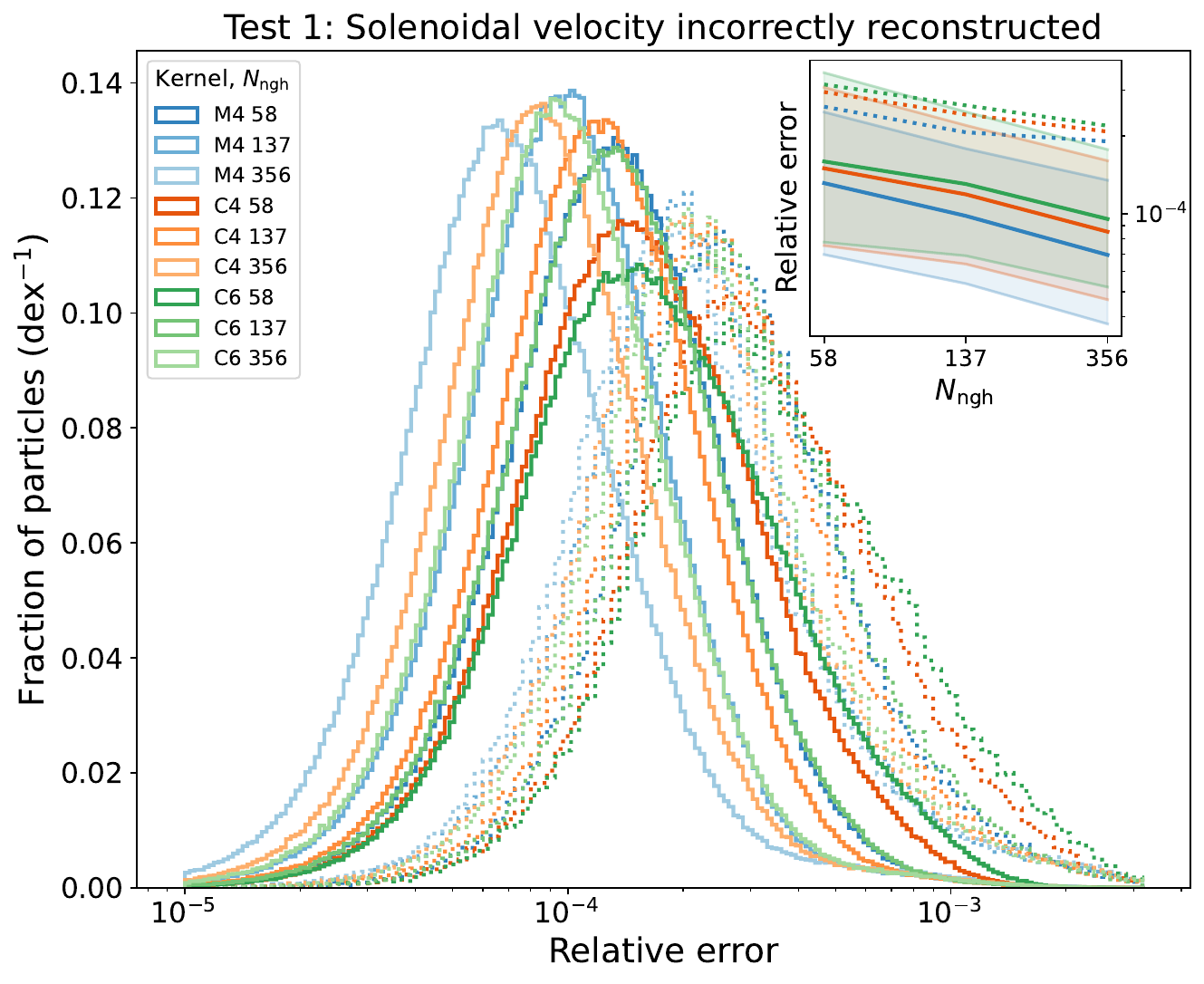}~ 
    \includegraphics[width=0.5\textwidth]{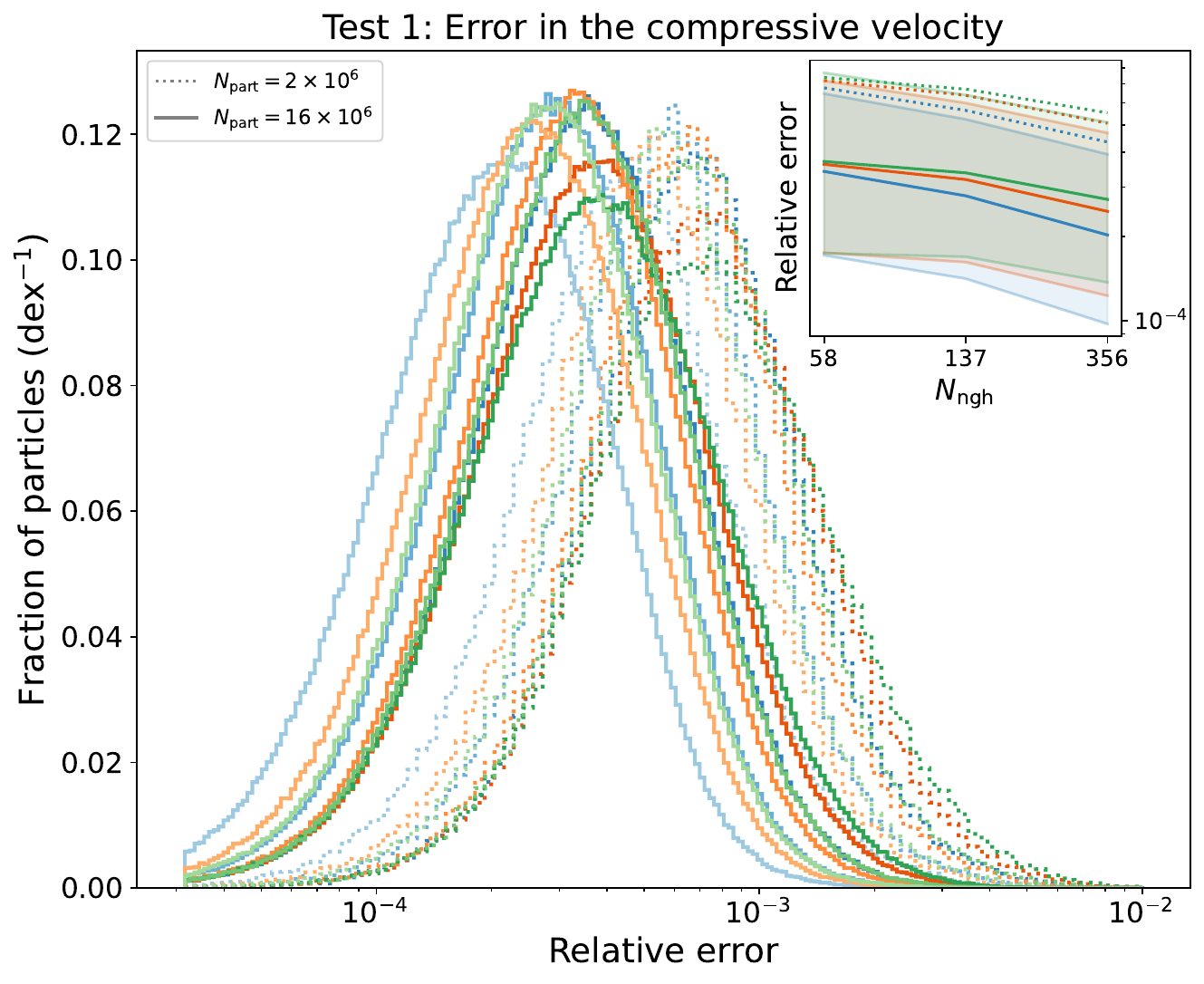}
    \caption{Summary results for Test 1. Within each panel, each line shows the distribution of a measure of the relative error amongst the particles (\textit{left panel}: fraction of solenoidal velocity magnitude incorrectly reconstructed; \textit{right panel}: relative error in the compressive velocity, with respect to the input velocity smoothed with the same kernel used in \vortexp{}). Solid (dotted) lines show the realisation with $N_\mathrm{part}=16 \times 10^6$ ($N_\mathrm{part}=2 \times 10^6$) particles. Blue, red and green hues correspond to the M$_4$, C$^4$ and C$^6$ kernels, with lighter tones corresponding to increasing $N_\mathrm{ngh}$. The insets summarise this information by showing the trends of the median and $(16-84)$ percentiles with $N_\mathrm{ngh}$ and kernel type. The confidence region is only shown in the 8$\times$ case for simplicity, being similar in the 1$\times$ case.}
    \label{fig:test1_results} 
\end{figure*}

A first test is aimed to assess the performance of the code in recovering a purely compressive vector field,

\begin{equation}
    \vb{v} = \omega_0 (x \vb{\hat u_x} + y \vb{\hat u_y} + z \vb{\hat u_z}),
    \label{eq:test1_input}
\end{equation}

\noindent which is spherically-symmetric and has constant divergence $\nabla \cdot \vb{v} = 3 \omega_0$ and null curl. The \textit{true} decomposition is, hence, $\vb{v_\mathrm{comp}}=\vb{v}$ and $\vb{v_\mathrm{sol}}= \vb{0}$. As it will be done with the rest of tests, Fig. \ref{fig:test1_results} presents the results of this test for $N_\mathrm{part}=2 \times 10^6$ (1$\times$; dotted lines) and $N_\mathrm{part}=16 \times 10^6$ particles (8$\times$; solid lines), and considering three of the available kernel shapes\footnote{For a summary on the test results also including the M$_5$ and M$_6$ kernels, at fix $N_\mathrm{ngh}$, see \ref{app:kM5M6}.} (M$_4$, blue; C$^4$, red; and C$^6$, green) with $N_\mathrm{ngh} = 58$, $137$ and $356$ neighbours (from darkest to lightest colours). The base grid is set to $N_x=128$ and $N_x=256$ for the 1$\times$ and 8$\times$ realisations, respectively. For each configuration, if $\vb{v}$ is the input velocity field and $\vb{\tilde v_\mathrm{comp/sol}}$ are the compressive/solenoidal components decomposed by \vortexp{}, we can determine two measures of the inaccuracies introduced by the method. The first one is the fraction of solenoidal velocity incorrectly reconstructed,

\begin{equation}
    \varepsilon(\vb{v_\mathrm{sol}}) = \frac{|\vb{\tilde v_\mathrm{sol}}|}{|\vb{v}|}.
    \label{eq:test1_error_absent}
\end{equation}

Secondly, we compute the relative error in the recovered compressive velocity, as done in \citet{Valles-Perez_2021_CPC}, from the propagation of errors from $\vb{v} = \sqrt{v_x^2 + v_y^2 + v_z^2}$ under the assumption of the three Cartesian components being uncorrelated,

\begin{equation}
\begin{aligned}
    \varepsilon(\vb{v_\mathrm{comp}}) = \frac{1}{\vb{v}^2} \Bigg[& v_x^4 \left( \frac{v_x - \tilde{v}_{\mathrm{comp},x}}{|v_x| + \epsilon} \right)^2 + v_y^4 \left( \frac{v_y - \tilde{v}_{\mathrm{comp},y}}{|v_y| + \epsilon} \right)^2  \\  
     + &  v_z^4 \left( \frac{v_z - \tilde{v}_{\mathrm{comp},z}}{|v_z| + \epsilon} \right)^2 \Bigg]^{1/2}
\end{aligned}
\label{eq:test1_error_present}
\end{equation}

\noindent with $\epsilon = 10^{-3} \max_N |\vb{v}|$. When computing these errors, we use, for $v_{x,y,z}$, the values of the velocity field smoothed with the same kernel used in \vortexp{}, so that we are comparing equivalent velocity fields. For more details on several possibilities for the error computation and their interpretation, see \ref{s:appendix.error_definition}.

The left-hand side panel of Fig.~\ref{fig:test1_results} shows the particle-wise distribution of incorrectly reconstructed solenoidal velocity, for all combinations of kernel, number of neighbours and number of particles. In the high-resolution realisation (8$\times$; solid lines), the amount of solenoidal velocity incorrectly decomposed by the algorithm corresponds to typical relative errors of $10^{-4}$, with the high-error tail extending up to $10^{-3}$. For a simpler visualisation, the inset shows the median and the $(16-84)$ percentiles of the error distribution as a function of the number of neighbours and the kernel shape. As a general trend, the longer the kernel effective size (higher $N_\mathrm{ngh}$, and lower-order), the lesser amount of solenoidal velocity the algorithm incorrectly reconstructs. In the low-resolution realisation (1$\times$; dotted lines), typical errors are about a factor of $2$ larger, keeping the same trends with kernel shape and size. This is a general trend in all tests, and is discussed in more detail in Sec. \ref{s:tests.convergence}.

The right-hand side panel contains the relative error in the recovered compressive velocity, whose median values range $(2-4)\times 10^{-4}$ in the 8$\times$ realisation and follow similar trends with kernel and neighbour choices. The fact that results improve when considering wider kernels (hence, a smoother interpolation) is expected, since damping higher-frequency noise makes the problem numerically easier.

\subsection{Test 2: constant curl field}
\label{s:tests.test2}

Parallel to Test 1, the second test aims to evaluate the performance of \vortexp{} when dealing with a purely solenoidal velocity field, 

\begin{equation}
    \vb{v} = \omega_0 (-y \vb{\hat u_x} + x \vb{\hat u_y}),
    \label{eq:test2_input}
\end{equation}

\noindent which has constant curl $\nabla \times \vb{v} = 2 \omega_0 \vb{\hat u_z}$ and null divergence. Therefore, the field can be decomposed as $\vb{v_\mathrm{comp}}=\vb{0}$ and $\vb{v_\mathrm{sol}}=\vb{v}$.

\begin{figure*}
    \centering 
    \includegraphics[width=0.5\textwidth]{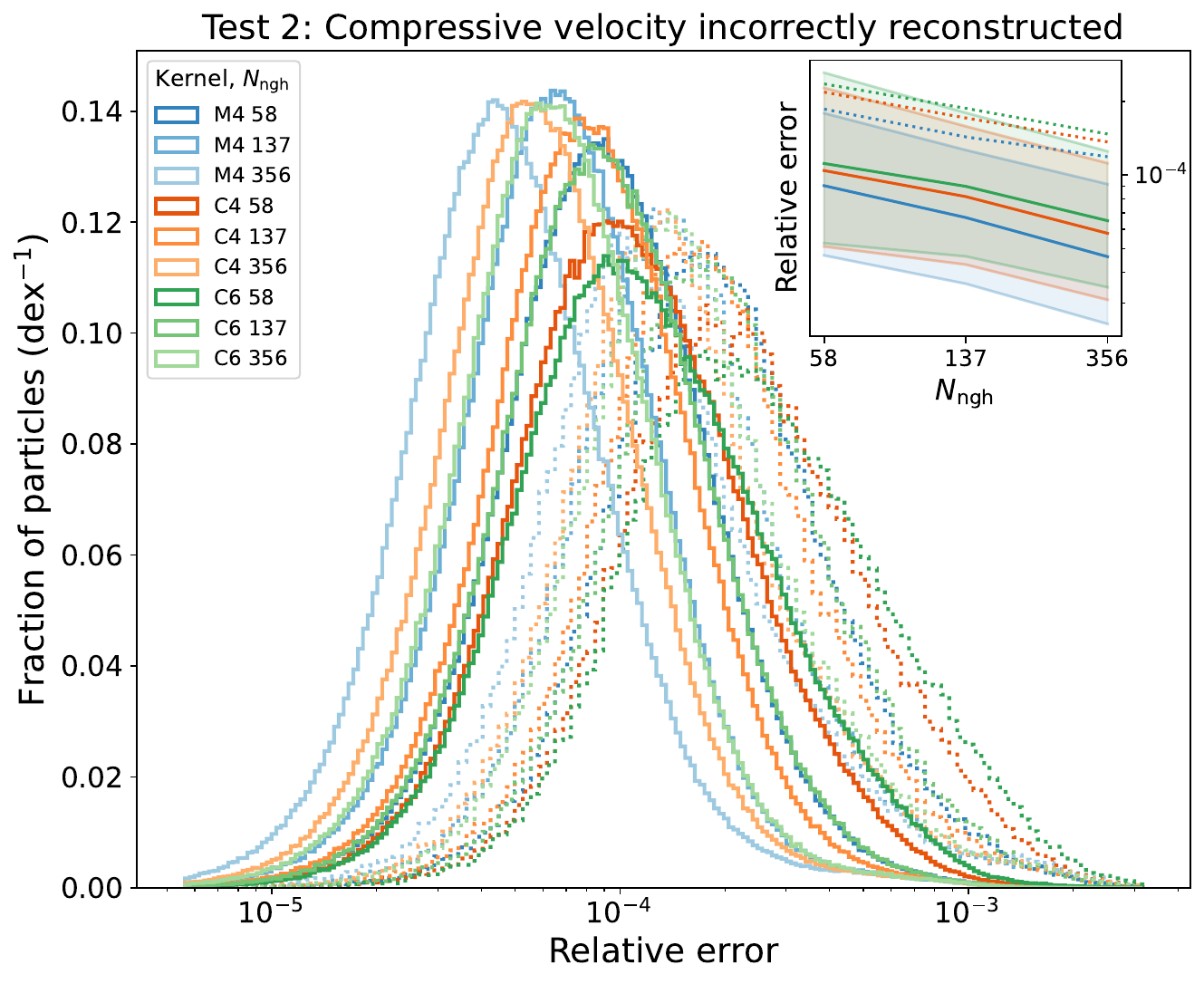}~ 
    \includegraphics[width=0.5\textwidth]{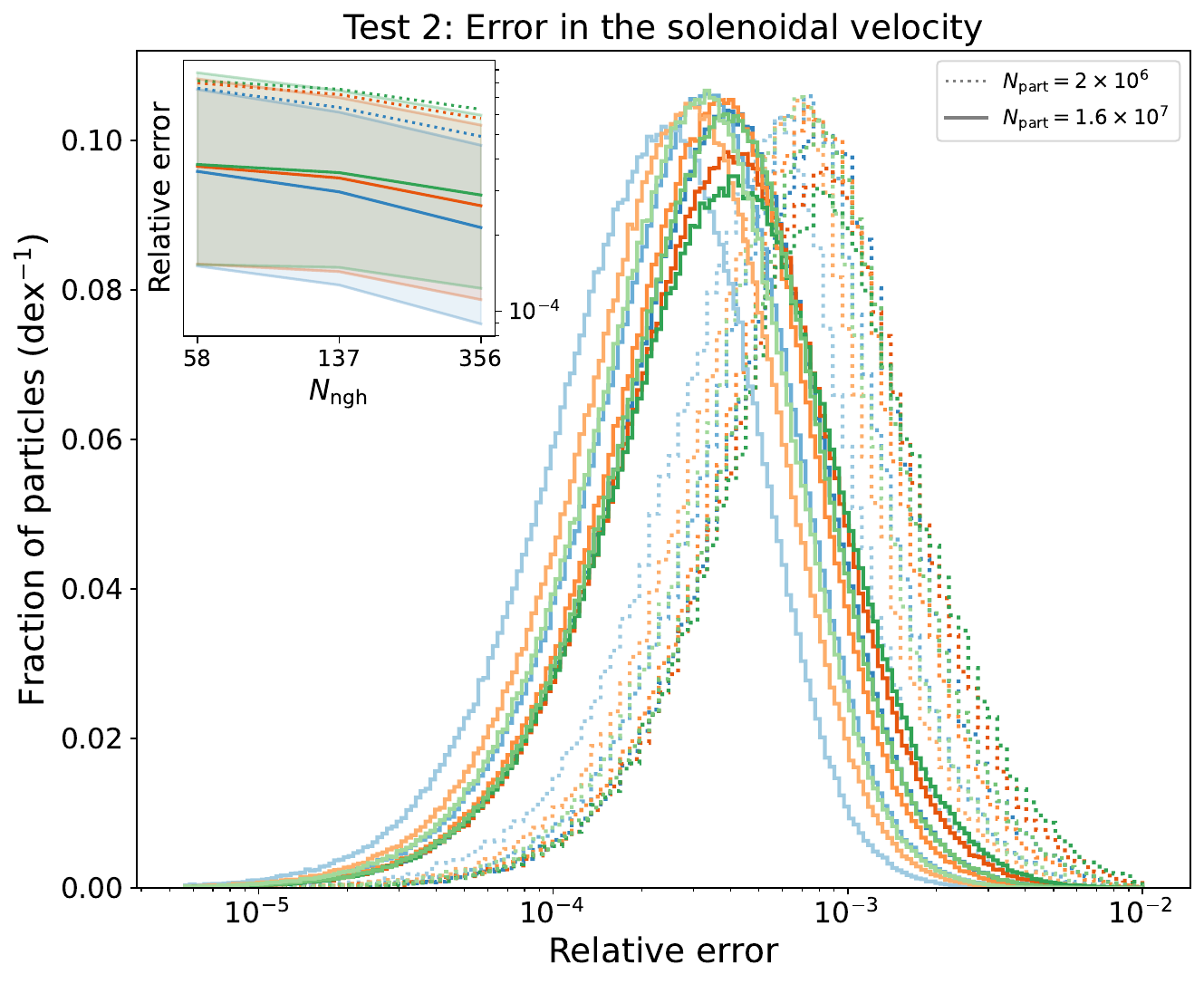}
    \caption{Summary results for Test 2. \textit{Left panel}: fraction of compressive velocity incorrectly reconstructed. \textit{Right panel:} relative error in the solenoidal velocity, with respect to the input velocity smoothed with the same kernel used in \vortexp{}. The elements in the panels are the same as the ones in Fig. \ref{fig:test1_results}.}
    \label{fig:test2_results}
\end{figure*}

The results are summarised in Fig. \ref{fig:test2_results} in a similar manner to the results of Test 1. The left panel describes the distribution of $|\vb{\tilde v_\mathrm{comp}}| / |\vb{v}|$, as in Eq. \ref{eq:test1_error_absent}, displaying a similar behaviour to the residual solenoidal component obtained in Test 1, albeit a factor $\sim 2$ smaller in magnitude. Regarding the errors in the solenoidal velocity (the one present in the input), its values and the trends with kernel type and number of neighbours are once again equivalent to those of Test 1.

Overall, Test 1 and Test 2 serve as a testbed for the following, more complex cases. The typical magnitudes of the errors associated to the cross-talk between the compressive and the solenoidal component ($|\vb{\tilde v_\mathrm{sol}}| / |\vb{v}|$ in Test 1, and $|\vb{\tilde v_\mathrm{comp}}| / |\vb{v}|$ in Test 2) involved in the HHD of particle-based data, although small (in the order of $\sim 10^{-4}$), are around four orders of magnitude above the ones shown in \citet{Valles-Perez_2021_CPC} for the grid-based version of \vortex{}. It is reasonable to expect this, since the velocity assignment onto the grid does not necessarily preserve the solenoidal/irrotational character of the input fields. However, as discussed in more detail in Sec. \ref{s:tests.convergence}, this is a limitation of the test design that progressively vanishes when considering increasing resolution of the input data.

\subsection{Test 3: mixed field}
\label{s:tests.test3}

The third test aims to gauge the level of cross-talk between the compressive and solenoidal components by considering a mixed field, composed of a superposition of low-frequency sinusoidal plane waves, with a compressive part,

\begin{equation}
    \vb{v_\mathrm{comp}} = \sin \left( \frac{2\pi x}{L}\right) \vb{\hat u_x} + \sin \left( \frac{2\pi y}{L}\right) \vb{\hat u_y} + \sin \left( \frac{2\pi z}{L}\right) \vb{\hat u_z}
    \label{eq:test3_input_comp}
\end{equation}

\noindent and a solenoidal part, 

\begin{equation}
    \begin{aligned}
    \vb{v_\mathrm{sol}} =
         & \Bigg[\sin \left( \frac{4\pi y}{L}\right) + \sin \left( \frac{6\pi z}{L}\right) \Bigg] \vb{\hat u_x} \\
       + & \Bigg[\sin \left( \frac{6\pi x}{L}\right) + \sin \left( \frac{4\pi z}{L}\right) \Bigg] \vb{\hat u_y} \\
       + & \Bigg[\sin \left( \frac{4\pi x}{L}\right) + \sin \left( \frac{6\pi y}{L}\right) \Bigg] \vb{\hat u_z}.
    \end{aligned}
    \label{eq:test3_input_sol}
\end{equation}

\begin{figure*}
    \centering 
    \includegraphics[width=0.5\textwidth]{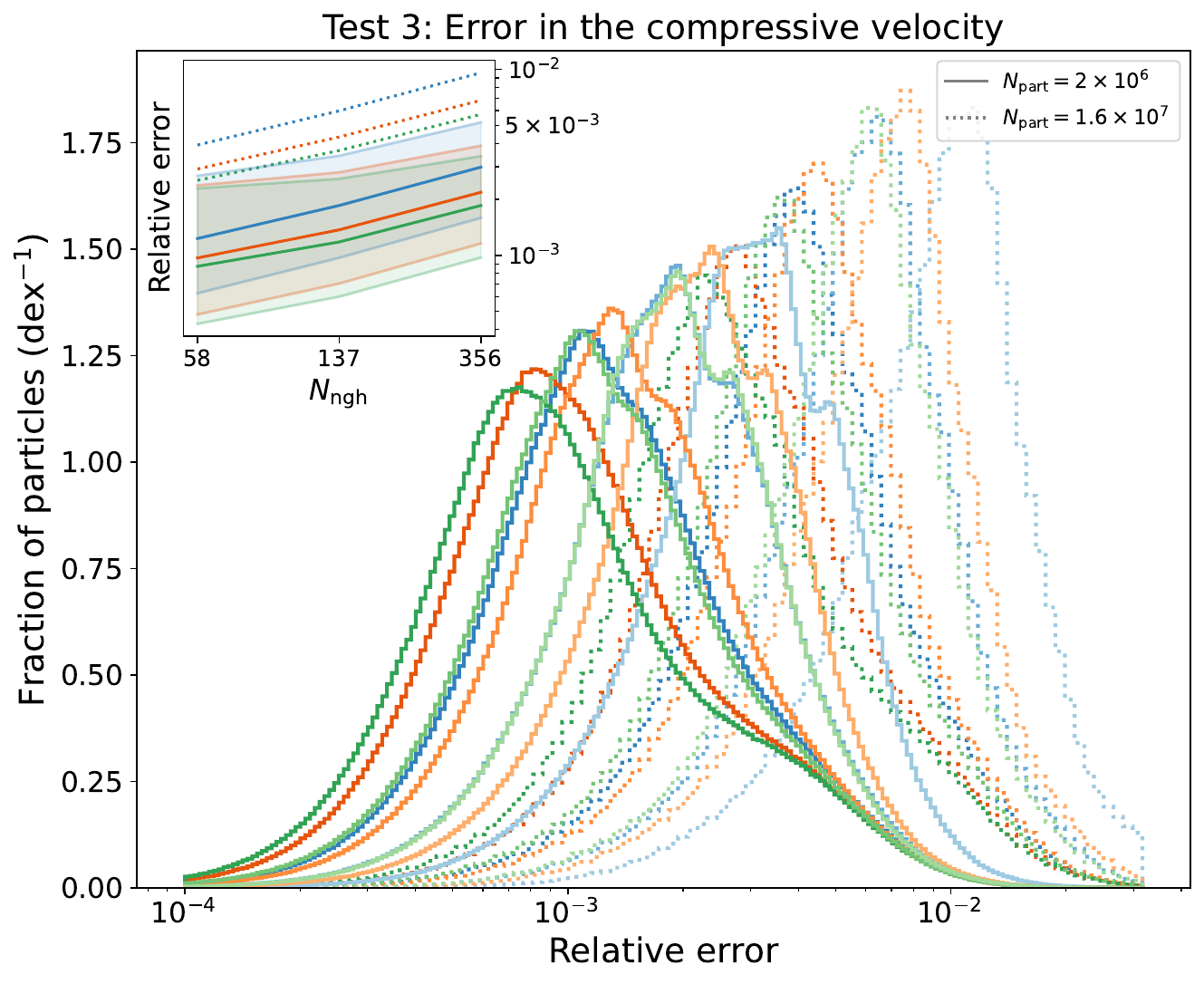}~ 
    \includegraphics[width=0.5\textwidth]{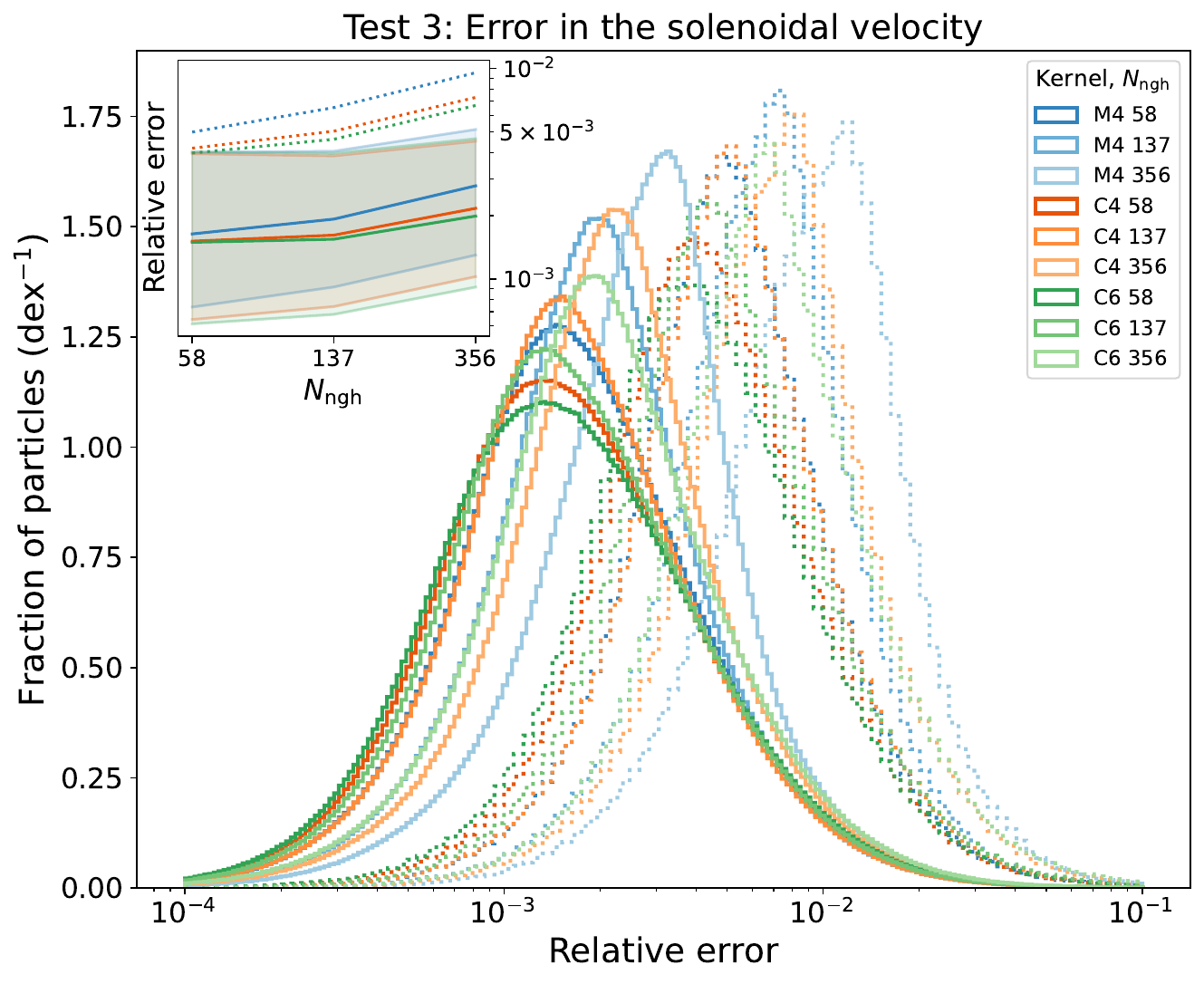}
    \caption{Summary results for Test 3. \textit{Left panel}: relative error in the compressive velocity. \textit{Right panel:} relative error in the solenoidal velocity. Both errors are computed with respect to the input velocity smoothed with the same kernel used in \vortexp{}. The elements in the panels are the same as the ones in Fig. \ref{fig:test1_results}.}
    \label{fig:test3_results}
\end{figure*}

Unlike the previous examples, the resulting input field, $\vb{v} = \vb{v_\mathrm{comp}} + \vb{v_\mathrm{sol}}$ is explicitly periodic in the input domain. In this case, we compute the error in recovering the compressive and the solenoidal velocity separately, following the definition in Eq. \ref{eq:test1_error_present}. The results are showcased in Fig. \ref{fig:test3_results} in a similar manner to the right-hand side panels of Figs. \ref{fig:test1_results} and \ref{fig:test2_results}.

In this mixed case, the HHD is brought up by \vortexp{} with relative accuracies around $\sim 10^{-3}$ for both components, which is around an order of magnitude larger than the results in the previous, pure cases, but similar to the performance of the grid-based version of \vortex{} on this same test. 

Interestingly, the behaviour of the median errors with kernel family and number of neighbours is the opposite to what was obtained in Tests 1 and 2. This difference is triggered by the fact that the velocity fields in Tests 1 and 2 are linear in the three Cartesian coordinates, implying that (in the ideal case of a homogeneous distribution), the average within a sphere coincides with the value of the velocity field at the centre. This is not the case in Test 3, which involves a non-linear velocity field. Since smoothing a vector field with non-constant resolution does not preserve its solenoidal/compressive character, it is reasonable that the test results show that, the smaller the extent of the kernel, the lesser amount of cross-talk between compressive and solenoidal fields due to the interpolation.

This fact signals an inherent limitation, not strictly of \vortexp{}, but of the tests themselves. Even though one inputs an analytically solenoidal vector field, this does not imply that the SPH divergence, $\sum_a (\vb{v_a} \cdot \nabla) W(\vb{r}-\vb{r_a})$, is null and hence the divergence-free condition is fulfilled in the SPH sense. Naturally, the differences get reduced when increasing the particle resolution, as discussed in Sec. \ref{s:tests.convergence}.

\begin{figure}
    \centering 
    \includegraphics[width=0.5\textwidth]{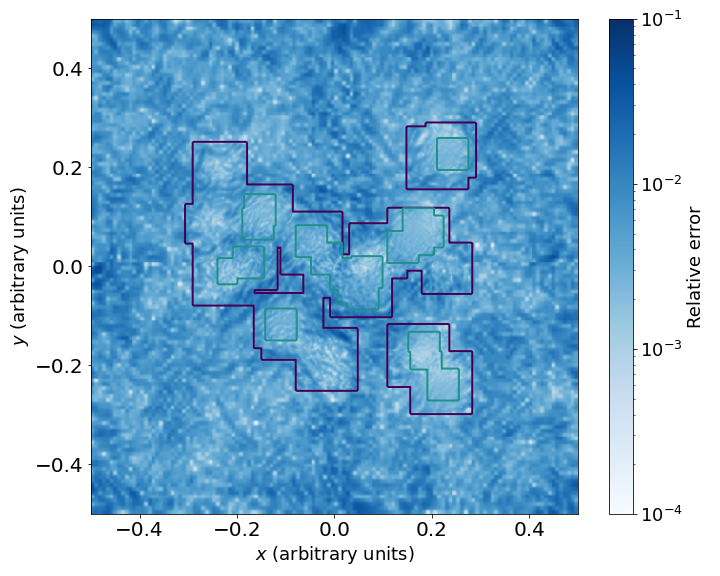}
    \caption{Thin slice through the mock domain, showing the spatial distribution of relative errors in the compressive component in Test 3, computed according to Eq. \ref{eq:test1_error_present}, for the particular run using the C6 kernel with $N_\mathrm{ngh}=58$ over the realisation with $N_\mathrm{part}=2 \times 10^6$ particles. Purple and turquoise contours indicate the regions refined up to level $\ell=1$ and $\ell=2$, respectively.}
    \label{fig:test3_map}
\end{figure}

To exemplify the spatial distribution of the errors committed by the algorithm, we present, in Fig. \ref{fig:test3_map}, a map showing the relative error in the compressive velocity through a thin slice of the domain. The relative error is computed as in Eq. \ref{eq:test1_error_present}, and we exemplify with the $N_\mathrm{part}=2 \times 10^6$ simulation, and running \vortexp{} using the C$^6$ kernel and 58 neighbours for the velocity assignment. When interpreting this map, one must bear in mind that, here, the error is computed in a cell-wise basis, instead of in a particle-wise basis, and therefore the results may not be identical to the ones shown in Fig. \ref{fig:test3_results}. The most substantial difference stems from the fact that, while evaluating the errors on a particle-wise basis is equivalent to a mass-weighting, doing it on a cell-wise basis is similar to a volume-weighting. Consequently, the latter makes the effect of low-density regions (which are, precisely, the ones with higher smoothing lengths and, therefore, the ones affected by the higher inaccuracies introduced by our artificially-imposed velocity field) more notorious. Hence, the error figures in this map are only to be interpreted as an upper limit.

When volume-weighted over the slice, typical (root-mean square) errors in the compressive velocity are below $\sim 2\%$ while, inside the most refined regions ($\ell=2$, in this case), typical errors fall in the order of $\sim 1 \text{\textperthousand}$. Additionally, looking at the error map, the boundaries between different resolution regions do not exhibit sharp features with respect to the error measures, implying that the results do not get artificially contaminated by the ad-hoc grid.

\subsection{Test 4: mock ICM-like field}
\label{s:tests.test4}

The last test considers a complex (in the sense of containing a superposition of oscillations spanning a wide range of spatial frequencies) mixed velocity field. This is operationally obtained by defining nine vector fields,

\begin{equation}
    \vb{v}_{ij} = \sum_{n=n_\mathrm{min}}^{n_\mathrm{max}} A_n^{ij} \sin \left(\frac{2 \pi n}{L} x_j + \psi^{(n)}_{ij} \right) \vb{\hat u_i}
    \label{eq:test4_field}
\end{equation}

\noindent where $\left\{x_i \right\}_{i=1}^{3} = \left\{x, \, y, \, z \right\}$ are the three Cartesian coordinates and $\left\{\vb{\hat u_i} \right\}_{i=1}^{3} = \left\{\vb{\hat u_x} , \, \vb{\hat u_y} , \, \vb{\hat u_z}  \right\}$ are their corresponding unit vectors. $n_\mathrm{min}$ and $n_\mathrm{max}$ are the lowest and highest-frequency modes, while $\left\{\psi^{(n)}_{ij}\right\}_{i,j=1}^{3}$ are nine sets of $n_\mathrm{max}-n_\mathrm{min}+1$ random phases uniformly sampled in the interval $[ 0, 2\pi [$. Lastly, the amplitudes are set by

\begin{equation}
    A_n^{ij} = \begin{cases}
        A_p \left( \frac{n}{p} \right)^{-1/2}  & \text{if } i=j \text{ (compressive)}\\
        A_p \left( \frac{n}{p} \right)^{-1/3} & \text{if } i\neq j \text{ (solenoidal)}
    \end{cases}
    \label{eq:test4_amplitudes}
\end{equation}

\noindent so that the compressive components, with $i=j$, obey a Burgers-like spectrum; while the solenoidal components, with $i \neq j$, follow a Kolmogorov spectrum.\footnote{To justify this point, let us consider that the amplitude of the components with spatial frequency $n/L$ scales as $A_n \propto n^{-\alpha}$. Then, the energy power spectrum (defined below, Eq. \ref{eq:power_spectrum}), $E(k)$, i.e. the specific kinetic energy per unit spatial frequency, scales as $E(k) \propto A_n^2 / k \propto k^{-(1+2\alpha)}$. Therefore, our compressive (solenoidal) amplitude scaling with $\alpha=1/2$ ($\alpha=1/3$) implies a Burgers (Kolmogorov) spectrum with $E(k) \propto k^{-2}$ ($E(k) \propto k^{-5/3}$).} The integer $p$ sets a pivot for the spectrum, or the mode where compressive and solenoidal amplitudes match each other. Once $p$ is fixed, $A_p$ sets the normalisation of the input vector field. From this description, the compressive field is straightforwardly obtained as

\begin{equation}
    \vb{v_\mathrm{comp}} = \sum_{i=x,y,z} \sum_{j=x,y,z} \delta_{ij} \vb{v}_{ij},
    \label{eq:test4_compressive}
\end{equation}

\noindent where $\delta_{ij}$ is the Kronecker delta, while

\begin{equation}
    \vb{v_\mathrm{sol}} = \sum_{i=1}^3 \sum_{j=1}^3 (1-\delta_{ij}) \vb{v}_{ij}
    \label{eq:test4_solenoidal}
\end{equation}

\noindent yields the solenoidal component. We have generated realisations of this mock velocity field with $N_\mathrm{part}=2\times 10^6$, $16 \times 10^6$ and $128 \times 10^6$ particles, introducing modes from $n_\mathrm{min}=1$ to $n_\mathrm{max}=25$ with $p=5$ as the mode where compressive and solenoidal amplitudes match. The choice of $n_\mathrm{max}$ is motivated by the effective resolution, $h = L \left( \frac{3 N_\mathrm{ngh}}{4 \pi N} \right)^{1/3}$ that can be resolved in each realisation. In the highest resolution realisation, with $N_\mathrm{ngh}=356$, this yields $h \approx 8.72 \times 10^{-3}$. If we require that the smallest wavelengths are sampled with, at least, four effective smoothing lengths, this sets an upper limit for the highest mode around $n_\mathrm{max} \sim 28$. In the lower resolution realisations, the highest modes may not be well represented in the most diffuse regions. This is why, for the purpose of assessing the accuracy of our decomposition algorithm, we will only consider the particles with $h < 5 \times 10^{-3}$.

\begin{figure*}
    \centering 
    \includegraphics[width=0.5\textwidth]{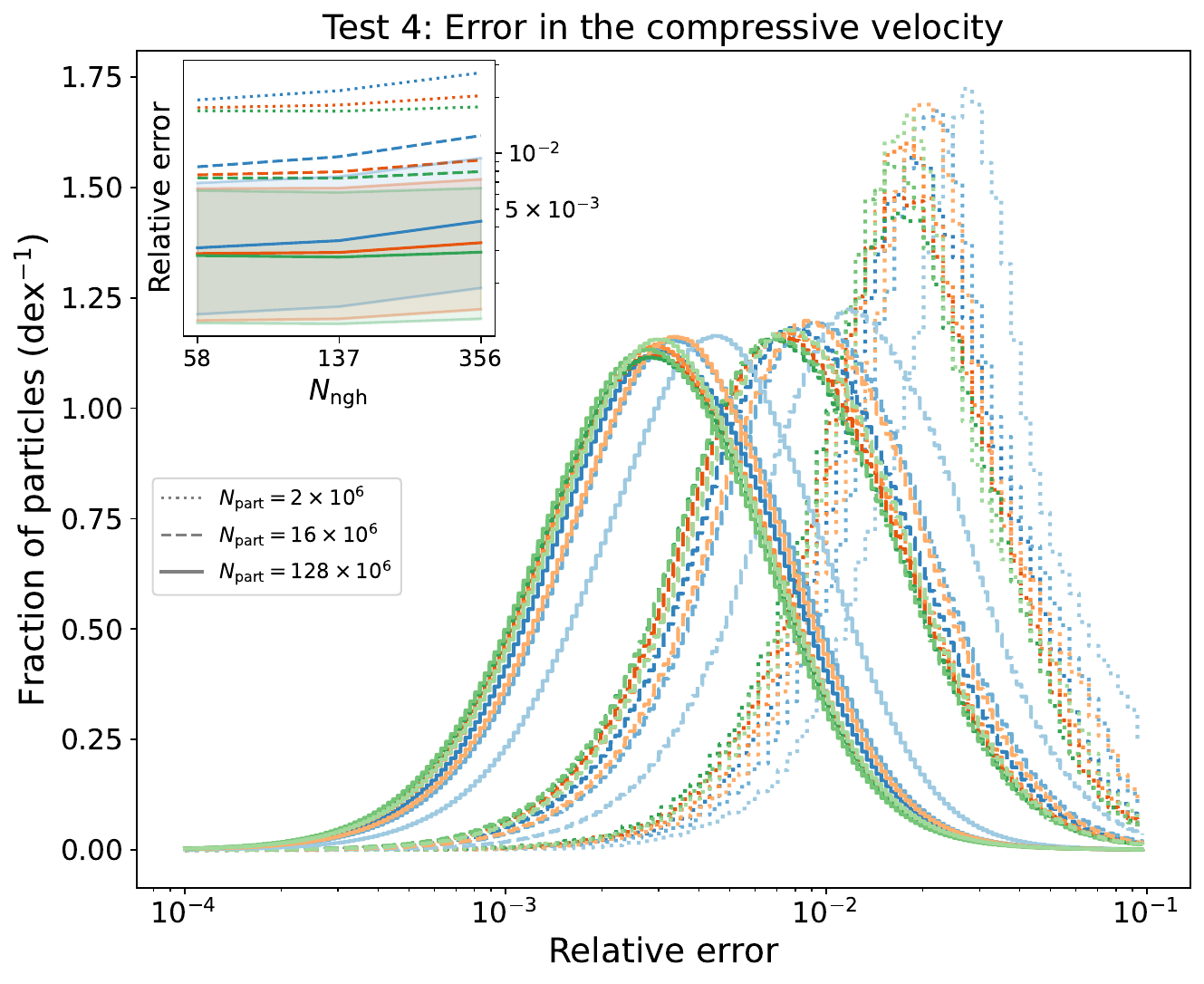}~ 
    \includegraphics[width=0.5\textwidth]{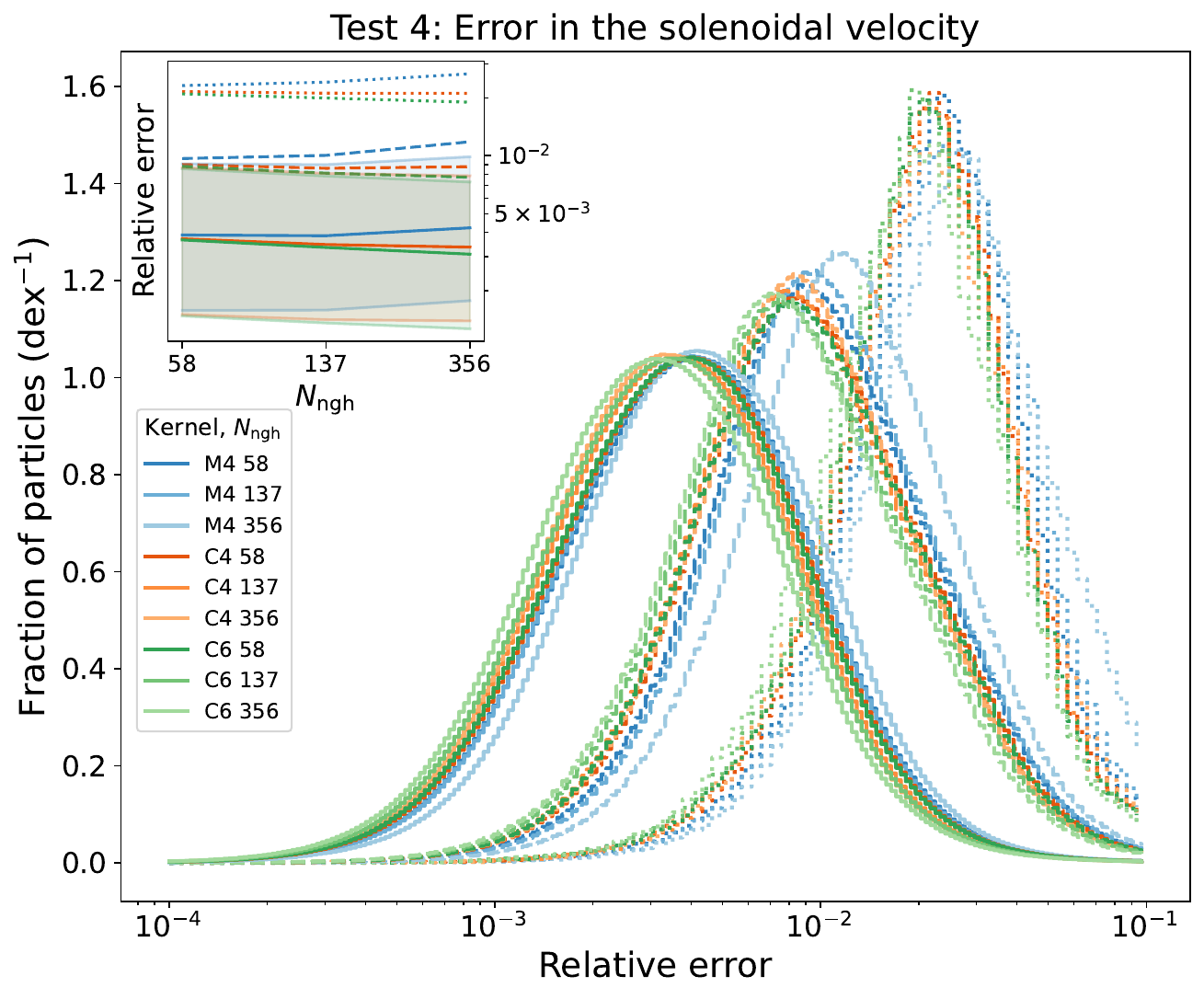}
    \caption{Summary results for Test 4. \textit{Left panel}: relative error in the compressive velocity. \textit{Right panel:} relative error in the solenoidal velocity. Both errors are computed with respect to the input velocity smoothed with the same kernel used in \vortexp{}. The elements in the panels are the same as the ones in Fig. \ref{fig:test1_results}.}
    \label{fig:test4_results}
\end{figure*}

The results are shown in Fig. \ref{fig:test4_results} in a similar manner to the previous tests, where now solid, dashed and dotted lines correspond to the $N_\mathrm{part}=128 \times 10^6$, $16 \times 10^6$ and $2 \times 10^6$ realisations, respectively. In this more complex situation, which resembles more closely than the previous tests a scenario of fully-developed turbulence, with a compressive and a solenoidal components spanning almost one and a half decades in scales with different spectral indices, the differences amongst kernel choices practically vanish, probably as a result of the competition of several effects. On the one hand, as seen in Test 3, the smoother the velocity assignment is, the larger the cross-talk between compressive and solenoidal modes is. On the other hand, shorter kernels imply more small-scale signal, making the decomposition less trivial.

\subsection{Convergence}
\label{s:tests.convergence}

\begin{figure}
    \centering 
    \includegraphics[width=0.5\textwidth]{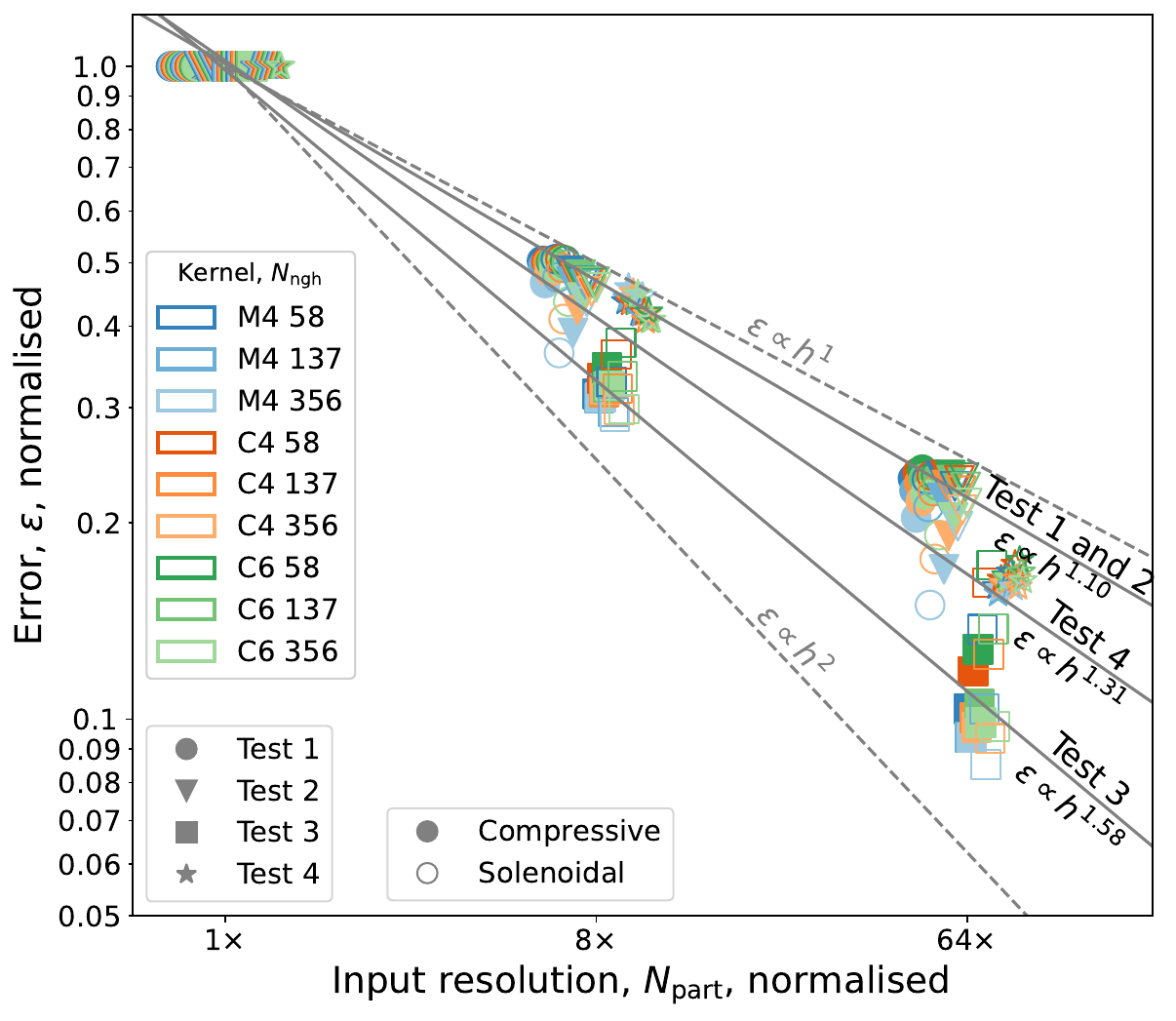}
    \caption{Convergence of the decomposition when increasing the number of particles, $N_\mathrm{part}$, at fixed input velocity field. The horizontal axis refers to the input resolution, $N_\mathrm{part}$, expressed as a factor of the lowest resolution one (1$\times$, 8$\times$ and 64$\times$ stand for $N_\mathrm{part}=2\times 10^6$, $16 \times 10^6$ and $128 \times 10^6$). The vertical axis shows the error in each run, normalised to the corresponding error of the 1$\times$ run. Different hues and lightness refer to different kernel functions and $N_\mathrm{ngh}$ as in the previous figures. Circles, triangles, squares and stars refer to Tests 1 to 4, respectively. Filled (empty) symbols indicate the compressive (solenoidal) component. Dashed, gray lines show the scalings $\varepsilon \propto h$ and $\varepsilon \propto h^2$ for reference, while solid lines show the fits for each test. Note that the $x$-coordinate of each dot has been slightly shifted just for visualisation purposes and, because of that, they may not match exactly the fit.}
    \label{fig:convergence}
\end{figure}

A general trend through Tests 1-4 is the fact that, when keeping the same analytical input velocity field, errors get reduced with increasing resolution ($N_\mathrm{part}$). The natural explanation for this fact is that, as $N_\mathrm{part}$ increases and, hence, the effective smoothing length, $h \propto N_\mathrm{part}^{-1/3}$, decreases, there is less cross-talk between the compressive and the solenoidal components due to the smoothing process itself.

Hence, it is interesting to assess this convergence through the different test cases explored above. This is shown in Fig. \ref{fig:convergence}, where we show the median of the error distributions for all kernels (M$_4$, C$^4$, and C$^6$), all values of $N_\mathrm{ngh}$ (58, 137 and 356), for both (compressive and solenoidal) components, and for the realisations with $N_\mathrm{part}=2\times 10^6$, $16 \times 10^6$ and $128 \times 10^6$ particles (1$\times$, 8$\times$ and 64$\times$, respectively).  Note that the $x$-coordinate of each dot has been slightly shifted just for visualisation purposes. These results are fit to a power-law for each test, in order to estimate a convergence order for each situation. In Tests 1 and 2, the results of \vortexp{} converge as $\varepsilon \propto h^{1.1}$. In the case of Test 3, which mixes solenoidal and compressive components, this convergence order is boosted to $\varepsilon \propto h^{1.58}$. This is expected, since in this case the cross-talk between compressive and solenoidal velocities is reduced as the typical smoothing length falls below the characteristic lengths of the input velocity field. In Test 4, which considers a more realistic, ICM-like, velocity field by further introducing a large range of spatial fluctuations, errors decay as $\varepsilon \propto h^{1.31}$.

It is worth noting that this test represents an idealised scaling situation, where the same input velocity field is realised with increasing resolutions. In an actual simulation, the velocity field will exhibit higher frequency modes with increasing resolution, and hence there is no limitation in applying \vortexp{} to arbitrarily low-resolution simulation data.

\subsection{Scalability}
\label{s:tests.scalability}

\begin{figure*}
    \centering 
    \includegraphics[width=\linewidth]{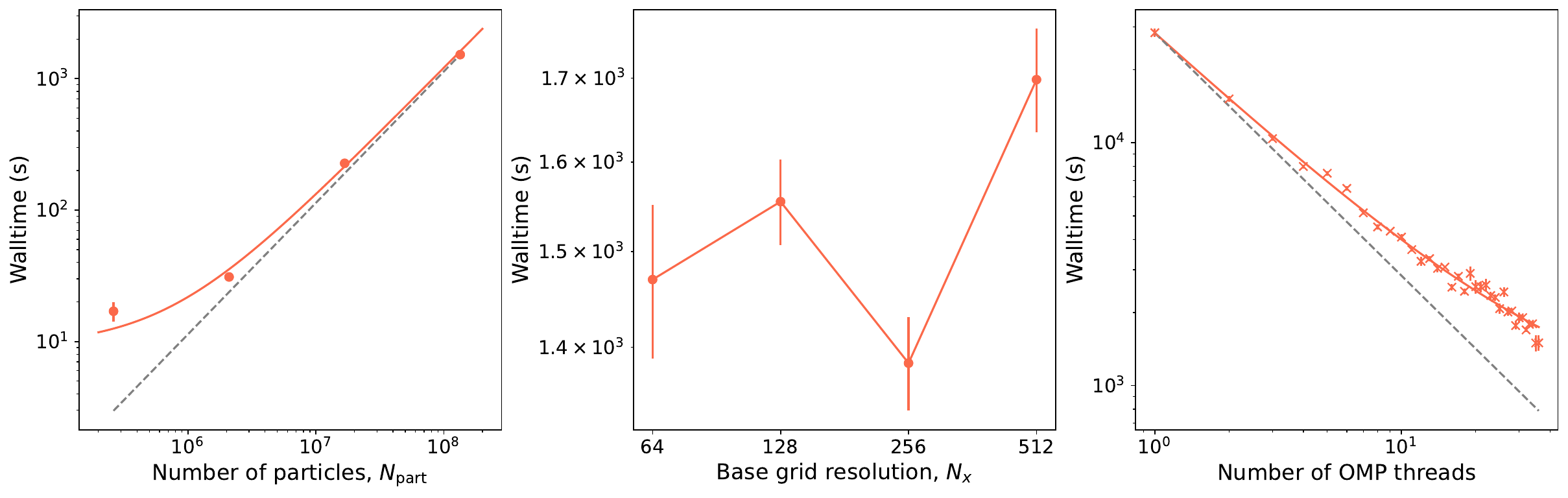}
    \caption{Scaling properties of \vortexp{}. In all plots, dots with vertical lines are the measured values and their uncertainty estimation by 5 repetitions of each test. \textit{Left-hand side panel}: scaling of the wall time required by the code with the number of particles. The red curve is a fit to these data. The dashed, grey line corresponds to an ideal, $(\Delta t)_\mathrm{wall} \propto N_\mathrm{part}$ situation. \textit{Middle panel}: scaling of the wall time required by \vortexp{} with the base grid resolution for the same number of particles. \textit{Right-hand side panel}: parallel scaling of the code, showing the wall time taken as a function of the number of OMP threads. The dashed, grey line corresponds to an ideal, $(\Delta t)_\mathrm{wall} \propto 1/n_\mathrm{threads}$ case.}
    \label{fig:scalability}
\end{figure*}

Last, we have assessed the computational requirements of \vortexp{} by means of some simple scalability tests, whose results are synthesised in Fig. \ref{fig:scalability}. All these tests have been run on a computing node equipping two 18-core CPU Intel$^\text{®}$ Xeon$^\text{®}$ Gold 6154 processors. The left-hand side panel informs about the wall time used by the \vortexp{} process when running realisations of Test 4 with $N_\mathrm{part}=250 \times 10^3$, $2 \times 10^6$, $16 \times 10^6$ and $128 \times 10^6$ particles. The scaling is consistent with an ideal scenario ($(\Delta t)_\mathrm{wall} \propto N_\mathrm{part}$, grey line), and can be fitted by 

\begin{equation}
    \frac{(\Delta t)_\mathrm{wall}}{\mathrm{s}} = (9.2 \pm 7.5) + (12.6 \pm 4.6) \left( \frac{N}{10^6} \right)^{0.99\pm0.10},
\end{equation}

\noindent which is represented by the red line in the corresponding panel. 

\begin{figure}
    \centering 
    \includegraphics[width=0.5\textwidth]{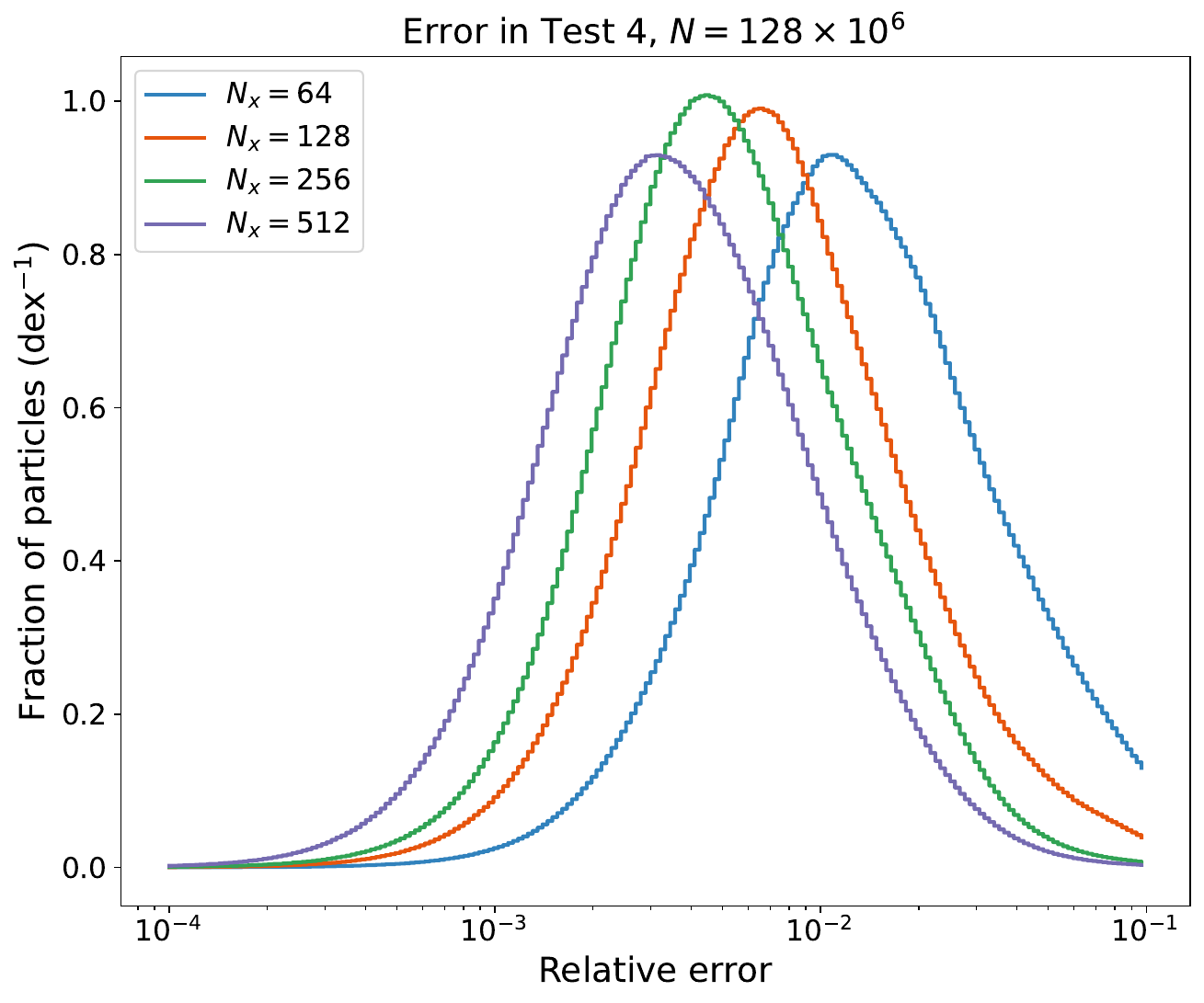}
    \caption{Distribution of particle-wise errors on the compressive velocity, for different runs of Test 4 with $N_\mathrm{part}=128 \times 10^6$ changing only the base grid resolution, $N_x$. The histograms in this figure are constructed analogously to those of the previous tests.}
    \label{fig:scalability_Nx}
\end{figure}

When keeping the number of particles constant and changing the resolution of the base grid (middle panel of Fig. \ref{fig:scalability}), there are not substantial differences in the execution time. This is the case because the mesh-creation strategy is efficient enough to create a larger number of refinement levels in the cases with smaller $N_x$, so that the number of resolution elements and the computational cost does not vary significantly. However, it is still interesting to check how the results vary for different choices of $N_x$, even when mantaining the refinement criterion on a fixed number of particles per cell (at any level) to flag it as refinable. These results are presented in Fig. \ref{fig:scalability_Nx} for the compressive velocity as an example. It turns out that, although the magnitude of the effect is small, the finer the base grid, the lesser amount of error. This might be understood from the fact that the refinement strategy is imperfect: not all refinable regions get refined, but only those which can be encompassed by refinement patches exceeding a certain minimum extension, $N_\mathrm{min}^\mathrm{patch}$ (see Sec. \ref{s:method.mesh}). Additionally, although on a more minor degree, the usage of coarser base grids implies that more job is relegated to the AMR elliptic solver. Unlike the FFT approach employed for the base grid, the AMR solver is iterative and errors may propagate to finer levels due to the interpolation of boundary conditions. Therefore, one can expect a slightly worsened performance with too coarse base grids, hence our recommendation to set $N_x$ to the closest power of 2 to $\sqrt[3]{N_\mathrm{part}}$.

Last, regarding the parallel scaling of the code, the right-hand side panel of Fig. \ref{fig:scalability} presents the wall time taken by \vortexp{} in Test 4 with $N_\mathrm{part}=128 \times 10^6$ particles with varying number of OMP threads. Generally, the behaviour departs only slightly from the ideal parallel scenario ($\Delta t \propto 1/n_\mathrm{threads}$). However, most of the degradation in this performance appears around $n_\mathrm{threads} \sim 18$, since at this point the job is not allocated in a single processor of our computing node and performance is penalised from the non-uniform memory access architecture. As a matter of fact, when reaching the right-hand side end of the graph ($n_\mathrm{threads} \sim 36$), the scaling is again approximately parallel to the ideal behaviour. All in all, the results can be fitted by the functional form

\begin{equation}
    \frac{(\Delta t)_\mathrm{wall}}{\mathrm{s}} = (760 \pm 100) + \frac{27590 \pm 210}{n_\mathrm{threads}^{0.931 \pm 0.016}},
\end{equation}

\noindent implying that the serial part of the code only accounts for $\sim 2.6 \%$ of the CPU time required by \vortexp{}.

\section{Applications}
\label{s:application}

\begin{figure*}
    \centering 
    {SPH} \hspace{6.4cm} {MFM}
    \includegraphics[width=0.4\textwidth]{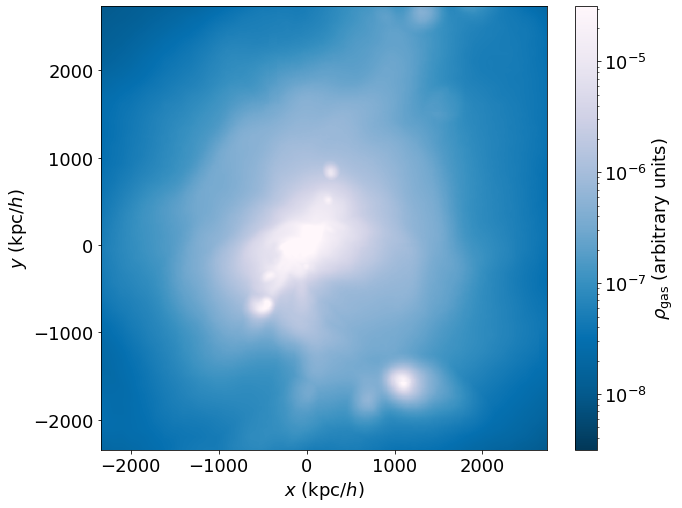}~ 
    \includegraphics[width=0.4\textwidth]{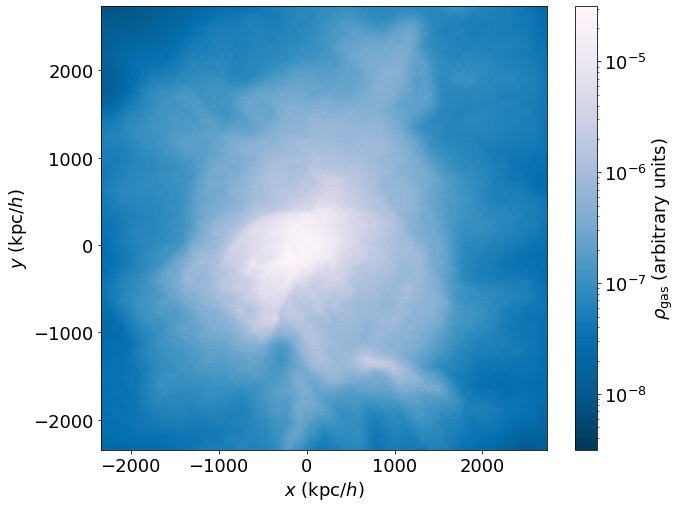}
    \includegraphics[width=0.4\textwidth]{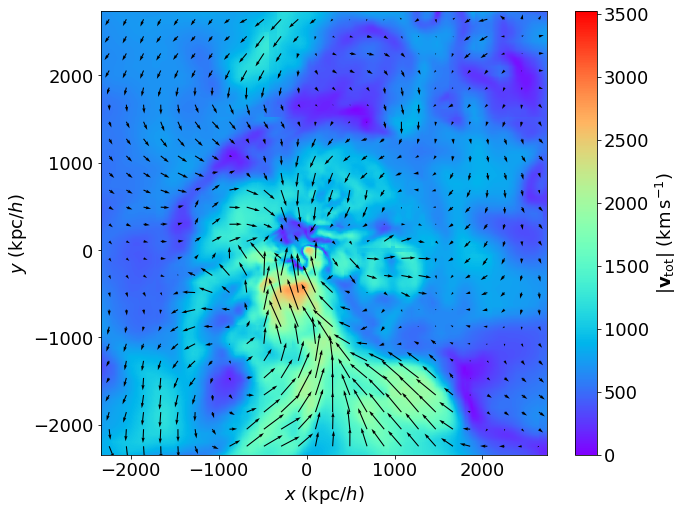}~
    \includegraphics[width=0.4\textwidth]{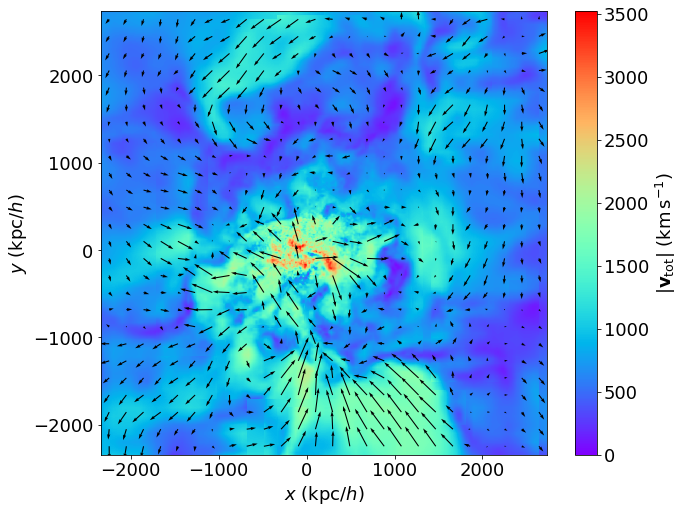}
    \includegraphics[width=0.4\textwidth]{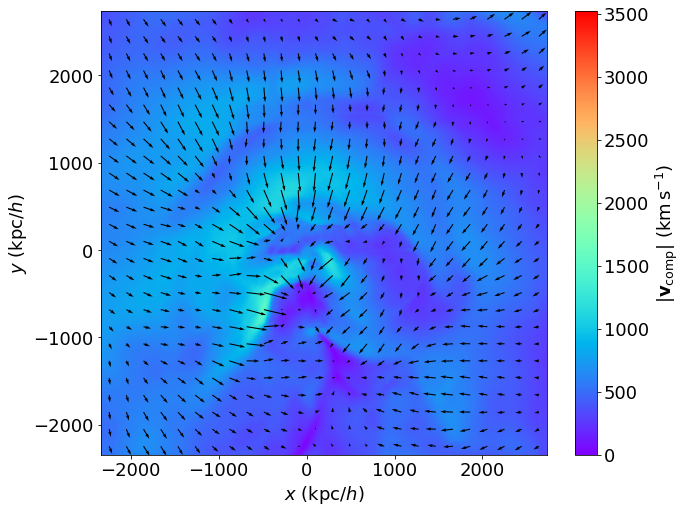}~
    \includegraphics[width=0.4\textwidth]{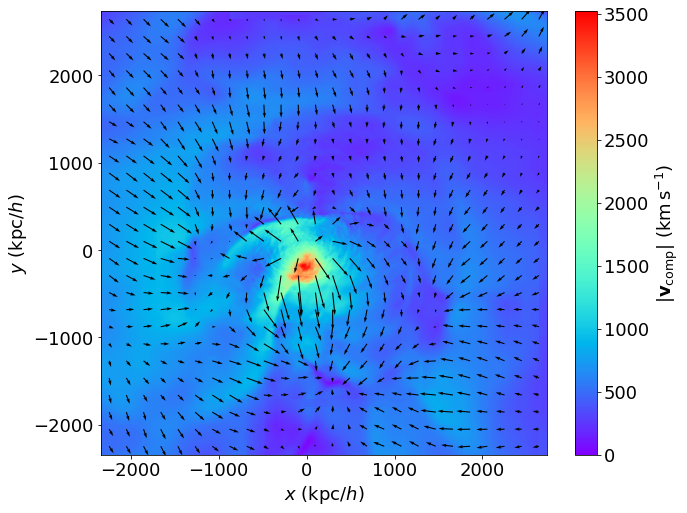}
    \includegraphics[width=0.4\textwidth]{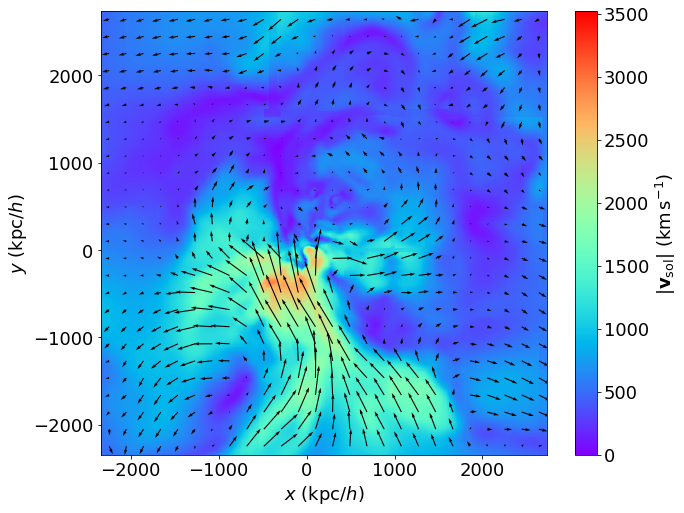}~
    \includegraphics[width=0.4\textwidth]{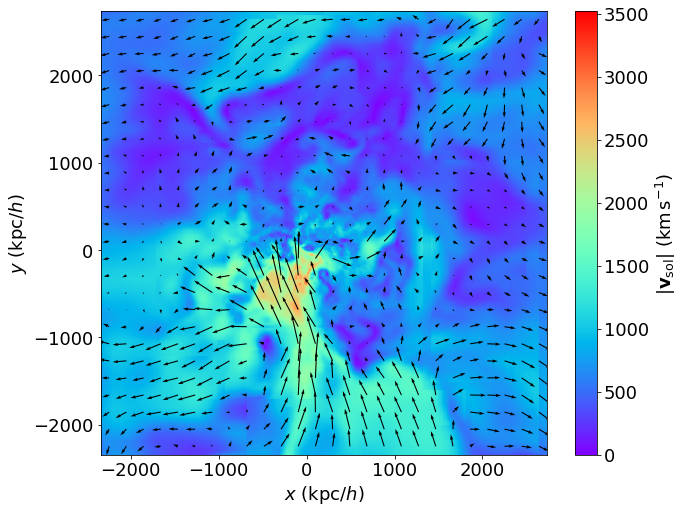}
    \caption{Comparison, performed with \vortexp{}, of the density and velocity structure of an SPH (left-hand side column) and an MFM (right-hand side column) simulation of the same cluster, with equivalent resolution, at $z=0$. From top to bottom, the different rows show thin slices of gas density, total velocity, compressive velocity and solenoidal velocity, respectively. In the velocity maps, the background colour shows the magnitude, while the arrows show the directions and magnitude of the $xy$ projection.}
    \label{fig:vortex_freddy}
\end{figure*}

This section is intended to show several applications of \vortexp{} to actual simulation data, without the aim of being exhaustive in its scientific discussion but just to exemplify different contexts where the code can be applied and qualitatively assess its performance. 

In particular, Fig.~\ref{fig:vortex_freddy} compares graphically (through $12 \, h^{-1} \, \mathrm{kpc}$-thick slices through the cluster centre) the velocity structure of an SPH (left) and an MFM (right) simulation of the same massive galaxy cluster, of mass $M_\mathrm{vir} \approx 1.34 \times 10^{15} \, h^{-1} M_\odot$ at $z=0$, with the same nominal resolution of $M_{\rm dm}=10^{9}M_{\odot}$, $M_{\rm gas}=1.6 \times 10^{8}M_{\odot}$. 

The cluster is taken from the Dianoga set of zoom-in simulations \citep{Bonafede+2011a} and ran with only gravity and hydrodynamical accelerations using the code \textsc{OpenGadget3}. A modern SPH implementation including artificial viscosity \citep{Beck_2016} and artificial conductivity \citep{Price2008} is used. The MFM implementation has been described by \citet{Groth_2023}. \vortexp{} was run in a box of $50 \, h^{-1} \, \mathrm{Mpc}$ centred on the cluster, with a base grid of $N_x=128$ and a maximum of $n_\ell=6$ refinement levels, yielding a peak resolution of $\Delta x_6 \approx 6 \, \mathrm{kpc}$. The refinement threshold was $n^{\rm refine}_{\rm part}=8$ and $N_{\rm min}^{\rm patch}=6$. Velocities were assigned to the grid within \vortexp{} using the same smoothing scheme employed for the evolution; that is, M$_4$ kernel with 32 neighbours for MFM, and C$^6$ with 295 neighbours for SPH.

Looking at the density maps (top row), while SPH seems to resolve more gaseous substructure, MFM captures more sharply the discontinuities associated to shocks (as, e.g., in the centre of the panel). These density maps, which are also interpolated by \vortexp{}, serve as a visual demonstration of the detail recovered by the assignment from particles to the grid.

Subsequent rows compare the total, compressive and solenoidal velocity magnitudes, respectively, with the arrows overplotted representing the velocity field direction in the projected plane. The difference between SPH and MFM is especially evident when comparing the compressive velocity field, where MFM shows a sharp description of shocks, which appear as thin surfaces of discontinuity, while in SPH they are smoothed out.

These differences can be quantified through the energy power spectra, $E(k)$,

\begin{equation}
    E(k) = 2 \pi k^2 P(k),
    \label{eq:power_spectrum}
\end{equation}

\noindent where $P(k)$ is the usual velocity power spectrum, built from the Fourier components of the velocity field, $\tilde {\vb{v}}(\vb{k})$ as

\begin{equation}
    P(k) = \frac{V}{(2\pi)^3} \langle |\tilde{\vb{v}}(\vb{k})|^2 \rangle_{\hat{\vb{k}}},
\end{equation}

\noindent $\langle \cdot \rangle_{\vb{\hat k}}$ denoting an average over the wavevectors such that $|\vb{k}| = k$. These spectra, computed on a cube $5 \,  h^{-1} \, \mathrm{Mpc}$ along each direction centred on the cluster, are shown in the top row of Fig. \ref{fig:vortex_powerspectra}. The left hand-side panel shows the total, compressive and solenoidal spectra for SPH (solid lines) and MFM (dashed lines). Although the trends and slopes are similar, as shown in the right-hand side panel, MFM shows an overall higher normalisation (up to $60\%$ higher) at all scales and for all components. This implies a higher kinetic energy budget in the MFM simulation, perhaps as a consequence of a greatly reduced numerical dissipation, with the effects on the compressive component being especially relevant at large ($\sim \mathrm{Mpc}$) scales associated to shock fronts, and the differences in the solenoidal contribution shifting towards slightly smaller scales (a few $\sim 100 \, \mathrm{kpc}$).

\begin{figure*}
    \centering 
    {Ideal} \hspace{6.2cm} {Viscosity}
    \includegraphics[width=0.4\textwidth]{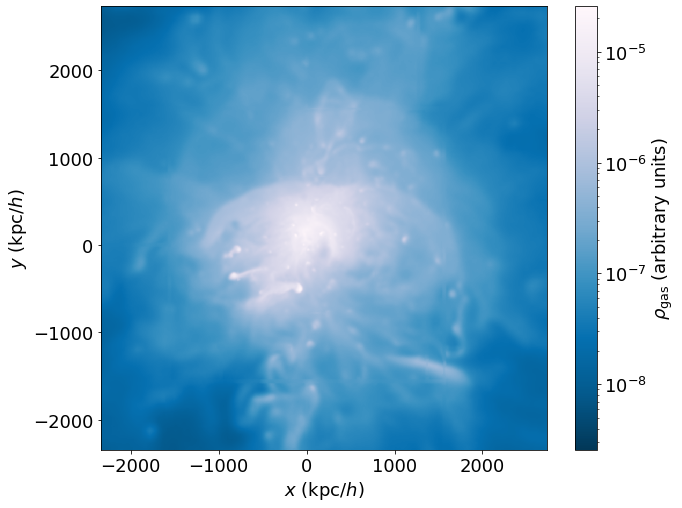}~ 
    \includegraphics[width=0.4\textwidth]{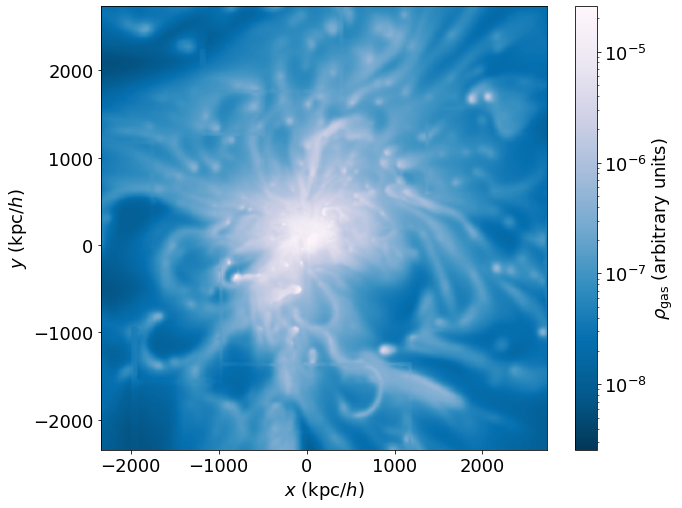}
    \includegraphics[width=0.4\textwidth]{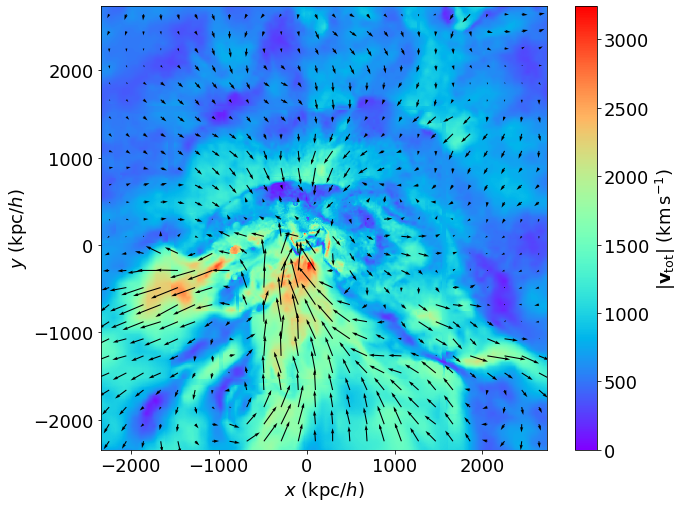}~
    \includegraphics[width=0.4\textwidth]{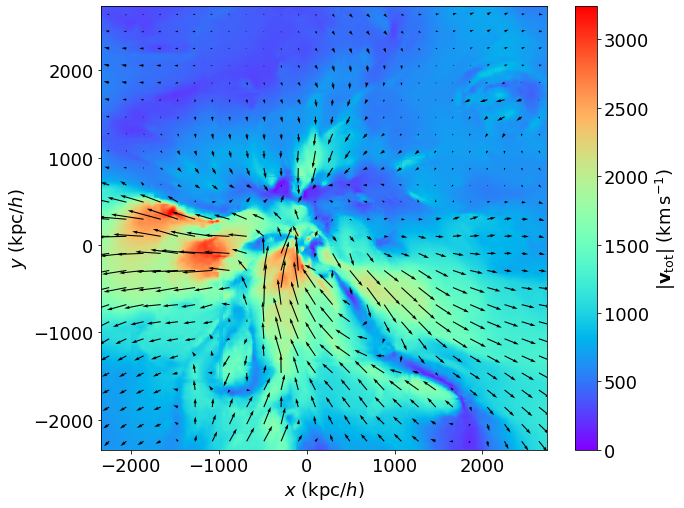}
    \includegraphics[width=0.4\textwidth]{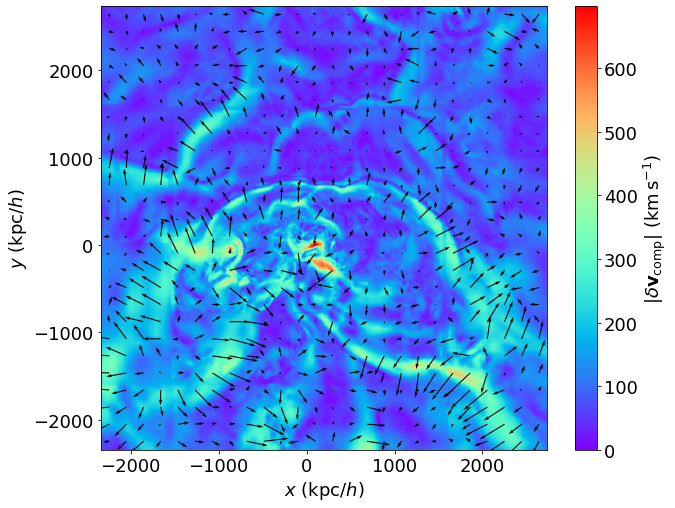}~
    \includegraphics[width=0.4\textwidth]{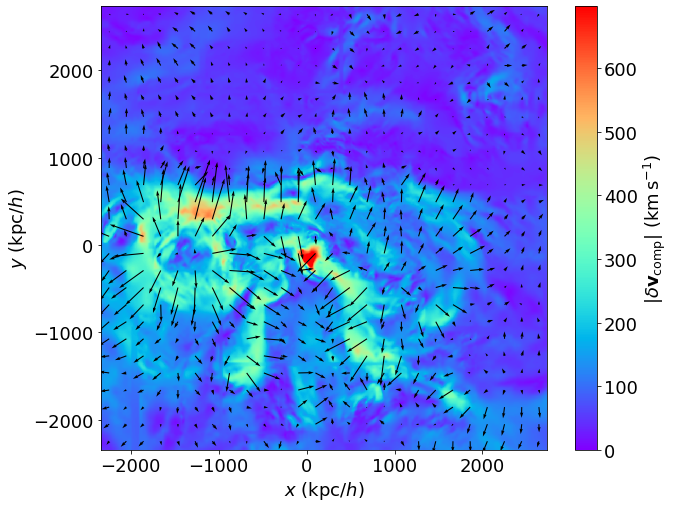}
    \includegraphics[width=0.4\textwidth]{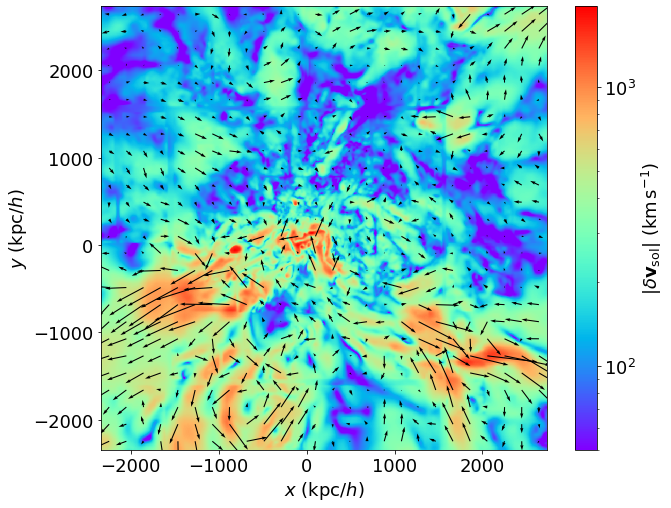}~
    \includegraphics[width=0.4\textwidth]{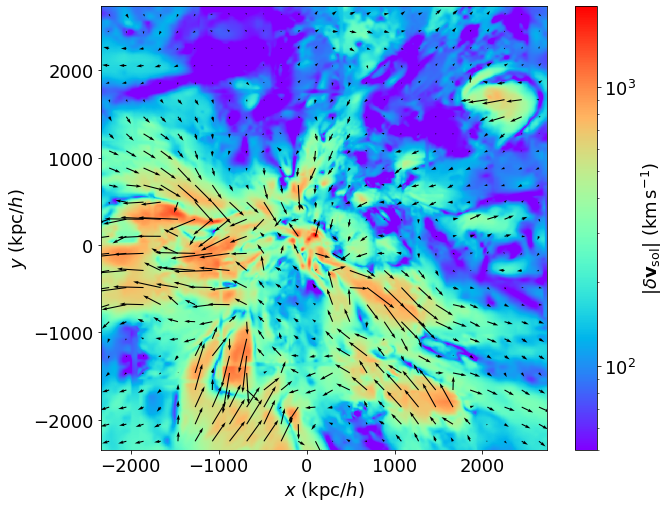}
    \caption{Comparison, performed with \vortexp{}, of the density and velocity structure of an SPH simulation without viscosity (left-hand side column) and with viscosity (right-hand side column) of the same cluster in Fig. \ref{fig:vortex_freddy}, with 10 times the mass resolution, at $z=0$. From top to bottom, the different rows show gas density, total velocity, compressive part of the turbulent velocity and solenoidal part of the turbulent velocity, respectively. In the velocity maps, the background colour shows the magnitude, while the arrows show the directions and magnitude of the $xy$ projection.}
    \label{fig:vortex_tirso}
\end{figure*}

In Fig. \ref{fig:vortex_tirso}, a similar comparison is shown for two SPH simulations of the same cluster, without (left) and with (right) physical viscosity. The physical viscosity implementation has been described by \citet{Marin_Gilabert_2022}. Additionally, magnetic field and an anisotropic thermal conduction were also included (see \citep{Steinwandel_2022} for details). Both simulations have numerical (mass) resolution $10\times$ with respect to the ones shown above. Due to the higher resolution, to properly capture the small scales, the maximum refinement levels was increased to $n_\ell=8$ and the velocity interpolation was done using a M$_4$ kernel with $58$ neighbours. 

The differences are striking even when looking at the density slices, with the simulation with physical viscosity showing a much more complex morphology of clumps and filamentary structures. While the second row shows the total velocity, as in Fig. \ref{fig:vortex_freddy}, the third and fourth rows show the compressive and solenoidal components of the turbulent (small-scale) velocity field, as extracted by the multi-scale filter in \vortexp{}, using $\mathcal{M}^\mathrm{thr} = 1.5$ as a stopping condition for the filter, $\Delta_\mathrm{tol} = 0.1$ and $\chi = 0.05$ as parameters for the filter, and a volume-weighting for the bulk velocity. While in the non-viscous simulation shock surfaces are clearly apparent as thin regions in the slice of the compressive turbulent velocity field, when physical viscosity is added, compressive velocities get strongly damped and the shock surfaces are smoothed out. Also, when considering the solenoidal velocity, it becomes apparent from the maps that viscosity is suppressing solenoidal turbulence on small scales, reducing the tangential motions of the gas and leading to a more predominantly radial movement.

\begin{figure*}
    \centering 
    {SPH vs. MFM (Figure \ref{fig:vortex_freddy})}\par
    \includegraphics[width=0.42\textwidth]{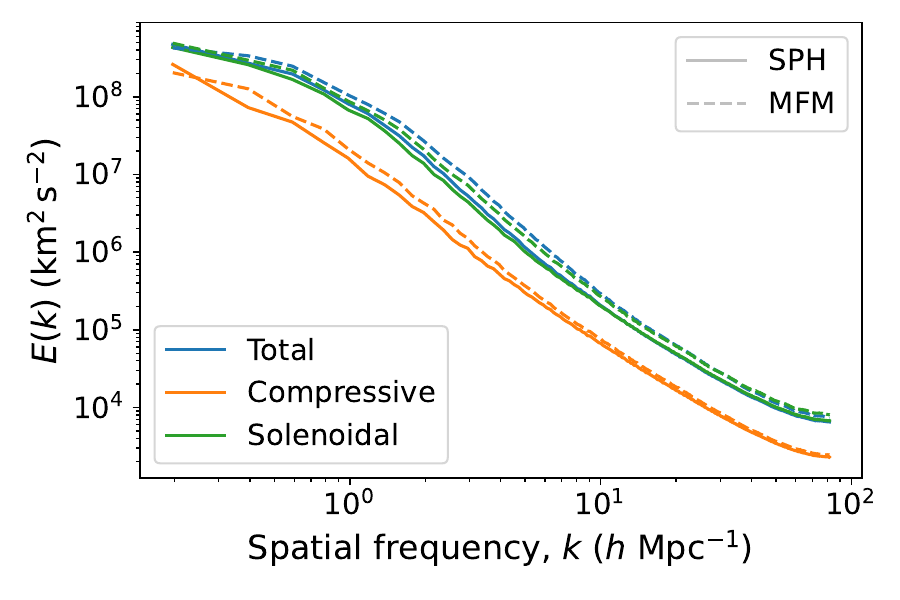}~
    \includegraphics[width=0.42\textwidth]{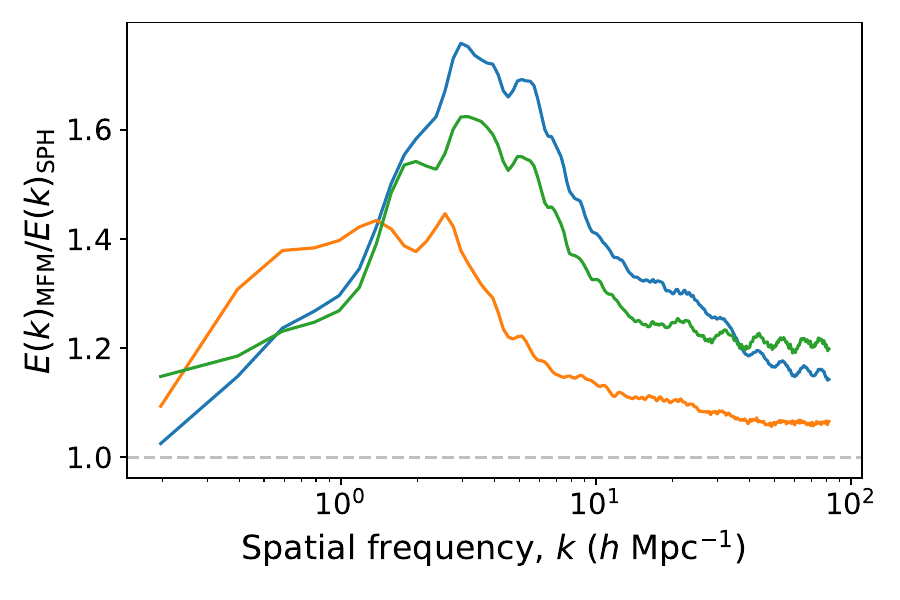}\par
    {Ideal vs. Viscosity (Figure \ref{fig:vortex_tirso})}\par
    \includegraphics[width=0.42\textwidth]{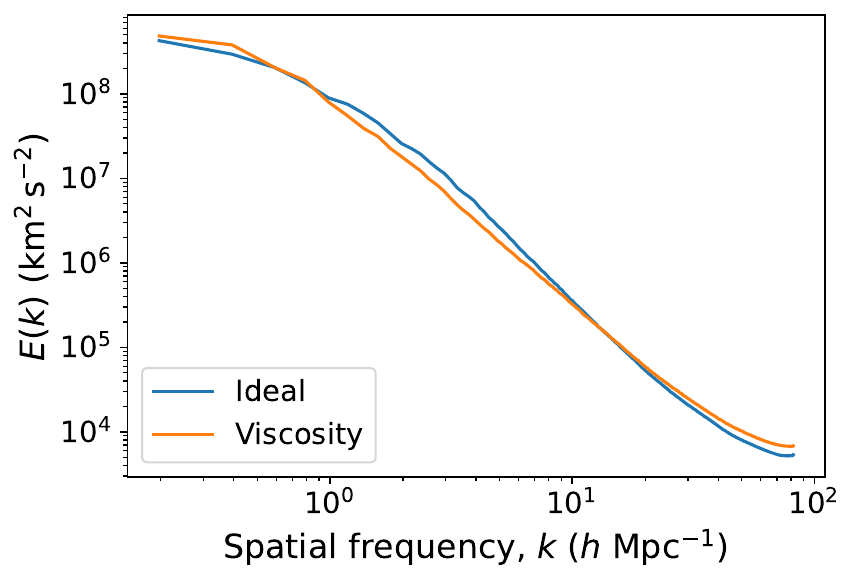}~
    \includegraphics[width=0.42\textwidth]{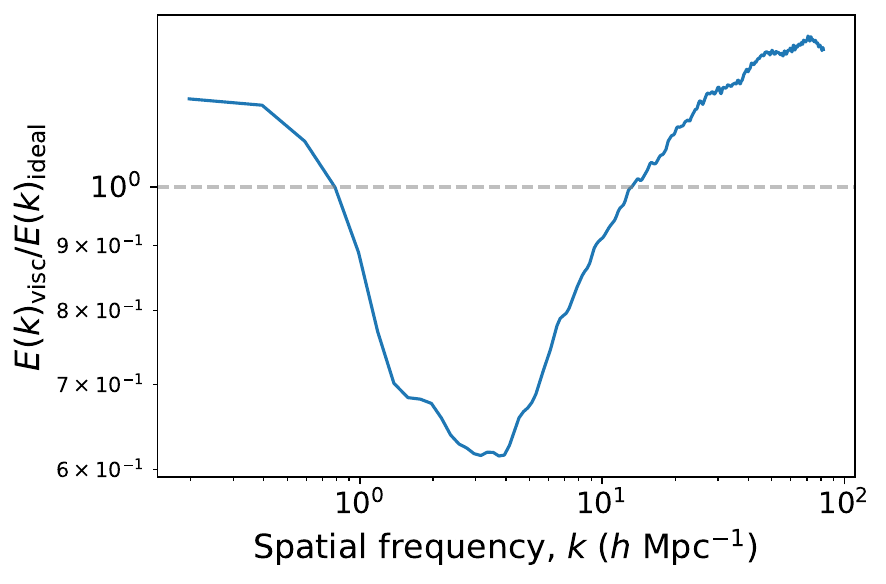}
    \caption{\textit{Top panels:} Comparison of the kinetic energy spectra (for the total velocity, in blue; and for the compressive and solenoidal, in orange and green, respectively) for SPH (solid lines) and MFM (dashed lines). For better comparison, the right-hand side panels show the ratio of the MFM to SPH spectra. \textit{Bottom panels:} Same as above, but for the ideal (solid lines) vs. viscosity (dashed lines) comparison, only for the total velocity.}
    \label{fig:vortex_powerspectra}
\end{figure*}

Again, this is shown more quantitatively through the energy spectra in the bottom panels of Fig. \ref{fig:vortex_powerspectra}. The left-hand side panel shows the energy spectra of the total velocity for the case without viscosity (blue) and with viscosity (orange). To better highlight the differences, the right-hand side panel shows the ratio between these two. Interestingly, while at medium scales ($1 \, h \, \mathrm{Mpc}^{-1} \lesssim k \lesssim 10 \, h \, \mathrm{Mpc}^{-1}$) viscosity is suppressing velocity fluctuations by up to $\sim 40\%$, it seems that the viscous case has higher velocity fluctuations on smaller scales ($k \gtrsim 10 \, h \, \mathrm{Mpc}^{-1}$). An interpretation for this apparently counter-intuitive result could be that, while viscosity is damping velocity fluctuations on small scales, the absence of turbulent motions suppresses gas mixing, generating the more extreme density fluctuations seen in the upper panels of Fig. \ref{fig:vortex_tirso}. In turn, this may be causing the velocity field to display higher fluctuations on these small scales, as a consequence of the more complex mass distribution. This will be analysed in more detail in Marin-Gilabert et al., in prep.

While giving a thorough physical interpretation of these results is beyond the scope of this short section, the examples presented above show the potential of \vortexp{} as a tool for analysing the velocity field in particle-based simulations.

\begin{figure}
    \centering 
    \includegraphics[width=0.5\textwidth]{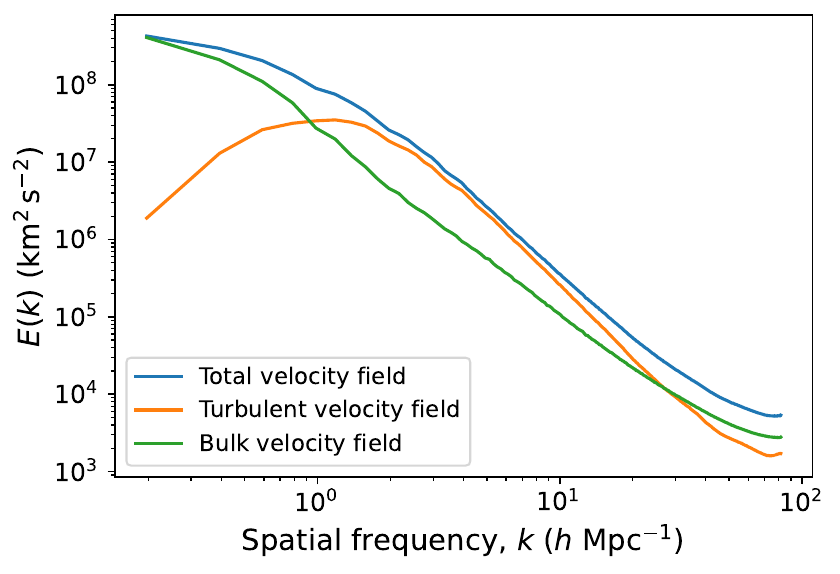}
    \caption{Kinetic energy spectra for the ideal SPH simulation (left panels of Fig. \ref{fig:vortex_tirso}), for the total velocity field (blue), and for the bulk and turbulent components (green and orange, respectively), as a quantifiable test of the multi-scale filter for the Reynolds decomposition.}
    \label{fig:powerspectra_reynolds}
\end{figure}

Finally, as a more quantitative test of the Reynolds decomposition performed by \vortexp{}, Fig. \ref{fig:powerspectra_reynolds} shows several energy spectra of the same simulation in the left panels of Fig. \ref{fig:vortex_tirso}. In particular, the blue line corresponds to the energy spectrum of the total velocity field, while the orange and green lines contain the same information about the turbulent and the bulk velocity fields, respectively. 
These spectra show that, while the bulk velocity field is contributed especially by the longest wavelengths, the turbulent part is strongly suppressed at scales $\gtrsim 1 h^{-1} \, \mathrm{Mpc}$, and dominates by a factor of $\sim 5$ in the range $1 h \, \mathrm{Mpc}^{-1} \lesssim k \lesssim 25 h \, \mathrm{Mpc}^{-1}$. At scales smaller than $40 h^{-1} \, \mathrm{kpc}$, the bulk component is again dominant. Far from being a physical effect or a feature of the filtering scheme, this only appears as a consequence of the finite resolution of the simulation. Indeed, the median SPH smoothing length within the box where the spectra are computed is $\sim 55 h^{-1} \, \mathrm{kpc}$, rendering the results on scales smaller than this --when averaged over the whole virial volume-- nonphysical due to being below the numerical dissipation scale. Still, at the scales that can be accurately probed in this test, the results confirm a correct qualitative behaviour of the filtering scheme.

\section{Conclusions}
\label{s:conclusion}

The analysis of turbulent velocity fields emerging from CFD simulations of astrophysical systems is highly non-trivial, amongst other reasons, because of the intrinsically multi-scale nature of astrophysical flows --exhibiting features spanning a wide range of scales--, the highly compressible nature of astrophysical plasmas, and the corresponding presence of strong gradients, shocks and other discontinuities. These analyses are further complicated when the fluid is discretised with non-constant spatial resolution, as it is the case of AMR, but also particle-based and moving-mesh codes, which are amongst the most widely used in the astrophysical community. 

In this work, we have introduced \vortexp{}, a novel tool for the analysis of the velocity fields of particle-based, meshless or moving-mesh simulation data in post-processing, that builds on the algorithm introduced for patch-based AMR data presented by \citet{Valles-Perez_2021_CPC} and extended in \citet{Valles-Perez_2021_MNRAS}. Our algorithm relies on the representation of the velocity field defined by a set of particles (or mesh-generating points) on an ad-hoc AMR set of nested grids. From this representation, we are able to perform the Helmholtz-Hodge decomposition using a combination of FFT and iterative solvers that can bring down the algorithmic complexity of the problem as low as $\mathcal{O}(N \log N)$. For the Reynolds decomposition, \vortexp{} implements a multi-scale filter that allows for the extraction of the turbulent and bulk components of the velocity field. While this strictly implies a $\mathcal{O}(N^2)$ computation, the use of the spatial information of the AMR representation enables a significant reduction of the computational cost.

The implementation of our algorithms has been tested through several idealised and more complex tests in a variety of configurations regarding the choice of free parameters that determine the mesh creation and the velocity assignment procedure. Typical relative errors in the different tests are found to be in the order of $\sim 10^{-3}$, with a satisfactory convergence behaviour when increasing the resolution of the input data, and clear trends in what regards the choices of free parameters for the velocity assignment. Furthermore, we have assessed the computational performance of the code, its scaling with the problem size (measured through the number of particles, at least up to $N \sim 10^8$), and the parallel scaling properties of its OMP implementation.

Finally, we have shown the results from several applications of \vortexp{} to actual simulation data from SPH and MFM simulations of galaxy clusters with different physics, in order to illustrate the potential of the code for the analysis of the velocity field in particle-based simulations. These applications have shown the ability of \vortexp{} to capture the differences in the velocity field structure between different simulation codes and different physical setups, and to provide a quantitative assessment of the Reynolds decomposition performed by the code.

The OMP-parallelised implementation of \vortexp{} is publicly available (see Sec. \ref{s:method}) and will enable the detailed study of different aspects related to astrophysical turbulence in galaxy cluster environments in the near future, among other applications.


\section*{Acknowledgements}
We warmly thank the two anonymous reviewers for their careful reading and thorough suggestions on our manuscript, which have enabled us to improve the presentation of our work. This work has been supported by the Agencia Estatal de Investigación Española (AEI; grant PID2022-138855NB-C33), by the Ministerio de Ciencia e Innovación (MCIN) within the Plan de Recuperación, Transformación y Resiliencia del Gobierno de España through the project ASFAE/2022/001, with funding from European Union NextGenerationEU (PRTR-C17.I1), and by the Generalitat Valenciana (grant CIPROM/2022/49). DVP acknowledges support from Universitat de València through an Atracció de Talent fellowship, and gratefully thanks the hospitality of the Universitäts-Sternwarte München, where part of this work was done during a research stay funded by Universitat de València. Part of the tests have been carried out using the supercomputer Lluís Vives at the Servei d'Informàtica of the Universitat de València. KD, FG and TMG are acknowledging support by the COMPLEX project from the European Research Council (ERC) under the European Union’s Horizon 2020 research and innovation program grant agreement ERC-2019-AdG 882679. KD and FG acknowledge support by the Deutsche Forschungsgemeinschaft (DFG, German Research Foundation) under Germany’s Excellence Strategy - EXC-2094 - 390783311.

\bibliographystyle{elsarticle-num-names} 
\bibliography{cpc_vortex.bib}

\clearpage
\begin{appendix}
\numberwithin{figure}{section}

\section{Assessment of the errors introduced by the imposition of periodicity}
\label{s:appendix.periodicity}

\begin{figure*}
    \centering 
    \includegraphics[width=0.5\linewidth]{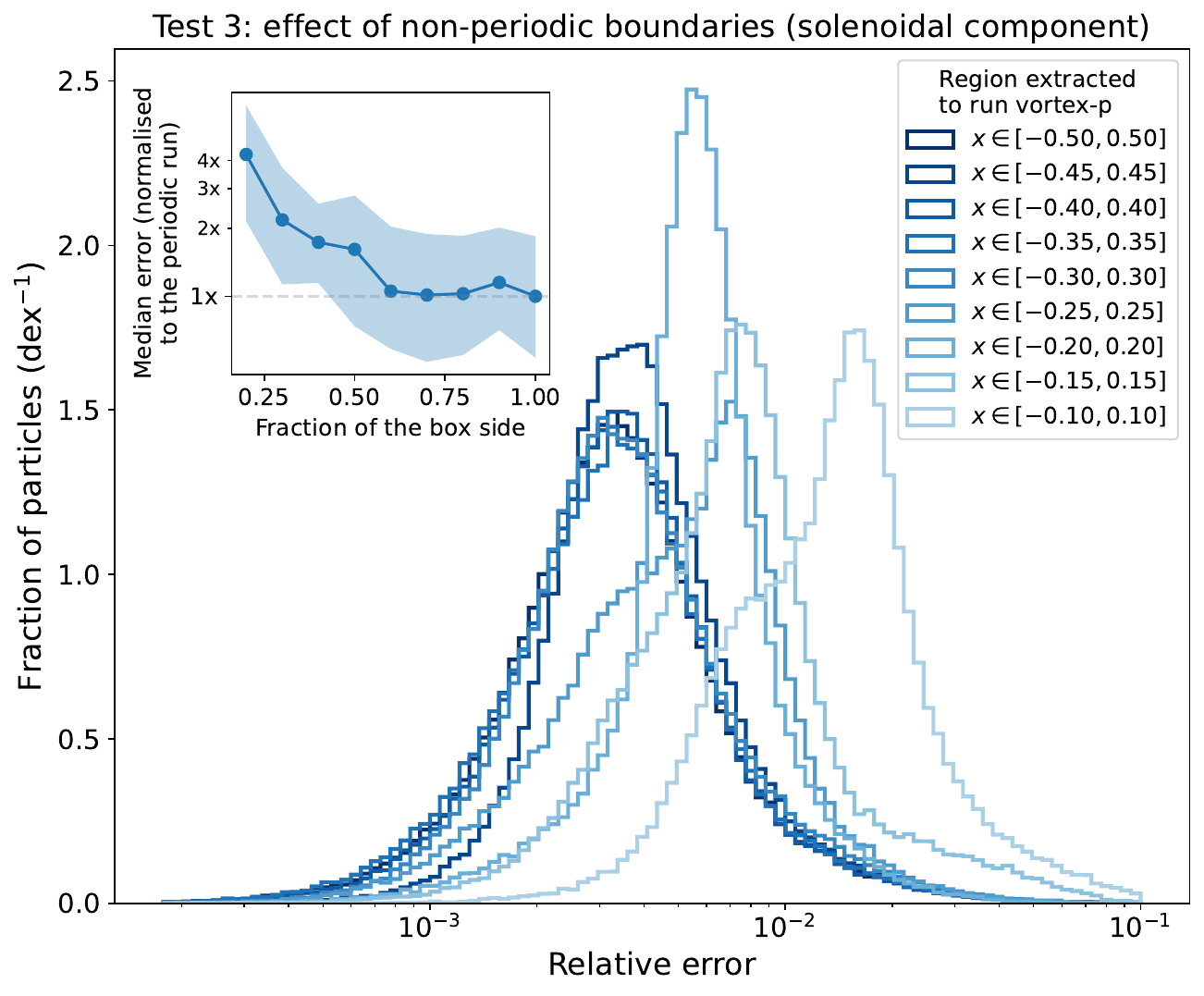}~
    \includegraphics[width=0.5\linewidth]{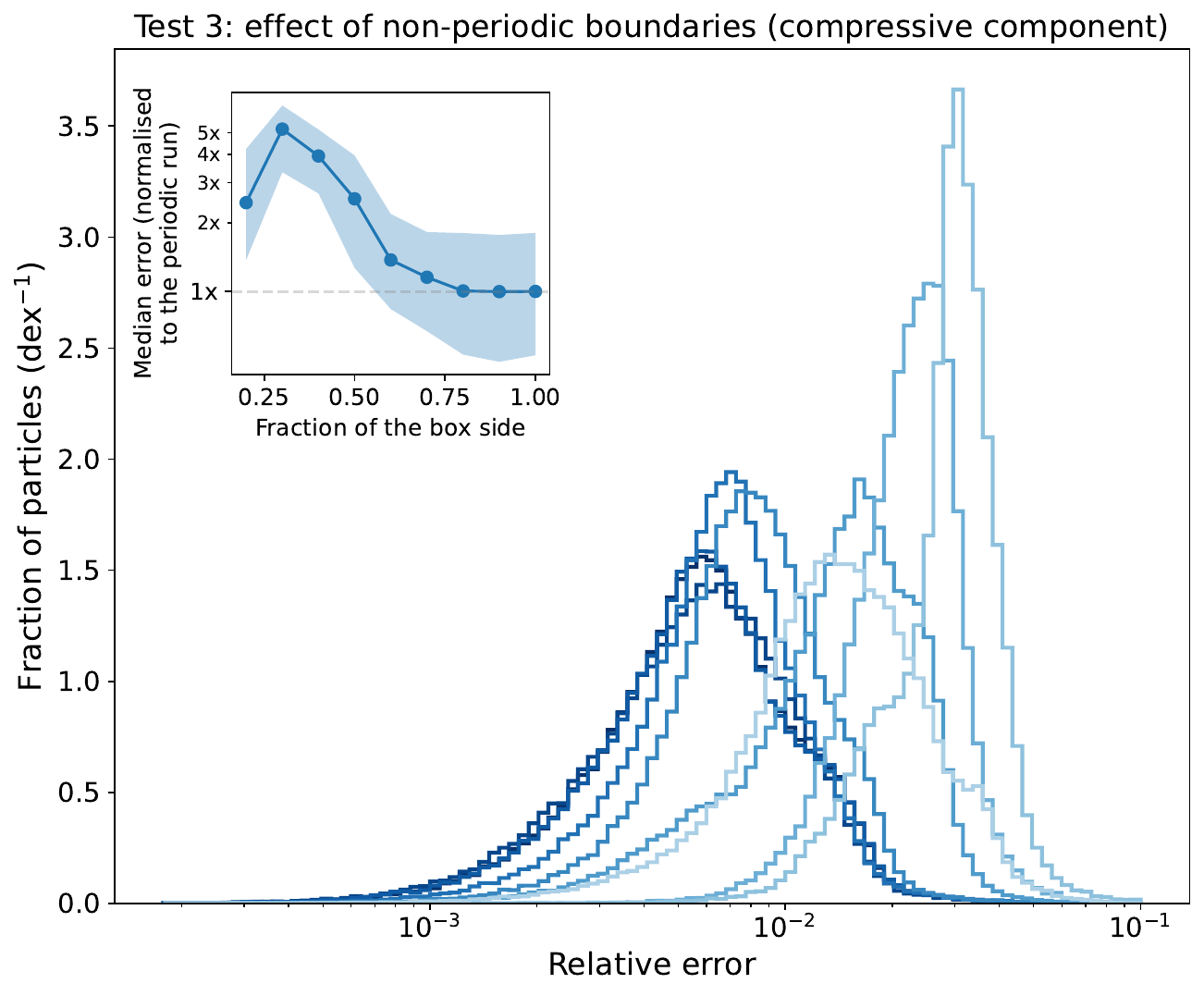}
    \caption{Study of the effect of non-periodicity on the results of the HHD. The left-hand side and right-hand side panels show the results, for the solenoidal and compressive components respectively, of Test 3 with the C$^4$ kernel and $N_\mathrm{ngh}=137$, when restricting the input domain to smaller regions, in such a way that the boundary conditions are non-periodic. The results are always assessed in the cubic reigon $x,y,z \in [-0.1, 0.1]$, so that the effect of the distance of the region of interest to the location of the non-periodic boundary can be studied. The inset summarises the median and $(16-84)$-percentiles of these distributions, normalised to the results for the periodic situation ($f=1$).}
    \label{fig:appendix.periodicity}
\end{figure*}

In general, a simulation domain may be non-periodic, or may correspond to a region much larger than the region of interest where \vortexp{} is to be applied. However, the usage of FFT for the base grid implies periodicity. While the computation of Fourier transforms of non-periodic regions is fairly common in the literature studies of velocity fields, e.g. to extract spectra, it is still interesting to explicitly validate the process and assess the errors introduced by the lack of periodicity. 

To do so, we have repeated Test 3 (Sec. \ref{s:tests.test3}), which was explicitly periodic in $x,y,z \in [-L/2, L/2]$ for the case using the C$^4$ kernel and $N_\mathrm{ngh}=137$, but restricting the input domain to increasingly smaller fractions $f$ of the domain, so as to generate a non-periodic situation by setting the domain to $x,y,z \in [-fL/2, fL/2]$. In particular, assuming $L=1$ for simplicity, let us consider that the region of interest (where the errors will be evaluated) corresponds to $x,y,z \in [-0.1, 0.1]$. In Fig. \ref{fig:appendix.periodicity}, we show the distribution of particle-wise errors (as it has been done in all the tests in Sec. \ref{s:tests}) for several input regions, with darkest colours corresponding to larger paddings around the region of interest (larger $f$).

The results can be summarised in the inset plot, where we show the errors as a function of the input domain sidelength (the smaller, the more restricted), normalised to the error when the whole, periodic domain is kept ($f=1$). This is shown for the solenoidal and the compressive component, respectively, in the left-hand side and right-hand side panels. As a general trend, as $f$ is reduced and the non-periodic boundaries get closer to the region of interest, there is no increase in the error figures at least until $f=0.6$. This implies that \vortexp{} can be safely applied to a restricted region of a larger simulation just by taking the precaution of choosing a domain around the object of interest $\sim 3$ times its size. 

While, in this test, the errors in the compressive component grow slightly faster than the ones in the solenoidal component as the input domain is shrunk, this might be dependent on the particular velocity field and we do not pursue a more in-depth examination here.

\section{Assessment of the level of conservativeness of the velocity assignment scheme}
\label{s:appendix.conservativeness}

\begin{figure}
    \centering 
    \includegraphics[width=0.5\textwidth]{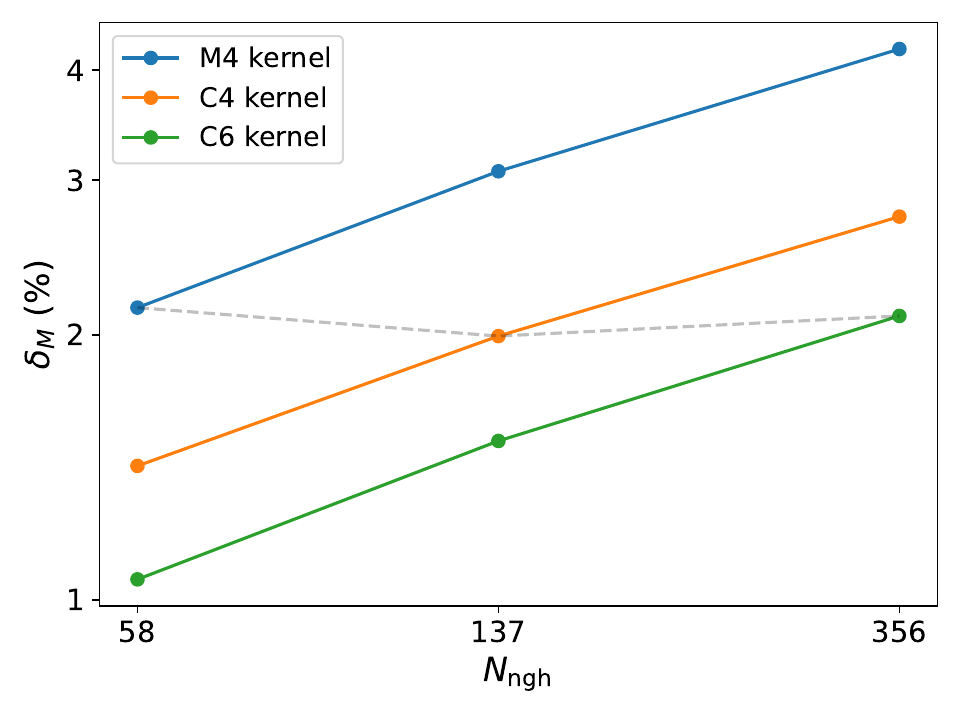}
    \caption{Assessment of the level of conservativeness of the grid-assignment procedure followed in \vortexp{}. Different lines correspond to different kernel functions and show, as a function of the number of neighbours within the kernel compact support, the error in the assignment of density to the AMR grid, computed according to Eq. \ref{eq:density_assignment}. The dashed line joins three runs with equivalent kernel effective extent (M$_4$ with $N_\mathrm{ngh}=58$, C$^4$ with $N_\mathrm{ngh}=137$ and C$^6$  with $N_\mathrm{ngh}=356$).}
    \label{fig:appendix.conservativeness}
\end{figure}

As mentioned in Sec. \ref{s:method.interpolation}, within the discussion of the interpolation procedure, it is by design non-conservative, in the sense that the volume integral of the gridded data does not necessarily equate to the sum of the same quantity over the particles. The reason for this stems from the fact that we are not performing a standard SPH sum, where each particle contributes according to its own smoothing length, but assigning a kernel length to each cell centre instead due to the requirements of our velocity assignment procedure. While there is no reason for this condition to hold on a non-extensive quantity such as velocity, it is still interesting to check to which extent does the interpolation violate the conservation of a given input field.

With that aim, we have used data from the SPH simulation also shown in the left-hand side panel of Fig. \ref{fig:vortex_freddy}, and assigned the density to each cell centre as

\begin{equation}
    \rho_\text{cell} = \frac{\sum_{\alpha \in \text{kernel}} m_\alpha}{\frac{4\pi}{3} h_\text{cell}^3},
    \label{eq:density_assignment}
\end{equation}

\noindent where the sum is performed over the particles in a sphere of radius $h_\text{cell}$ around the cell centre. Then, we estimate the global violation of conservativity as:

\begin{equation}
    \delta_M = \frac{\sum_\text{cells} \rho_\mathrm{cell} \Delta V_\mathrm{cell} - \sum_\alpha m_\alpha}{\sum_\alpha m_\alpha},
\end{equation}

\noindent which is the quantity shown in Fig. \ref{fig:appendix.conservativeness} as a function of the kernel family and the number of neighbours used for the interpolation. As a general trend, conservativeness is better preserved when the extent of the kernel used for the assignment is smaller: that is to say, the lower $N_\mathrm{ngh}$ is, and the lower the standard deviation of the kernel, $\sigma/h$, is. 

As a consistency check, we finally note that the results obtained with the M$_4$ kernel with 58 neighbours, C$^4$ kernel with 137 neighbours and C$^6$ kernel with 356 neighbours, which have the same kernel extent, are roughly equal, as represented by the dashed line in the figure.

\section{Discussion on the error estimation for the tests}
\label{s:appendix.error_definition}

\begin{figure*}
    \centering 
    \includegraphics[width=0.5\linewidth]{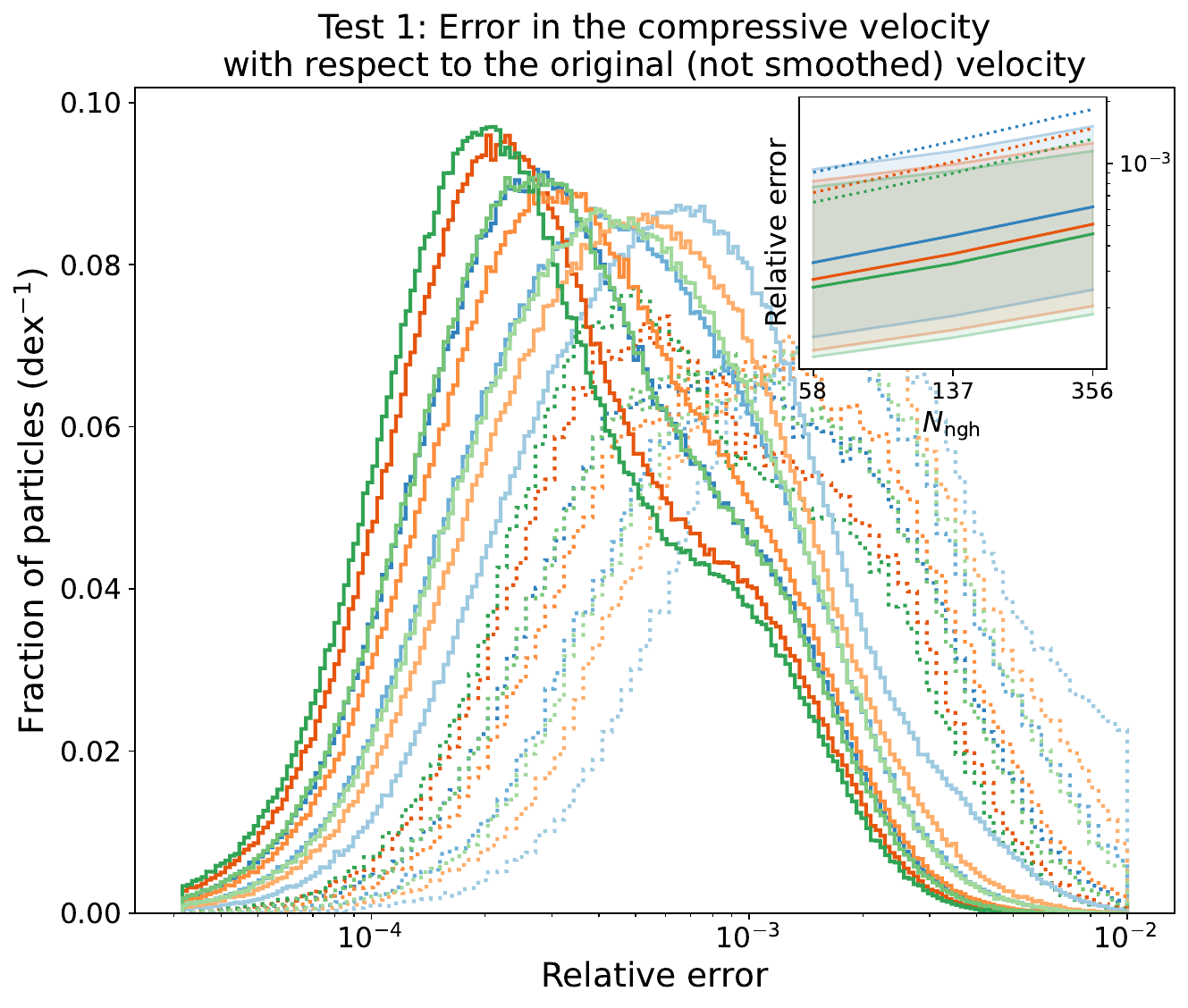}~
    \includegraphics[width=0.5\linewidth]{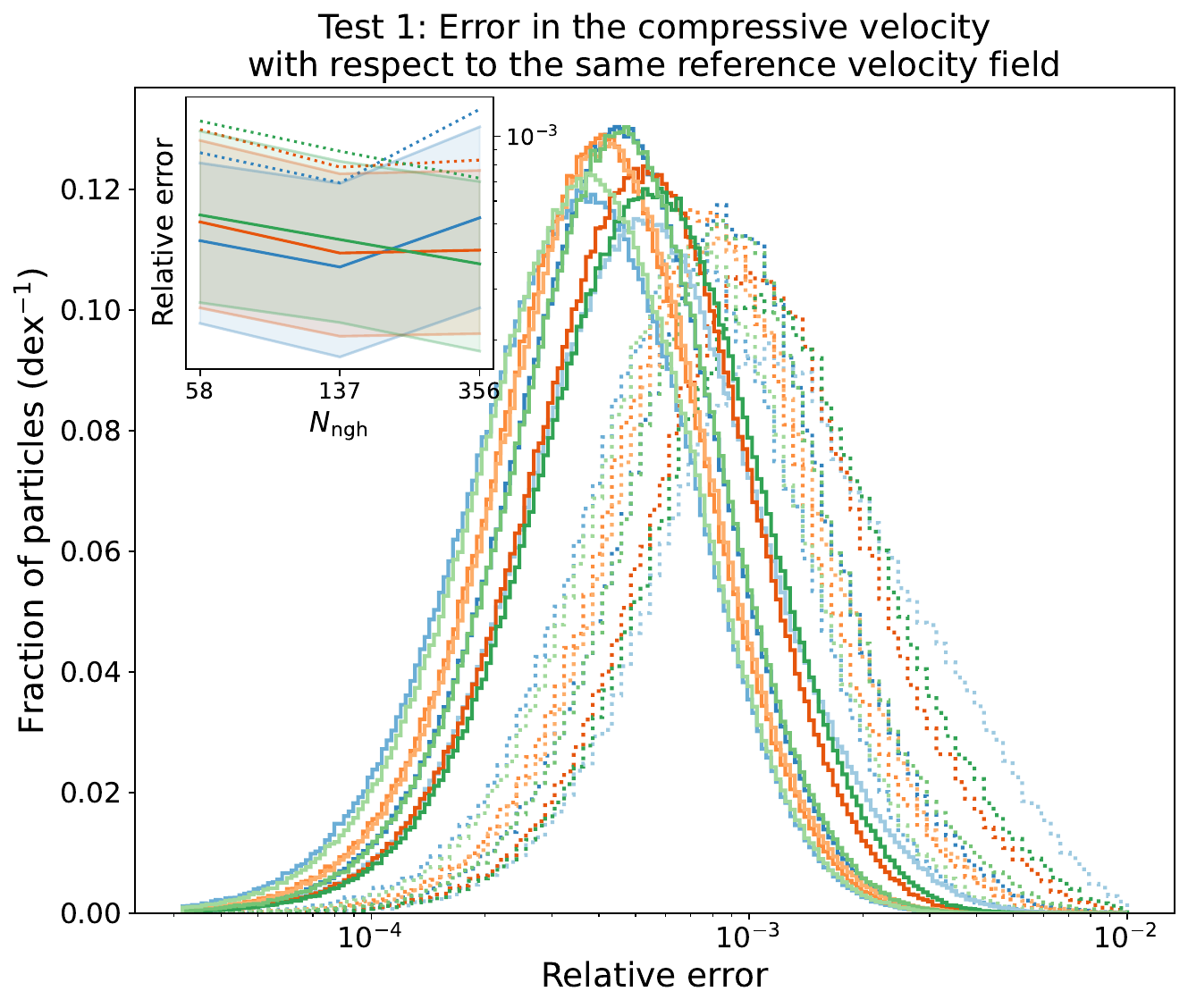}
    \caption{Discussion on the error estimation for the tests. Both panels are equivalent to the right-hand side panel of Fig. \ref{fig:test1_results}, but with different error definitions. \textit{Left-hand side panel}: error computed with respect to the original input velocity. \textit{Right-hand side panel}: error computed with respect to the same reference velocity, i.e., the one smoothed with the C$^4$ kernel comprising $N_\mathrm{ngh}=137$ neighbours.}
    \label{fig:appendix.error_definition}
\end{figure*}

Through the tests in Sec. \ref{s:tests}, we have chosen to compute the errors as a particular measure (see Eqs. \ref{eq:test1_error_absent} and \ref{eq:test1_error_present}) of the difference between the results decomposed by \vortexp{} and the known decomposition, smoothed with the same kernel function and number of neighbours as used when running the code in each case. It would be possibly to argue, however, that this approach simplifies the problem for more extended kernels, since it gets rid of increasingly higher frequency modes. Nevertheless, neither comparing the decomposed velocities with the input velocities is strictly fair, since \vortexp{} represents the smoothed velocity field, and this is the one which the decomposition is brought up on. This would imply that, if compared to the input, non-smoothed velocity field, the more compact kernels ought to perform comparatively better just due to the fact that the input and output velocity field differ by a lesser amount due to the smoothing operation. This is checked empirically in the left-hand side panel of Fig. \ref{fig:appendix.error_definition}, which is equivalent to the right-hand side panel of Fig. \ref{fig:test1_results} but with this alternative definition of error.

Alternatively, one could perform the tests by stating that the ground smoothed velocity field is a particular one (e.g., the one computed  with the C$^4$ kernel with $N_\mathrm{ngh}=137$), for instance because it is assumed that this velocity field has been the product of a simulation using this particular configuration for evolving the SPH equations, and compare all the results to this smoothed solution. This is precisely what we show in the right-hand side panel of Fig. \ref{fig:appendix.error_definition}. The resulting error distributions, which are better interpreted in the inset showing the behaviour of the mean and $(16-84)\%$ quantiles, are rather non-trivial, in the sense that there is not a monotonic trend with $\sigma/h$ nor with $N_\mathrm{ngh}$.

Naturally, this discussion is only relevant due to the artificiality introduced by the tests. In an actual simulation output, the best choice is to use the same smoothing operation as used for the evolution, since this is the only fully physically meaningful one. In the tests, we have introduced a mock velocity field which has a known decomposition. However, the grid assignment with non-uniform smoothing length does not preserve the solenoidal/compressive behaviour of these fields. Unfortunately, we are not aware of any simple method capable of assigning a non-trivial velocity field onto a particle distribution in such a way that a given divergence/curl (in the SPH sense) is obtained. Therefore, we have chosen to present these tests which, despite the fact that they need to be interpreted with caution due to their artificiality, inform precisely about the amount of cross-talk between the velocity components introduced by our grid assignment strategy.

\section{Further notes on the velocity assignment scheme}

In full consistency with SPH, velocities should be assigned to a cell centre $\vb{x}$ as

\begin{equation}
    \vb{v}(\vb{x}) = \frac{\sum_i m_i \vb{v_i} W(|\vb{x}-\vb{x_i}|, h_i)}{\sum_i m_i W(|\vb{x}-\vb{x_i}|, h_i)}
    \label{eq:velocity_SPH}
\end{equation}

\noindent for a mass-weighted assignment, where the sum is restricted to the particles with $W(|\vb{x}-\vb{x_i}|, h_i) > 0$  and each particle contributes according to its individual smoothing length $h_i$. 

\begin{figure*}
    \centering
    \includegraphics[width=0.5\textwidth]{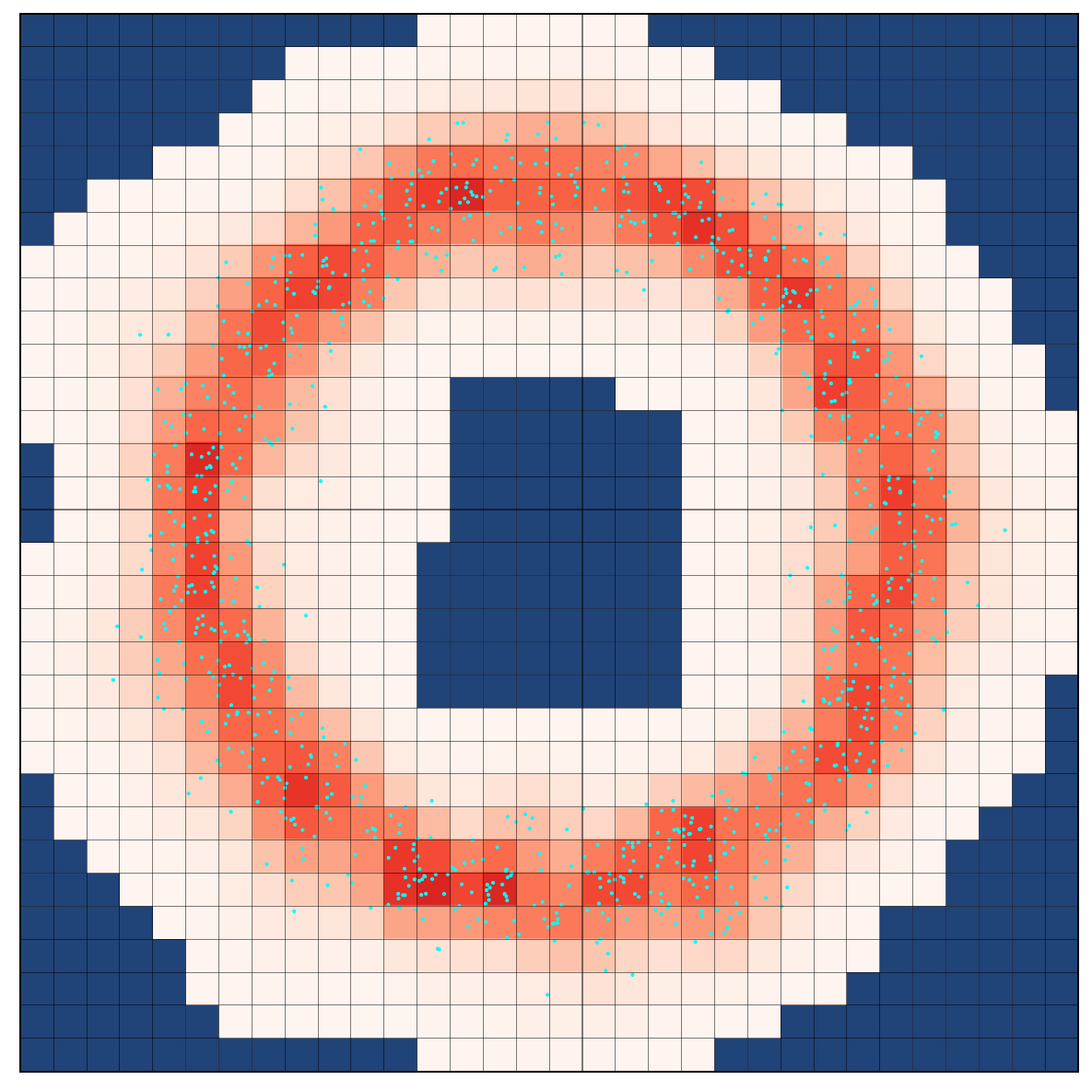}~ 
    \includegraphics[width=0.5\textwidth]{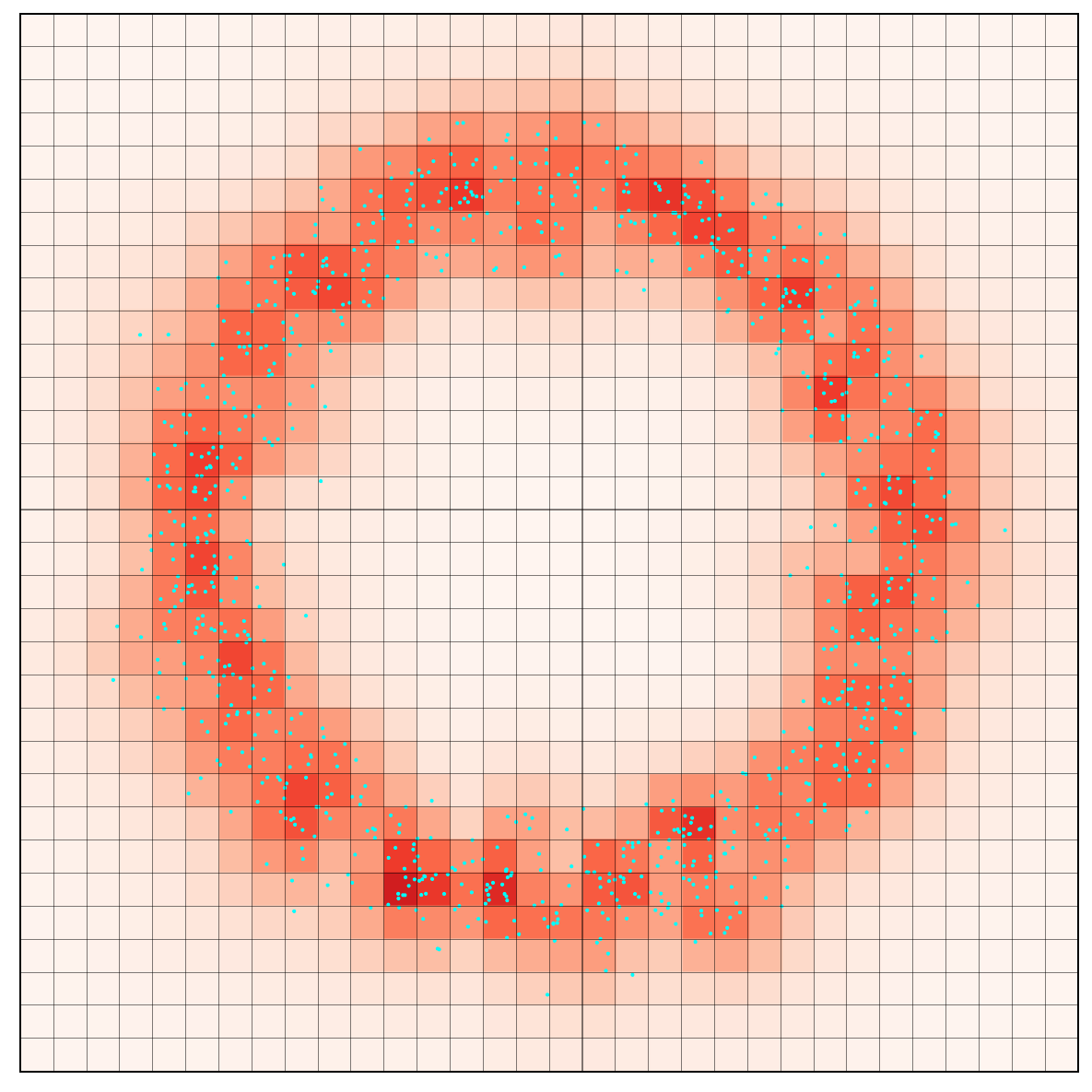}
    \caption{2-dimensional test comparison between a standard, SPH density interpolation (left-hand side panel) and the interpolation scheme involved in \vortexp{} (right-hand side panel). In both panels, cyan dots represent a particle distribution leaving a hole in the central region of the plot. Red background colours show the grid-assigned density, darkest being higher density. In the left-hand side panel, corresponding to the standard SPH interpolation where each particle does not contribute outside their smoothing length, there is a region, represented in blue, which is not contributed by any particle. While this just implies `hole' regions filled by zeros in the density interpolation (or any extensive quantity), when applied to the interpolation of an intensive quantity (e.g., velocity), indeterminate forms ($0/0$) arise. This is prevented by the grid-assignment scheme implemented in \vortexp{}.}
    \label{fig:SPH_density}
\end{figure*}

However, if considering an arbitrary grid, there could be cells that end up being contributed by no particles. To better exemplify this, let us consider a two-dimensional case for simplicity. Let us consider a ring of particles, such as the one described by the cyan dots in the left-hand side panel of Fig. \ref{fig:SPH_density}, and an arbitrary grid covering the domain as represented by the black lines (in this case, uniform for simplicity).  We have assigned the density field to this two-dimensional grid using the standard SPH sum, where we have considered $N_\mathrm{ngh}=58$ and the cubic spline kernel where each particle contributes according to their smoothing length. This is represented in the figure by the red background colour, the darkest being the densest. The cells coloured in blue correspond to grid points where no particle has contributed, so that the assigned mass is 0. Therefore, when assigning the velocity to the grid in the usual SPH manner, according to Eq. \ref{eq:velocity_SPH}, indeterminate forms ($0/0$) would arise in all this blue region.

By contrast, our grid assignment scheme presented in Sec. \ref{s:method.interpolation} prevents this downsampling by design, by ensuring that each cell is contributed by at least $N_\mathrm{ngh}$ particles, in such a way that a local smoothing length, $h(\vb{x})$ is computed on each cell centre, and that is the smoothing length used in the velocity assignment scheme, according to

\begin{equation}
    \vb{v}(\vb{x}) = \frac{\sum_i m_i \vb{v_i} W(|\vb{x}-\vb{x_i}|, h(\vb{x}))}{\sum_i m_i W(|\vb{x}-\vb{x_i}|, h(\vb{x}))}.
    \label{eq:velocity_SPH_my}
\end{equation}

\noindent This is shown in the right-hand side panel of Fig. \ref{fig:SPH_density}, where no cell would remain with an undefined velocity according to our assignment scheme.

In the standard SPH assignment, the regions shaded in blue (see left-hand side panel of Fig. \ref{fig:SPH_density}) do not belong to the domain of the solution, which therefore is not simply connected, preventing the application of Helmholtz's theorem. However, it is worth keeping in mind that the fully-consistent interpolation should be that in Eq. \ref{eq:velocity_SPH} and, therefore, it is worth testing to which extent do the velocities assigned by both methods coincide in high-density regions (where there is no undersampling in the standard SPH assignment).

\begin{figure*}
    \centering
    \includegraphics[width=0.5\textwidth]{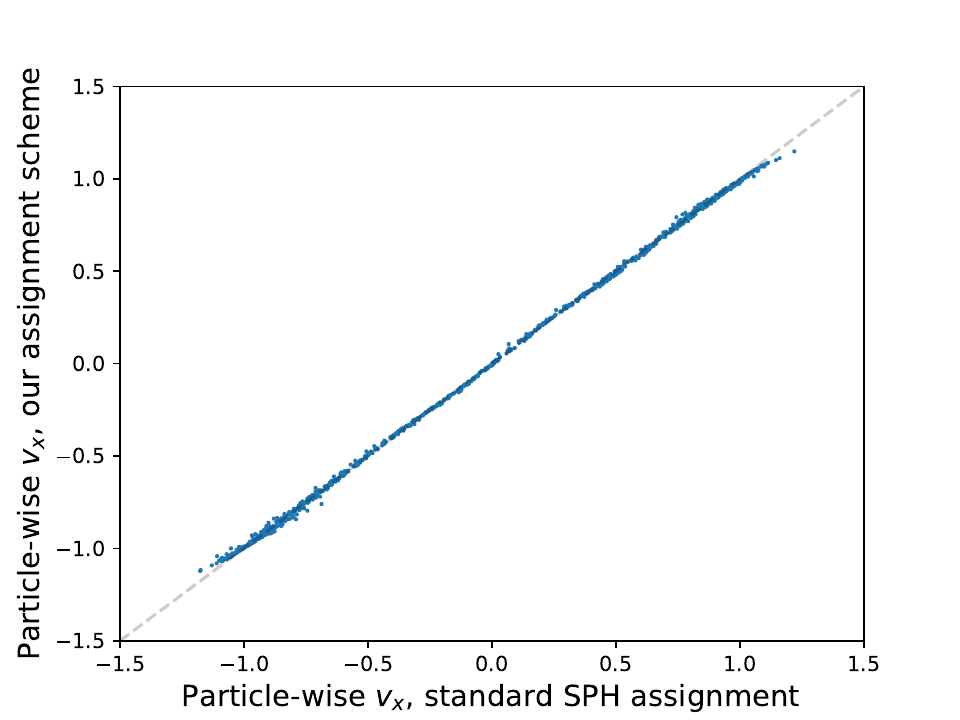}~
    \includegraphics[width=0.5\textwidth]{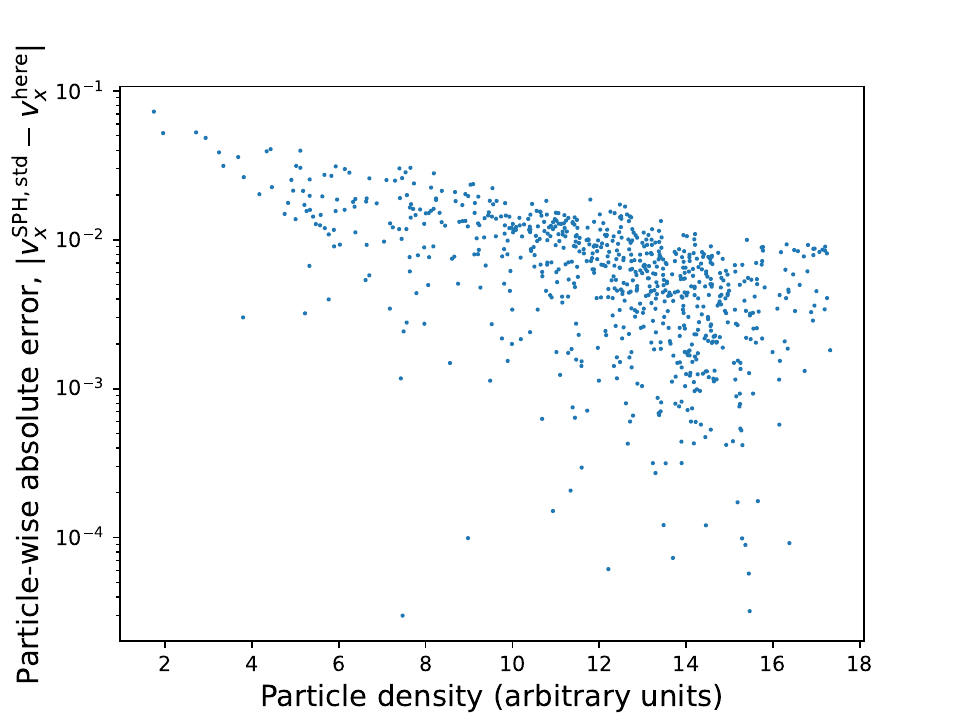}
    \caption{Quantitative comparison between the standard, SPH interpolation and the one implement in \vortexp{}. \textit{Left-hand side panel}: direct relation between the values of $v_x$ assigned according to the standard SPH interpolation (horizontal axis) and our scheme (vertical axis). The gray, dashed line is the identity relation. \textit{Right-hand side panel:} relation between the absolute discrepancy between both assignment schemes (vertical axis) and particle number density (in arbitrary units; horizontal axis).}
    \label{fig:errors_interp}
\end{figure*}

Using the same bidimensional set-up, we have assigned a simple velocity field, $\vb{v_i} = x_i \vb{\hat u_x} + y_i \vb{\hat u_y}$, to the particle distribution. We have proceeded with the purely SPH-like assignment and the one we use in \vortexp{}, and compared the values of these interpolated velocity fields at the particle positions. The left-hand side panel of Fig. \ref{fig:errors_interp} shows the relation between the SPH-assigned velocities (horizontal axis) and the velocities assigned according to our procedure, Eq. \ref{eq:velocity_SPH_my} (vertical axis). Velocities accurately follow the identity line, with small scatter and with deviations only visible for the particles with largest $|x_i|$, which is just an effect of the boundary conditions for this particular set-up.

In the right-hand side panel of Fig. \ref{fig:errors_interp}, we show the relation between the absolute discrepancy in the $x$ component of the velocity field according to both methods, and particle density (in arbitrary units). Interestingly, it is apparent from the figure that both definitions of the velocity field converge in high-resolution regions, while the largest errors only appear in low-density regions (where the SPH velocity field is, by construction, ill-defined).

\section{Tests on the M$_5$ and M$_6$ kernels at fix $N_\mathrm{ngh}$}
\label{app:kM5M6}

\begin{figure*}
    \centering
    \vspace{-2cm}
    \includegraphics[width=0.4\textwidth]{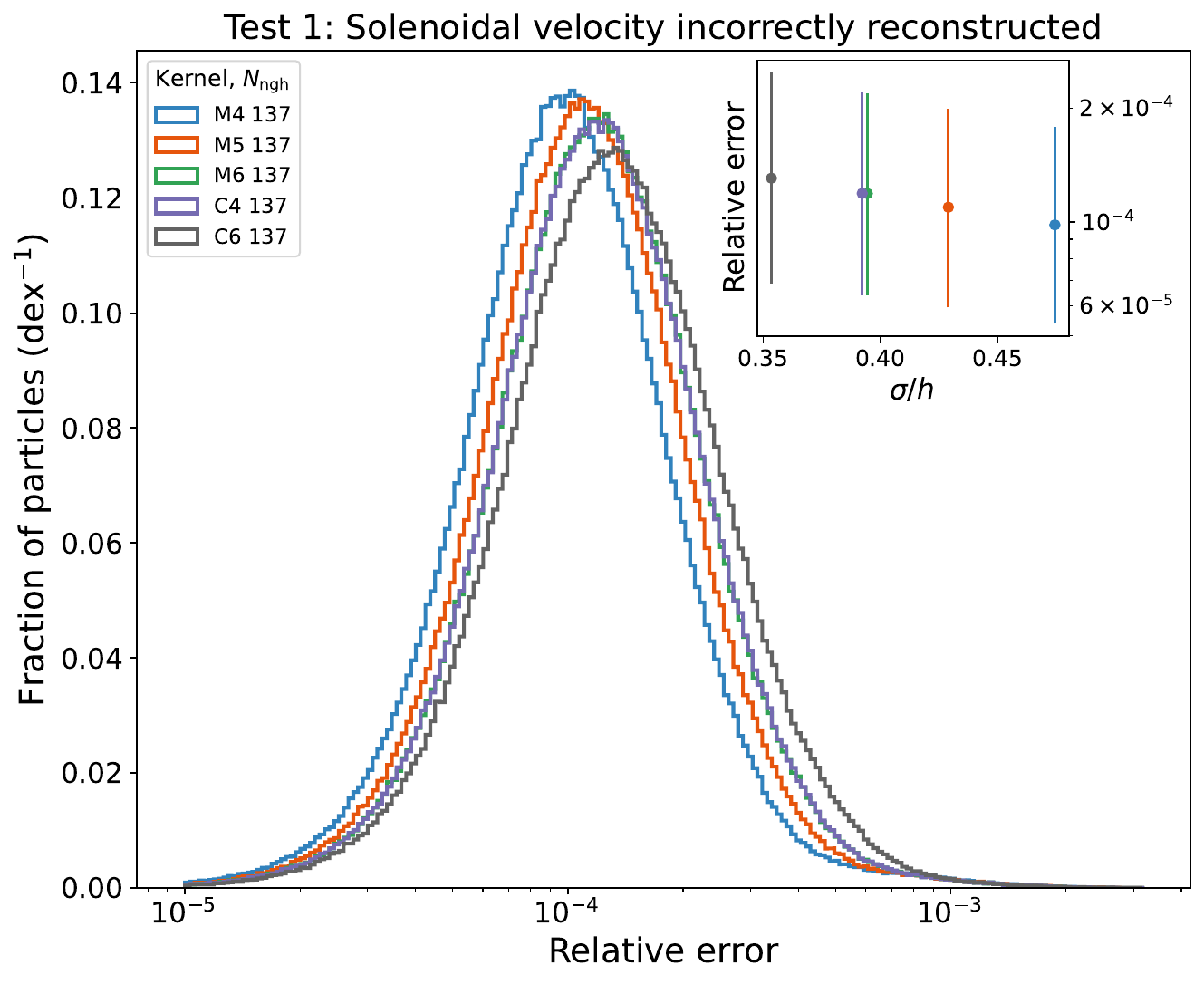}~
    \includegraphics[width=0.4\textwidth]{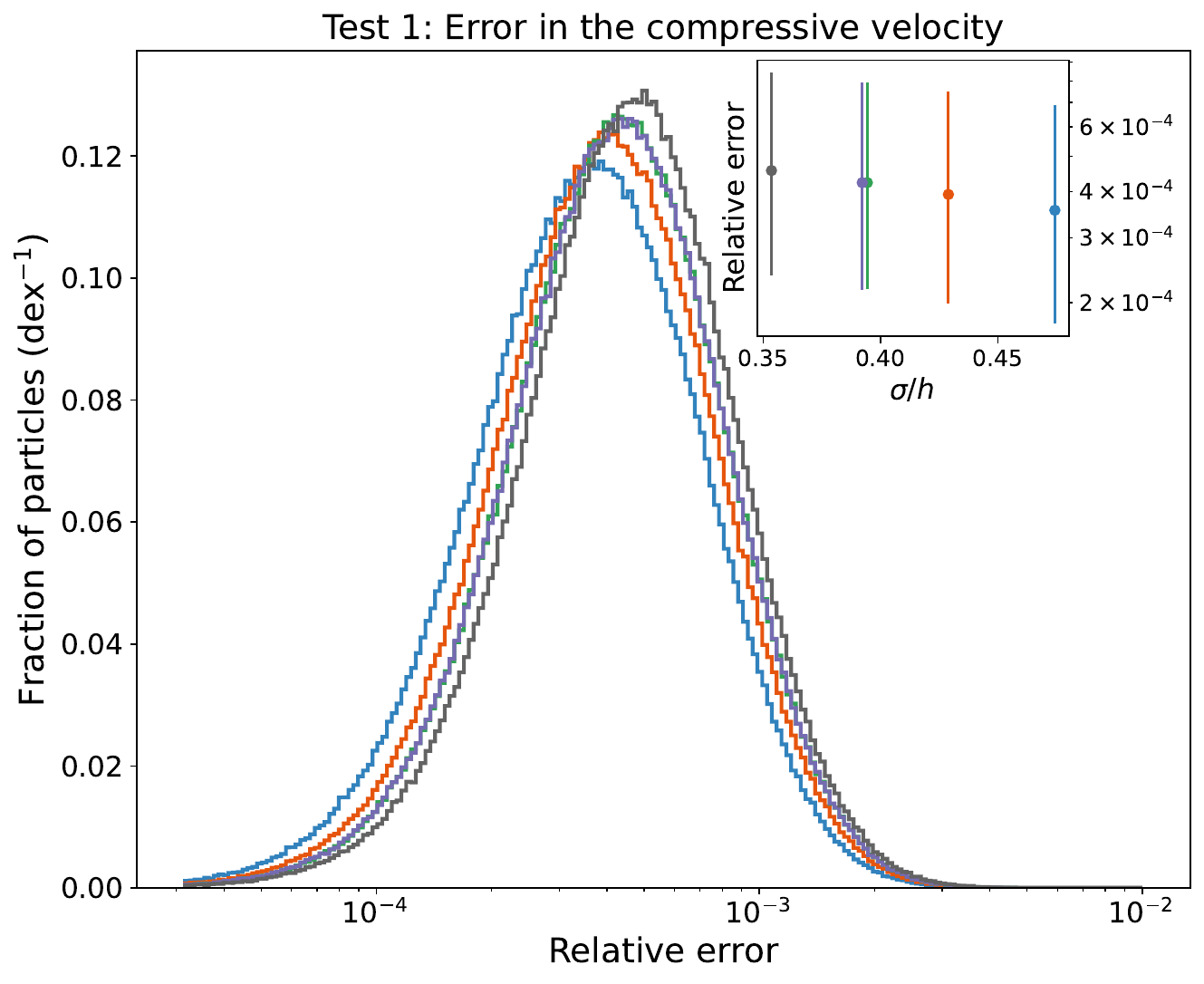}
    \includegraphics[width=0.4\textwidth]{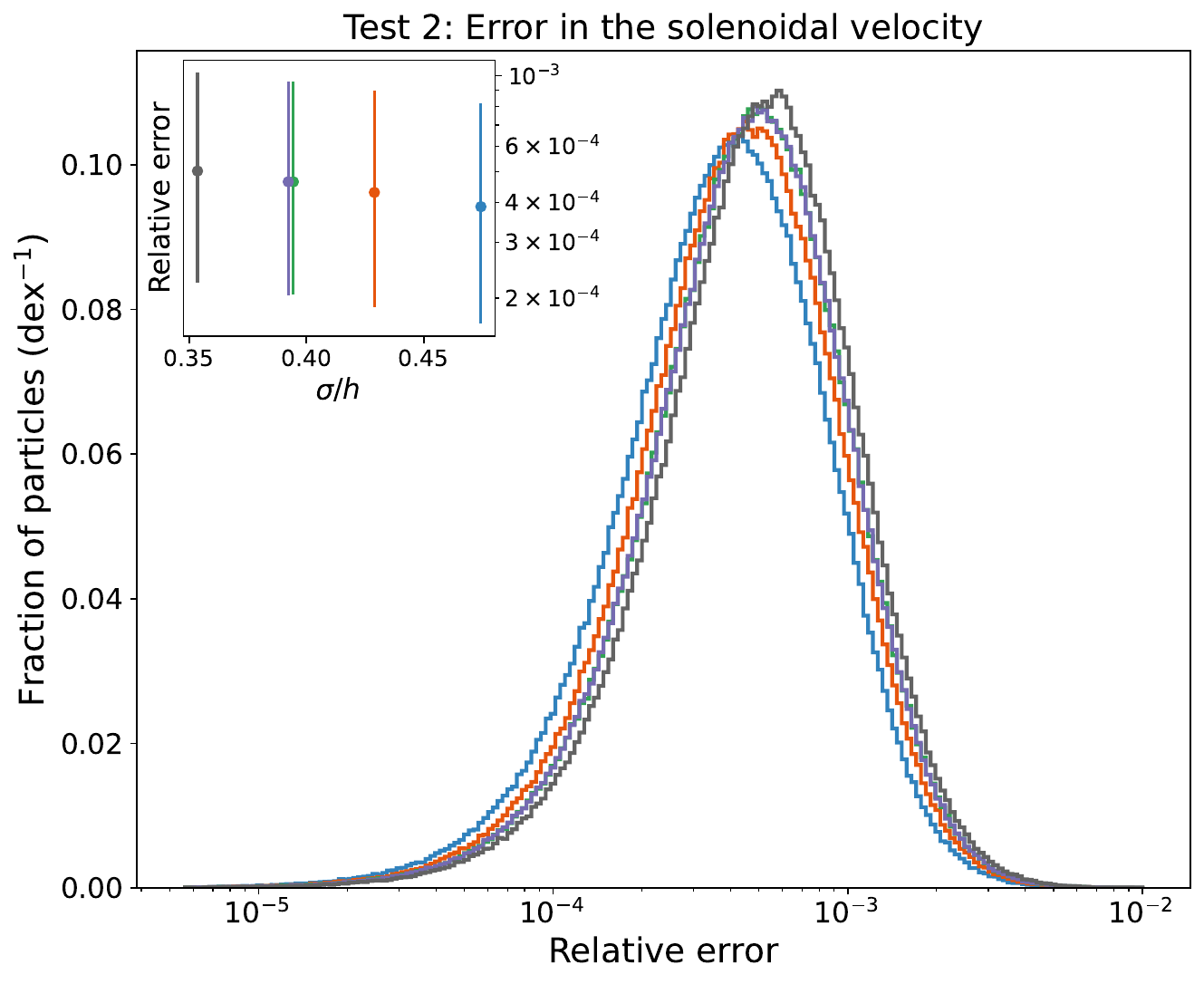}~
    \includegraphics[width=0.4\textwidth]{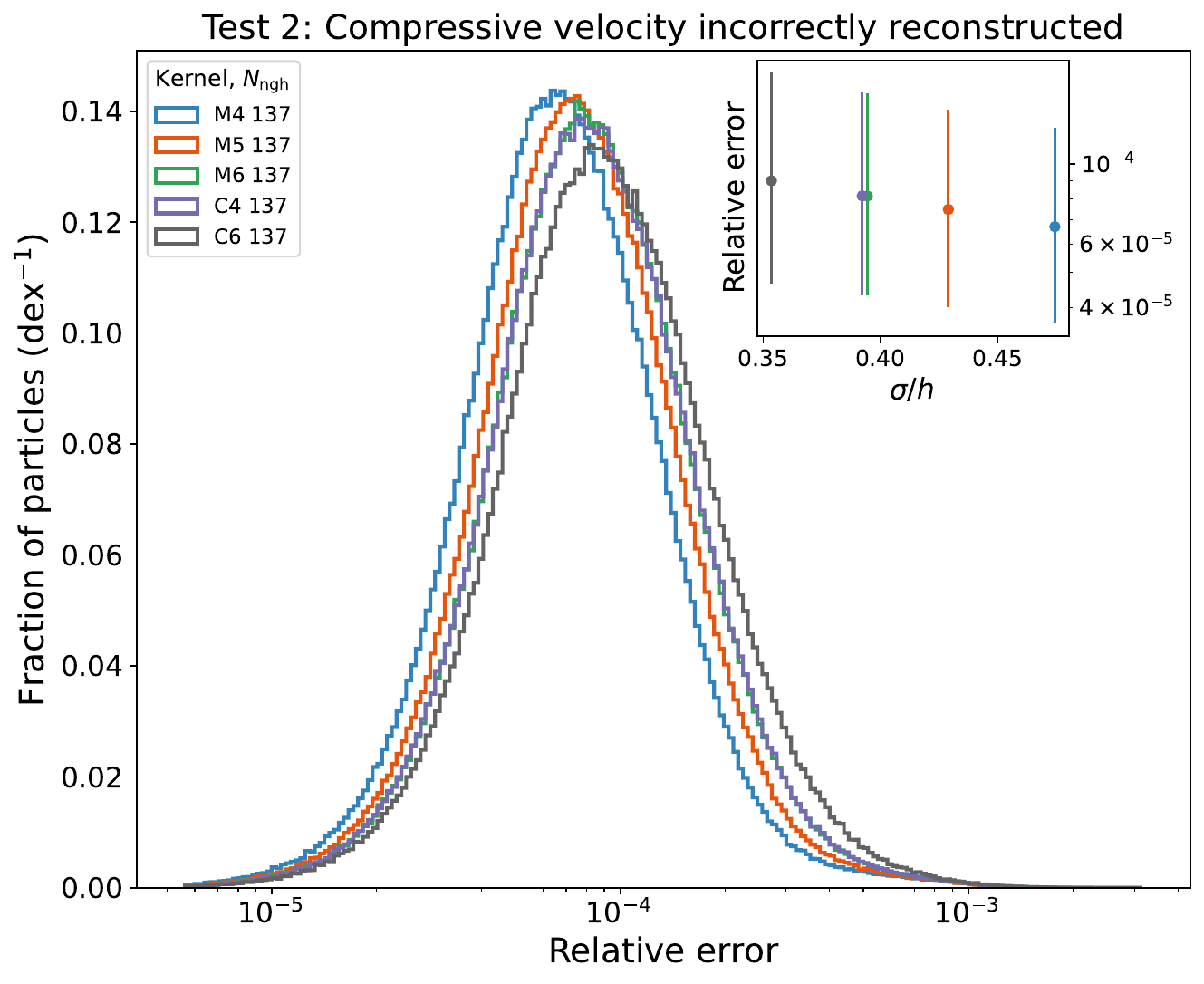}
    \includegraphics[width=0.4\textwidth]{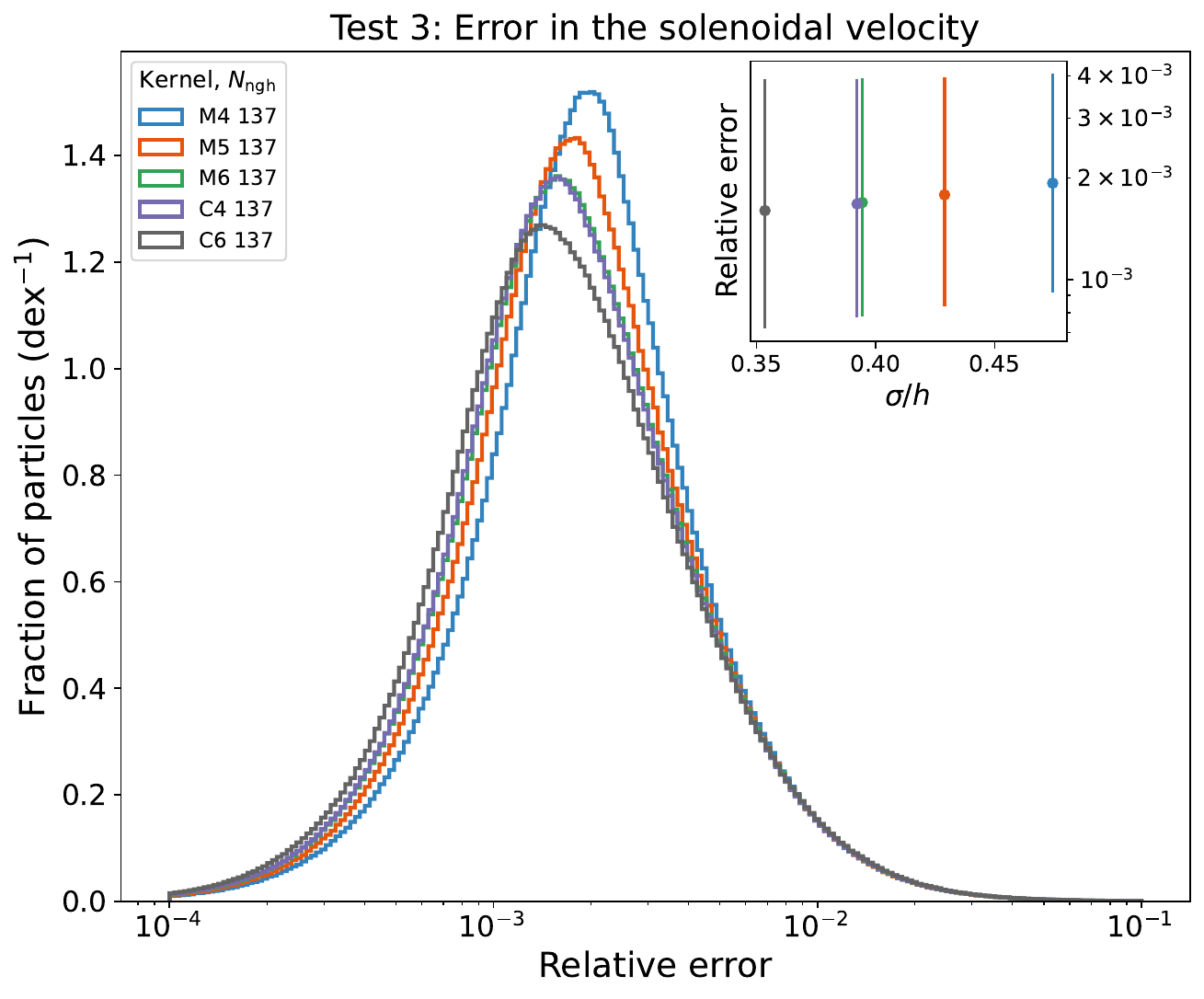}~
    \includegraphics[width=0.4\textwidth]{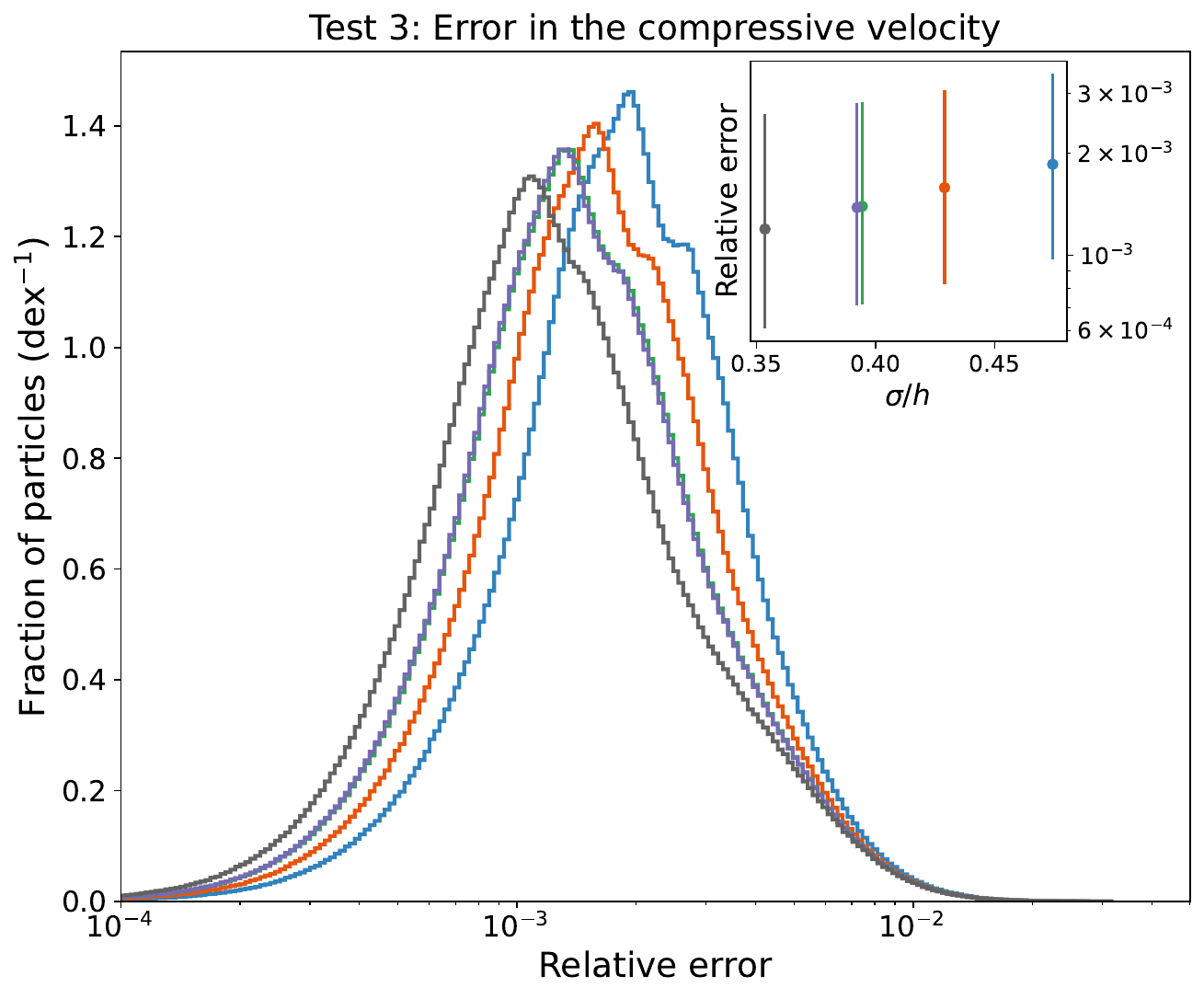}
    \includegraphics[width=0.4\textwidth]{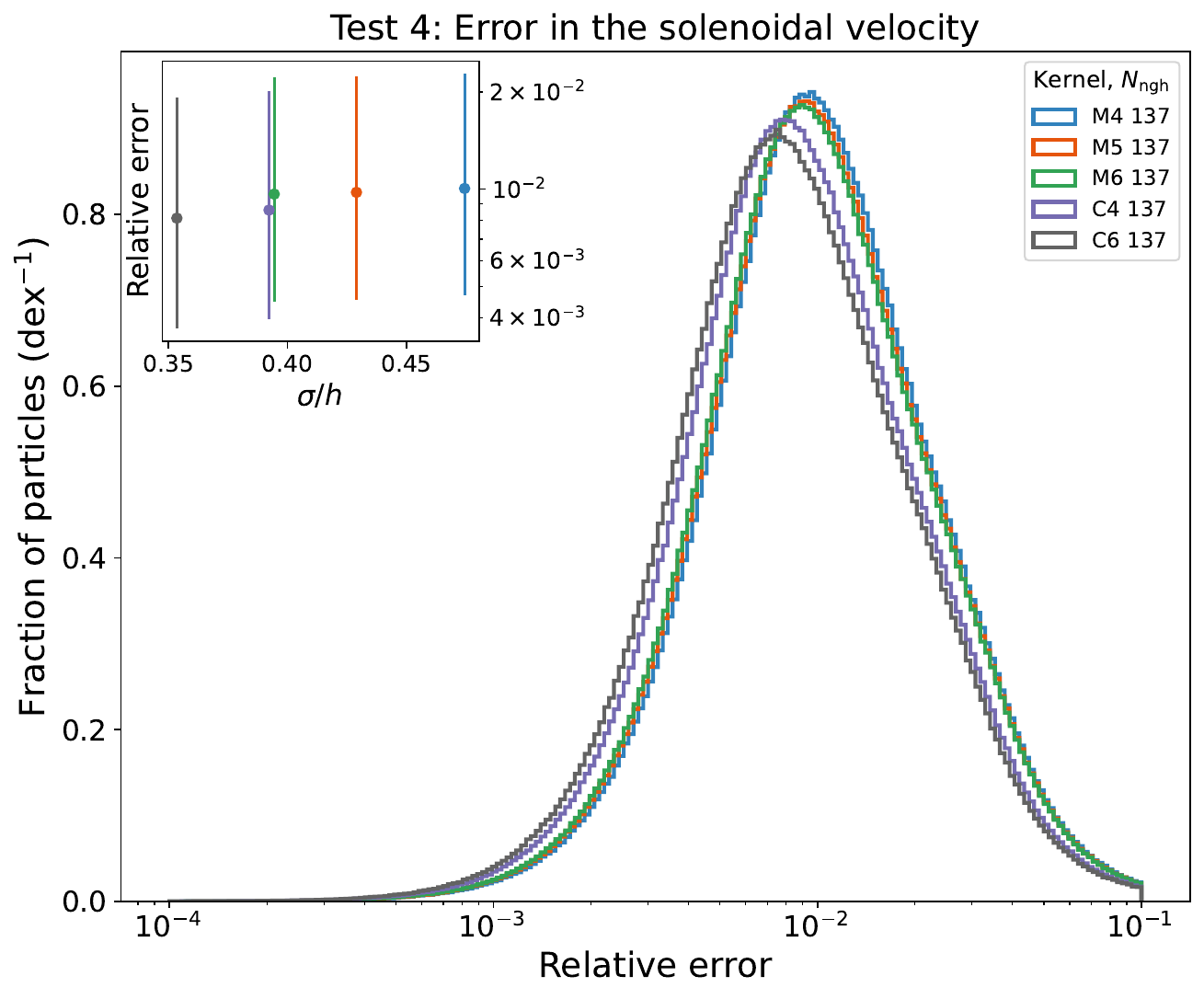}~
    \includegraphics[width=0.4\textwidth]{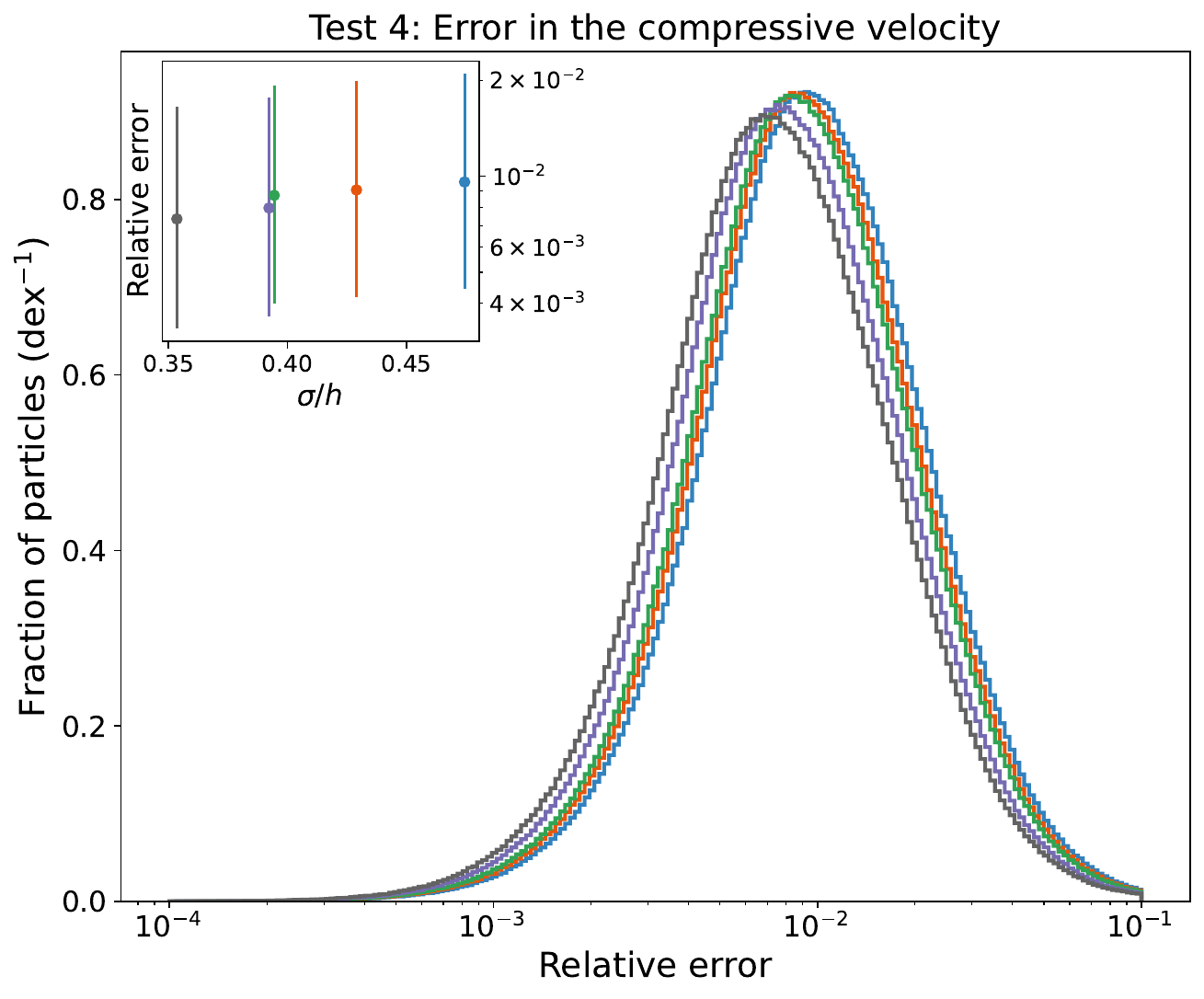}
    \caption{Results of Tests 1 to 4 (subsequent rows) at fix $N_\mathrm{ngh}=137$, when considering the five kernels implemented according to the colour legend. Each panel is equivalent to the ones in Figs. \ref{fig:test1_results}, \ref{fig:test2_results}, \ref{fig:test3_results} and \ref{fig:test4_results}. Lines represent the histogram of values of the relative errors, computed according to Eqs. \ref{eq:test1_error_absent} and \ref{eq:test1_error_present}. The insets shown the median and $(16-84)$ percentile of the error distributions as a functions of the effective width of the kernel, $\sigma / h$, i.e. the standard deviation of the kernel in units of its total compact support radius.}
    \label{fig:kM5M6}
\end{figure*}

In Sec. \ref{s:tests}, we have concentrated on the cubic spline (M$_4$) and the two C$^4$ and C$^6$ Wendland kernels, for the sake of a concise and clear discussion of the uncertainties introduced by the decomposition. It has been shown that higher-order kernels from the B-spline family, i.e. the quartic (M$_5$) and quintic (M$_6$) spline, outperform Wendland kernels as density estimators at a given number of neighbours, the latter needing large number of neighbours to converge \citep{Dehnen_2012}. Hence, it is worth showing the performance of these additional kernels in our battery of tests. This is shown in Fig. \ref{fig:kM5M6}, where we summarise the results of Tests 1 to 4, when \vortexp{} is run with each of the implemented kernels, at fix $N_\mathrm{ngh}=137$. Each panel contains equivalent information to those in Figs. \ref{fig:test1_results}, \ref{fig:test2_results}, \ref{fig:test3_results} and \ref{fig:test4_results}.

The insets show the median (and 16-84 percentiles) of these error distributions as a function of the effective width of the kernel, $\sigma / h$. Here, $\sigma$ is the standard deviation of the kernel, which we can compute as 

\begin{equation}
    \sigma = \frac{\int_0^R x^4 W(x)}{\int_0^R x^2 W(x)}
\end{equation}

\noindent and its values for all the kernels involved here are given, for instance, in table 1 of \cite{Price_2018}. On the other hand, note that here, by $h$ we are denoting the full compact support radius of the kernel, i.e. $W(h)=0$. Interestingly, the kernels added to this test (M$_5$, in orange lines; and M$_6$, in green lines) gently fall in the trends already outlined by the ones in Sec. \ref{s:tests}, with the only slightly peculiar case being that of Test 4, where the trends are still followed albeit with considerably more scatter.

This highlights that the effective width of the kernel appears to be the single parameter driving the differences amongst kernel choices at fix $N_\mathrm{ngh}$. This is not necessarily at odds with the findings of \citep{Dehnen_2012}. While the high-order B-splines have been shown to be better density estimators, the errors in the HHD are not dominated by the convergence of the density estimator, but instead by the cross-talk between compressive and solenoidal component introduced by a particles-to-grid smoothing scheme with non-constant smoothing length. In this regard, it seems that, in the complex (non-linear) Tests 3 and 4, the smaller the kernel effective width (and hence, the smaller the overlap and the more local the interpolation), the lesser amount of error appears in the decomposition. The magnitude of these differences, however, is in any case small (less than a factor of 2 between the broadest and narrowest kernel), but consistent.

\section{Further tests on the multi-scale filter}
\label{app:filter}

Here we present two additional considerations regarding the implementation of the multi-scale filter in \vortexp{} for performing an RD. \ref{app:filter.weight} discusses the differences in the resulting RD depending on whether the bulk velocity has been computed mass-weighted or volume-weighted. An assessment on the validity of the approximate scheme to flag strong shocks is presented in \ref{app:filter.flag}.

\subsection{Mass-weighted and volume-weighted bulk velocity determination}
\label{app:filter.weight}

\begin{figure*}
    \centering
    \includegraphics[width=0.5\textwidth]{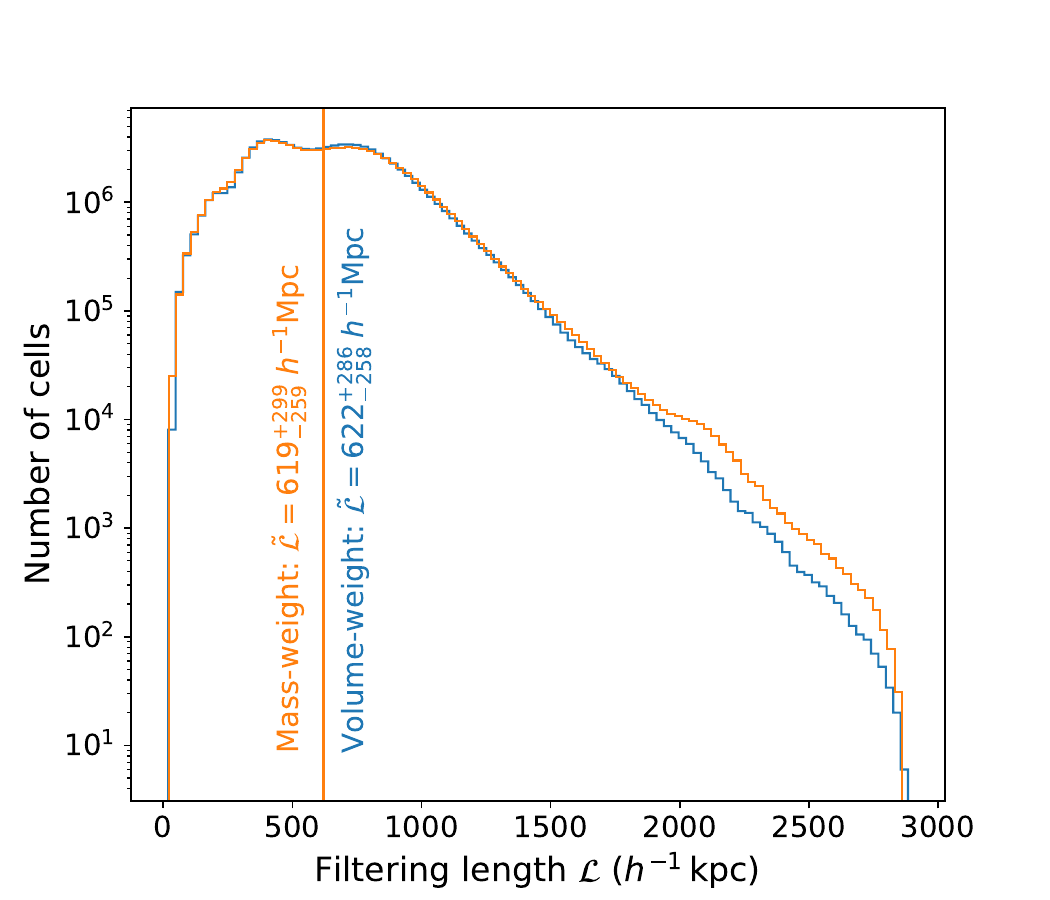}~
    \includegraphics[width=0.5\textwidth]{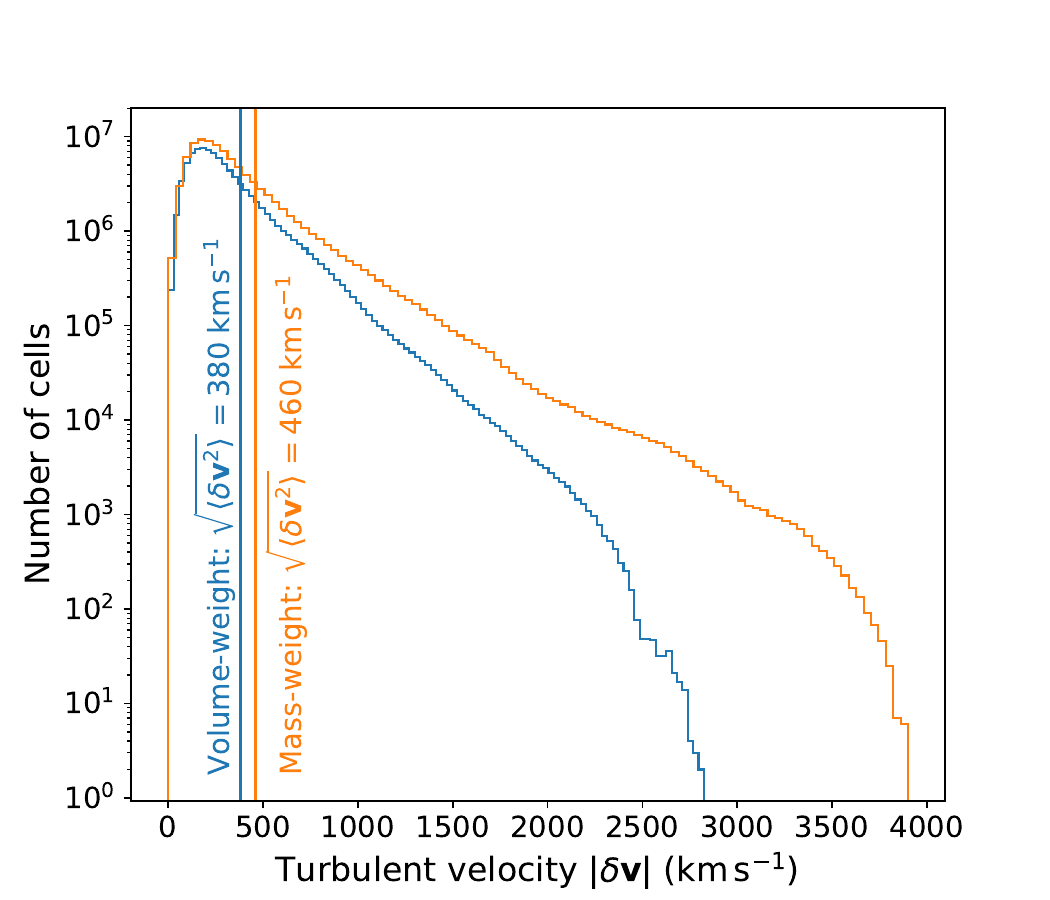}
    \caption{Comparison of the volume-weight (blue) and mass-weight (orange) for the bulk velocity computation in the iterative multi-scale filter. In both panels, the histogram represents the cell-wise distribution of the considered value. Vertical lines present summary statistics of these distributions. \textit{Left-hand side panel}: distribution of filtering lengths. \textit{Right-hand side panel}: distribution of turbulent velocity magnitudes.}
    \label{fig:app.filter.weight}
\end{figure*}

In \citet{Valles-Perez_2021_MNRAS}, we used gas density as the weight for the bulk velocity in the multi-scale filter, $w(\vb{x}) = \rho(\vb{x})$, in Eq. \ref{eq:bulk_velocity}. In \vortexp{}, there is the possibility to compute the bulk velocity around each cell as a volume-weighted or a mass-weighted average. It might thus be interesting to check to which extent the results on the RD from both schemes differ. Note that the election of weighting scheme affects, not only the final bulk velocity, but also the filtering lengths, $\mathcal{L}(\vb{x})$, since it can impact the convergence of the iterative algorithm.

We have run the multi-scale filter on the same simulation output shown in the left-hand side panel of Fig. \ref{fig:vortex_tirso}, using the volume-weighted and the mass-weighted scheme. The results are summarised in Fig. \ref{fig:app.filter.weight}, where blue histograms correspond to the volume-weighted bulk velocity, and orange histograms are for the mass-weighted case. The histograms are shown for a $5 \, h^{-1} \, \mathrm{Mpc}$ box around the cluster centre (roughly containing the virial volume), and histogram cell counts correspond to volume sampled at $12 \, h^{-1} \, \mathrm{kpc}$ resolution.

Regarding filtering length (left-hand side panel), both methods agree on the vast majority of cells, so that the median (and [16-84] percentiles) of $\mathcal{L}$ broadly coincide. However, there are notorious differences between these two schemes at the high $\mathcal{L}(\vb{x})$ tail (corresponding to low-density regions where the velocity field tends to be smoother). Nevertheless, this corresponds to a very small fraction of cells (note that counts are shown in logarithmic scale in the histograms). Regarding turbulent velocities (right-hand side panel), consistently with previous results, the peaks at small turbulent velocities (associated to small filtering lengths) coincide, while the mass-weighted scheme exhibits a higher number of high turbulent velocity cells, that consequently drive the root mean square (rms) turbulent velocity magnitude up by $\sim 20\%$. Consequently, even though the differences are not striking, care must be taken when considering the RD of inhomogeneous matter distributions, since the choice may be application dependent.

\subsection{Approximate scheme to flag strong shocks in absence of Mach number data}
\label{app:filter.flag}

The shock-identification scheme presented in Sec. \ref{s:method.shock} had, as its only purpose, to flag the volume elements hosting a strong shock, to act as an additional stopping condition for the iterations of the multi-scale filter. This is advisable, because when averaging velocities in between the pre-shock and the post-shock region, the velocity discontinuity may hinder convergence and thus produce nonphysically high filtering lengths. As discussed in Sec. \ref{s:method.shock}, our recommendation is to use a more accurate shock finder and feed this information to \vortexp{}, and only use this approximate shock-identification scheme as a fallback for the cases where it is not possible to provide Mach number information.

\begin{figure*}
    \centering
    \includegraphics[width=0.5\textwidth]{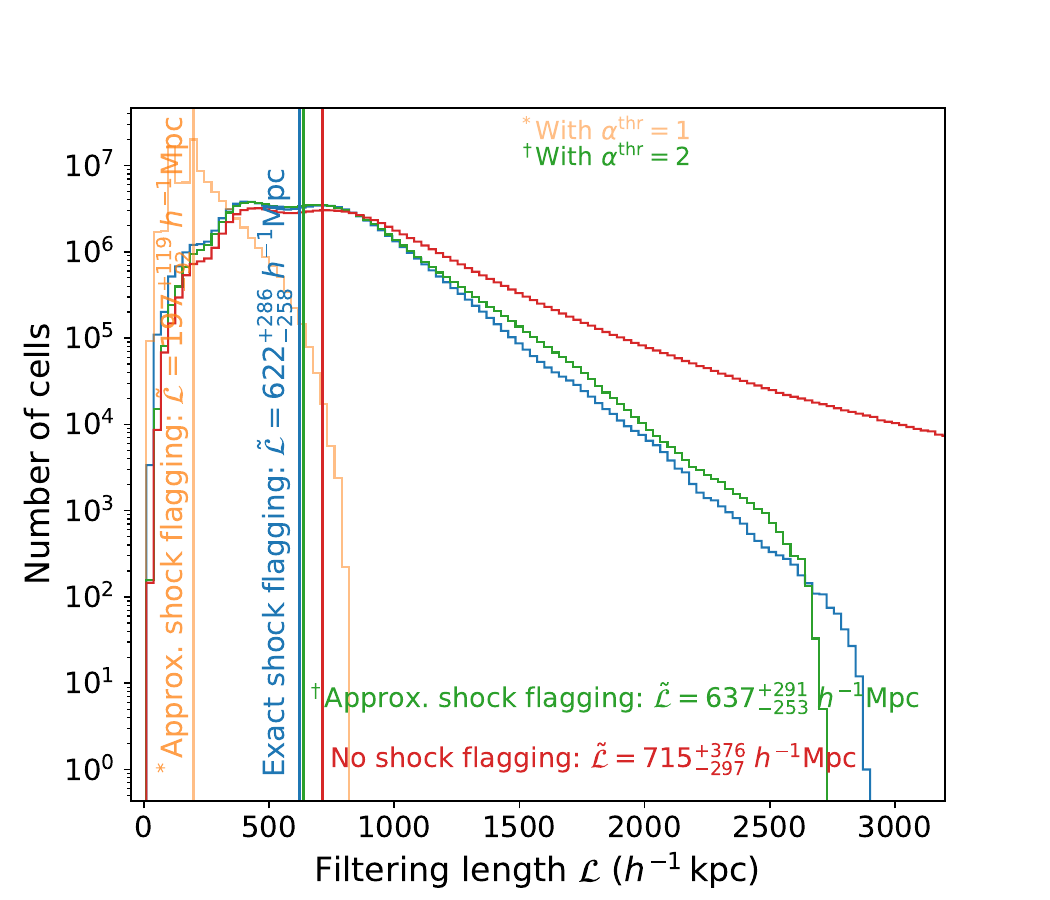}~
    \includegraphics[width=0.5\textwidth]{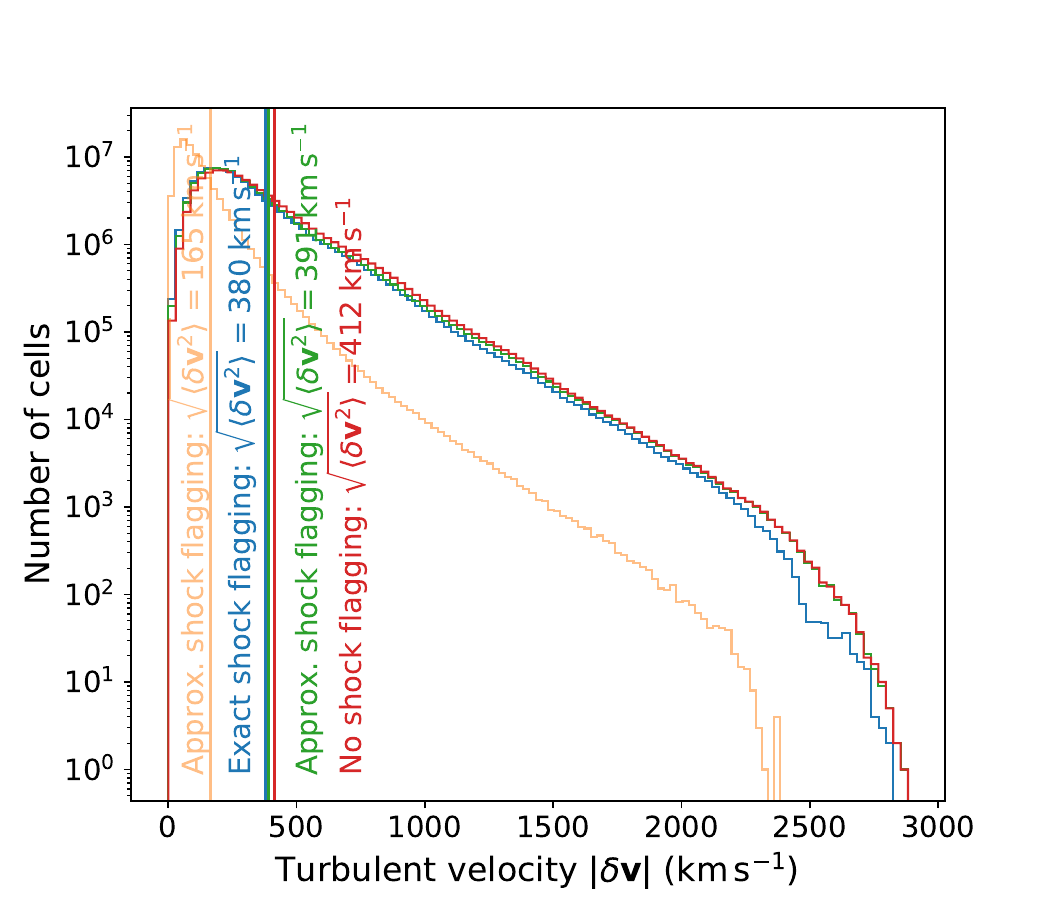}
    \caption{Comparison of the distribution of filtering lengths (left-hand side panel) and turbulent velocity magnitudes (right-hand side panel) as recovered by the multi-scale filter with the exact shock flagging scheme (blue), the approximate shock flagging scheme (orange and green, respectively, with $\alpha^\mathrm{thr}=1$ and $2$), and no stopping condition based on shocks (purple line).}
    \label{fig:app.filter.flag}
\end{figure*}

However, in this Appendix section we provide a simple analysis of the behaviour of this strategy of flagging strong shocks based on simple thresholds on velocity divergence and artificial viscosity, when applied to the multi-scale filter. In the results shown in the left panel of Fig. \ref{fig:vortex_tirso}, we had performed the RD by applying the multi-scale filter, with shocks identified based on a Mach number threshold (where Mach numbers had been identified in the simulation with the on-the-fly algorithm of \citet{Beck_2016_shock}, and hence this was input data for \vortexp{}). We have compared these results, where the multi-scale filter has used Mach numbers obtained from a robust shock finder, with the results of the multi-scale filter with our simplified scheme for flagging shocked volume.

This is precisely what we show in Fig. \ref{fig:app.filter.flag}. The elements of the figures are equivalent to those in Fig. \ref{fig:app.filter.weight}. Here, the blue histogram shows the distribution of filtering lengths (left-hand side panel) and turbulent velocity magnitudes (right-hand side panel) when applying the multi-scale filter with the exact shock-flagging approach (which corresponds to what is shown in Figs. \ref{fig:vortex_tirso} and \ref{fig:vortex_powerspectra}). For comparison, the orange and green lines correspond to the same quantities, as recovered by the algorithm using the approximate shock flagging. In both cases, we have set $(\nabla \cdot \vb{v})^\mathrm{thr} = - 1000 \, \mathrm{km \, s^{-1} \, Mpc^{-1}}$, consistent with typical values of the velocity divergence of galaxy cluster strong accretion shocks. The orange (green) histogram corresponds to a threshold on artificial viscosity $\alpha^\mathrm{thr} = 1$ ($\alpha^\mathrm{thr} = 2$).

While the test run with $\alpha^\mathrm{thr} = 1$ is unable of reproducing the distribution of filtering lengths and turbulent velocities, the test with $\alpha^\mathrm{thr} = 2$ does so, with mutually consistent median values of the filtering length and rms turbulent velocities. The striking similarity between the blue (exact) and green (approximate shock flagging) curves is a consequence of the fact that the shock limiter is acting on relatively few cases and, instead, the filtering length is more often determined by convergence of the turbulent velocity field with increasing filtering length (indicating that the outer scale of turbulence is reached).

However, even though the shock limiter might not be acting in the determination of the filtering length and turbulent velocity around most cells, it is still worth keeping it, since it prevents mixing pre- and post-shock velocities around strong shocks which can prevent or delay convergence in a (relatively small) number of cells. This is exemplified through the red lines in the two panels, which correspond to the multi-scale turbulent filter run with no additional stopping condition based on shocks. In this case, a tail with high $\mathcal{L}$ develops. While this does not bias significantly the turbulent kinetic energy budget (this occurs most typically in low-density regions where turbulent velocities are small, i.e. at the left-hand side of the right-hand side panel of Fig. \ref{fig:app.filter.flag}), it would impact the determination of other quantities such as the turbulent energy flux, $f_\mathrm{turb} \propto (\delta v)^3 / \mathcal{L}$.

\end{appendix}

\end{document}